# An Entropy-based Coefficient of Determination with Adjustment of Optimization Bias

Longhai Li[1,*]

[1]Department of Mathematics and Statistics, University of Saskatchewan

2026-08-05

**Abstract**

Classical likelihood-ratio tests and ΔAIC exacerbate the statistical significance crisis by scaling with sample size, routinely flagging negligible improvements as highly significant. While causal estimands like the average treatment effect (ATE) attempt to resolve this by quantifying practical magnitude, their reliance on the expectation operator inextricably ties them to the data's original coordinate scale. Furthermore, existing pseudo-$R^2$ metrics remain inadequate for this purpose: variance-based measures ignore higher-order distributional improvements, while both variance-based and entropy-based $R^2$ formulations lack invariance to monotone transformations. In this paper, we resolve these limitations by regarding Entropic Variance (EV) as a rigorous, scale-independent generalization of the error variance in ordinary least squares. We define the population EV-based parameter, $\rho_V^2$, which projects unbounded cross-entropy onto the standardized $[0, 1]$ scale, and establish that the EV-based $F_V$ statistic asymptotically follows an $F$-distribution. Building on the distributional properties of empirical EV, we propose two estimators: the empirical population $R^2_{\text{SV}}$ and the out-of-sample predictive $R^2_{\text{SVP}}$, derived by exponentiating the per-observation cross-entropy and incorporating a degrees-of-freedom correction for training optimism. Leveraging the $F_V$-distribution, we derive refined $p$-values and confidence intervals for $\rho_V^2$ without requiring intractable Fisher information matrices. Both simulation studies and a Parkinson's disease microbiome application demonstrate the superiority of conducting variable selection via these EV-$R^2$ metrics. For example, evaluating $R^2_{\text{SVP}}$ of a LASSO path via a data-splitting method reduces false discovery rates from 80% to 6% while preserving signal recall in a simulation study.

[*]**Corresponding author**: Longhai Li, Department of Mathematics and Statistics, University of Saskatchewan, 106 Wiggins Road, Saskatoon, S7N 5E6, Canada (Email: longhai.li@usask.ca)

**List of Abbreviations: CI** (Confidence Interval); **DGP** (Data Generating Process); **EV** (Entropic Variance); **EV-SNR** (Entropic Variance Signal To Noise Ratio); **FDR** (False Discovery Rates); **KL** (Kullback Leibler); **LOOCV** (Leave One Out Cross Validation); **LRT** (Likelihood Ratio Test); **MAE** (Mean Absolute Error); **MD** (Mahalanobis Distance); **RSS** (Residual Sum Of Squares); **SNR** (Signal To Noise Ratio); **SSR** (Sum Squares Reduction).

# 1 Introduction

The over-reliance on statistical significance to evaluate scientific findings has precipitated a well-documented reproducibility crisis in applied research (Billheimer, 2019; Wasserstein et al., 2019; Wasserstein & Lazar, 2016). Classical hypothesis testing procedures, such as likelihood ratio tests (LRTs), evaluate the probability of observing the data under a null model rather than quantifying the practical magnitude of an effect. Because the underlying non-centrality parameters of these tests increase with sample size, they routinely flag practically negligible improvements as highly "significant" in large datasets (Billheimer, 2019; Burnham & Anderson, 2002; Chatterjee, 2021; Guo & Shah, 2025; Lin et al., 2013; Vehtari et al., 2017; Wasserstein et al., 2019; Wasserstein & Lazar, 2016). Information criteria like AIC (Akaike, 1973) penalize model complexity. However, information criteria scale with the sample size; consequently, in massive cohorts, trivial deviations from the null can easily overcome the penalty term, leading to the endorsement of models with minimal practical utility (Burnham & Anderson, 2002). Indeed, as we elaborate in Sec. D, the drops in information criteria is dominated by the deviance, hence, they essentially function as measures of statistical significance rather than effect sizes.

The shift toward estimating causal effect sizes, such as the average treatment effect (ATE), exemplifies an attempt to resolve this statistical significance crisis. By quantifying differences in conditional means, researchers aim to capture practical importance rather than mere statistical evidence against a null. However, relying strictly on the expectation operator ($\mathbb{E}$) inextricably ties these estimands—and even recent distributional extensions Chernozhukov et al. (2023); Deuber et al. (2024)—to the data's original coordinate scale. Because they are vulnerable to non-linear transformations and lack a unified, dimensionless interpretation (Ding et al., 2019; Pepe et al., 2004), these metrics cannot serve as a universal standard. This underscores the persistent need for a generalized, dimensionless measure of effect size that bypasses $\mathbb{E}$ to quantify population-level magnitude independently of scale and sample size, while rigorously correcting for training optimism.

Historically, in ordinary least squares (OLS), the coefficient of determination ($R^2$) and its adjusted counterpart have fulfilled this role by quantifying the proportional reduction in residual variance. Efforts to generalize this interpretable metric to broader likelihood inference have largely bifurcated into variance-based and entropy-based paradigms. Variance-based extensions port the OLS formula directly by partitioning the response variance using predictive means and variance functions (Efron, 1978; Gelman et al., 2019). This includes adaptations for generalized linear and mixed models (Jaeger et al., 2017; Nakagawa et al., 2017; Rights & Sterba, 2019; Zhang, 2017; 2022) and domain-specific indices for binary outcomes (Tjur, 2009). However, their critical limitation is that they measure variation solely through the mean trajectory, leaving them insensitive to higher-order distributional improvements. A model capturing complex heteroscedasticity, zero-inflation, or heavy tails can sharply improve distributional accuracy even if the predictive mean remains unchanged. Furthermore, these metrics lack invariance under non-linear monotone transformations of the response, making the estimated effect size dependent on arbitrary coordinate scales (Clements & Hendry, 1993).

The second paradigm constructs pseudo-$R^2$ measures directly from the negative log-likelihood (**deviance, log-loss, or empirical cross-entropy**), but these encounter severe geometric and algebraic paradoxes in continuous domains. McFadden's $R^2$ (McFadden, 1974), defined as the proportional reduction in empirical entropy, breaks down for continuous data because differential entropy is not bounded below by zero. A sufficiently small scale parameter makes the baseline entropy negative, yielding metrics that wildly violate the $[0,1]$ bounds. The root problem is that this linear ratio of empirical cross-entropy lacks scale invariance: changing the units of measurement adds a scalar constant to the entropy, and this shifting baseline arbitrarily alters the ratio even though the absolute information gain is preserved. Attempts to avoid negativity, such as Cameron and Windmeijer's anchoring of $R^2$ to a saturated model (Cameron & Windmeijer, 1996), are ill-defined for continuous data beyond categorical groupings and still fail to resolve the lack of scale invariance. The Cox and Snell $R^2$ (Cox & Snell, 1989; Nagelkerke, 1991), which this paper generalizes, resolves scale invariance by evaluating the drop in exponential empirical cross-entropy. However, it relies on unpenalized empirical entropy, leaving it susceptible to finite-sample bias and overfitting without a rigorous inferential framework. These historical challenges stem from an insufficient characterization of finite-sample entropy distributions: the field has long mapped the residual sum of squares (RSS) to the deviance—a false equivalence motivated by Wilks' theorem and their shared asymptotic $\chi^2$ behavior.

This paper breaks from that equivalence to demonstrate that the **Entropic Variance** (EV), rather than entropy itself, provides the mathematically rigorous generalization of the OLS RSS for general regular parametric families. Building on this parallel, we define a EV-based metric of predictive effect size. Let $h^*$ denote the true population cross-entropy per observation. The population Entropic Variance is defined as $\nu^* = \exp(2h^*)$. The population effect size $\rho^2_{\mathrm{V}}$ is the proportional reduction in EVs:

$$\rho^2_{\mathrm{V}} = 1 - \frac{\nu^*_1}{\nu^*_0} = 1 - \exp(-2\Delta h^*), \tag{1}$$

where $\Delta h^* = h^*_0 - h^*_1$ is the cross-entropy difference between a restricted baseline model ($\mathcal{M}_0$) and a fuller model ($\mathcal{M}_1$). Within a continuous location-scale family, $\rho^2_{\mathrm{V}}$ is interpretable as the proportional reduction in squared kernel volumes and in the familiar conditional variances of the response. Crucially, unlike both variance and entropy $R^2$ metrics, it is strictly invariant to monotonic transformations of the response.

Direct substitution of the sample-based empirical EV, $\hat{v}_j = \exp(2\hat{h}_j)$, into Eq. (1) yields an optimistically biased estimator. To correct this finite-sample bias, we introduce a degrees-of-freedom adjustment $n - p$, directly analogous to the OLS residual variance correction, where $n$ and $p$ are sample size and parameter counts respectively. Normalizing the empirical EVs by their residual degrees of freedom gives our adjusted estimator:

$$R^2_{\mathrm{V}} = 1 - \frac{\hat{v}_1/(n-p_1)}{\hat{v}_0/(n-p_0)} = 1 - \left(\frac{n-p_0}{n-p_1}\right)\exp(-2\Delta\hat{h}). \tag{2}$$

We further extend this framework to out-of-sample prediction with $R^2_{\mathrm{VP}}$, an estimator that applies a squared complexity penalty. This arises from a refined characterization of the empirical Entropic Variance Signal-to-Noise Ratio (**EV-SNR**), $\widehat{\mathrm{SNR}}_{\mathrm{V}} = (\hat{v}_0 - \hat{v}_1)/\hat{v}_1 = \exp(2\Delta\hat{h}) - 1$. Building on Wilks' theorem and Bartlett's correction, we establish that the empirical SNR maps to a ratio of independent $\chi^2$ random variables, yielding a non-central $F$-distribution for $F_{\mathrm{V}}$:

$$F_{\mathrm{V}} \equiv \left(\frac{n-p_1}{k}\right)\widehat{\mathrm{SNR}}_{\mathrm{V}} \quad \dot{\sim} \quad F_{k,n-p_1}(\lambda_{\mathrm{V}}), \quad \text{with} \quad \lambda_{\mathrm{V}} = n\frac{\rho^2_{\mathrm{V}}}{1-\rho^2_{\mathrm{V}}}, \tag{3}$$

where $k = p_1 - p_0$ is the difference in parameter counts. Inverting this distribution provides a refined finite-sample $p$-values and continuous confidence intervals (CIs) for $\rho^2_{\mathrm{V}}$, successfully extending the inference machinery of OLS to broader regular parametric families.

The theoretical unification yields crucial new tools for both statistical inference and practical data analysis. On the inferential side, the generalized sampling distribution of the EV-SNR delivers more accurate $p$-values and superior Type I error control over the classical Wilksian $\chi^2$ approximation. While Bartlett corrections (Bartlett, 1937; Cordeiro, 1983; Lawley, 1956) and high-dimensional scaling limits (Sur et al., 2019) improve the finite-sample likelihood-ratio distribution, they depend on complex, model-specific Fisher information matrices. In contrast, the $F_{\text{V}}$ distribution applies universally across regular parametric models without requiring intractable analytic derivations.

Furthermore, this framework bridges the gap between abstract information theory and practical model evaluation. By exponentiating the per-observation cross-entropy, $R^2_{\text{V}}$ projects the unbounded log-likelihood onto a standardized $[0, 1]$ scale. This mapping elucidates the distinction between total and per-observation empirical entropy, yielding an interpretable quantification of predictive effect size. Consequently, it provides researchers with a robust new criterion for variable selection—one grounded in practical predictive utility rather than solely on classical statistical significance or with parameter shrinkage.

The remainder of the paper is organized as follows. Sec. 2 defines the population EV and $\rho^2_{\text{V}}$, detailing their geometric properties, invariance, and structural connections to OLS. Sec. 3 turns to empirical estimation, introducing the bias-corrected $R^2_{\text{V}}$ and predictive $R^2_{\text{VP}}$, and establishing the distribution of the empirical EV-SNR that underpins formal hypothesis tests and CIs. Sec. 4 validates these distributions and estimators, and evaluates the superiority of conducting variable selection via predictive effect size, followed by a real-world data application in Sec. 5. The *Supplementary Materials* contain asymptotic proofs, empirical comparisons with alternative metrics, and an elaborated recovery of classical OLS theory using the $R^2_{\text{V}}$ framework. We conclude in Sec. 6 with a discussion of future work.

# 2 Population $\rho^2$ based on Entropic Variance

## 2.1 Definitions of EV, Population $\rho^2_{\text{V}}$, and Predictive $\rho^2_{\text{VP}}$

Our predictive metrics are based on the cross-entropy between the true data-generating process (DGP), denoted $f^*$, and the parametric model family $\mathcal{M}_j$ with parameter $\boldsymbol{\theta}_j$. **To formally define these metrics, we adopt a consistent notational convention.** Throughout this paper, the superscript $*$ (e.g., $h^*(\cdot)$ or $v^*(\cdot)$) designates quantities evaluated in expectation with respect to the true DGP, $f^*$, whereas a circumflex $(\hat{\cdot})$ indicates their empirical counterparts calculated from the training data. The parameter vector $\boldsymbol{\theta}$ of the **evaluated predictive density** $f_j(y_i \mid \mathbf{x}_i, \boldsymbol{\theta})$ is included as the argument of these metric functions to indicate it is the metric of the evaluated model with parameter vector $\boldsymbol{\theta}$.

For a given parameter vector $\boldsymbol{\theta}$, we first define the **individual expected cross-entropy** function of $\boldsymbol{\theta}$ for model $\mathcal{M}_j$ given the $i$-th individual with $\mathbf{x}_i$ as: $h^*_{j,i}(\boldsymbol{\theta}; \mathbf{x}_i) = \mathbb{E}_{Y_i \sim f^*}\big[ -\log f_j(Y_i \mid \mathbf{x}_i, \boldsymbol{\theta}) \big]$. Correspondingly, we define the individual **EV** function of $\boldsymbol{\theta}$ given $\mathbf{x}_i$ as the exponentiated twice cross-entropy: $v^*_{j,i}(\boldsymbol{\theta}; \mathbf{x}_i) = \exp(2h^*_{j,i}(\boldsymbol{\theta}; \mathbf{x}_i))$. The expected mean-cross-entropy of the model for a given $\boldsymbol{\theta}$ is the mean of these individual entropies:

$$h^*_j(\boldsymbol{\theta}) = \frac{1}{n}\sum_{i=1}^{n} h^*_{j,i}(\boldsymbol{\theta}; \mathbf{x}_i). \tag{4}$$

The **EV function** $v^*_j(\boldsymbol{\theta})$ is the **geometric mean (GM)** of the individual EVs:

$$v^*_j(\boldsymbol{\theta}) = \exp(2h^*_j(\boldsymbol{\theta})) = \left(\prod_{i=1}^{n} v^*_{j,i}(\boldsymbol{\theta}; \mathbf{x}_i)\right)^{1/n} = \mathrm{GM}\left( \,[v^*_{j,i}(\boldsymbol{\theta}; \mathbf{x}_i)]_{i=1}^{n} \,\right), \tag{5}$$

where GM $([a_i]_{i=1}^{n})$ is the geometric mean of $a_1, ..., a_n$, used throughout this paper

We now define the population metrics at the **true** parameter vector—$\boldsymbol{\theta}^*_j$—the projected parameter vector of $\mathcal{M}_j$ that minimizes the total expected cross-entropy: $\boldsymbol{\theta}^*_j = \arg\min_{\boldsymbol{\theta}} h^*_j(\boldsymbol{\theta})$. Evaluating $h^*_{j,i}(\boldsymbol{\theta}^*_j)$, we obtain the **individual population cross-entropy** $h^*_{j,i}(\boldsymbol{\theta}^*_j)$. The **population EV** $\nu^*_j$ is defined as the EV function at $\boldsymbol{\theta}^*_j$:

$$\nu^*_j = v^*_j(\boldsymbol{\theta}^*_j) = \exp(2h^*_j(\boldsymbol{\theta}^*_j)) = \mathrm{GM}\left( \,[v^*_{j,i}(\boldsymbol{\theta}^*_j)]_{i=1}^{n} \,\right). \tag{6}$$

When comparing two nested models $\mathcal{M}_0 \subset \mathcal{M}_1$, we define the **population Entropic Variance Signal-to-Noise Ratio (pop. EV-SNR)** as the relative inflation of the restricted model's EV over the fuller model's EV:

$$\phi_{\mathrm{V}}^2 = \frac{\nu_0^* - \nu_1^*}{\nu_1^*} = \frac{\nu_0^*}{\nu_1^*} - 1 = \exp(2\Delta h^*) - 1, \tag{7}$$

where $\Delta h^* = h_0^*(\boldsymbol{\theta}_0^*) - h_1^*(\boldsymbol{\theta}_1^*)$. Analogous to the proportion of variance explained in OLS, the **EV-based population partial effect size** $\rho_{\mathrm{V}}^2$ is derived from this ratio as:

$$\rho_{\mathrm{V}}^2(\mathcal{M}_1 \mid \mathcal{M}_0) = \frac{\phi_{\mathrm{V}}^2}{1 + \phi_{\mathrm{V}}^2} = 1 - \frac{\nu_1^*}{\nu_0^*} = 1 - \exp(-2\Delta h^*). \tag{8}$$

As an illustration, applying Eq. (8) to OLS, we obtain

$$\rho_{\mathrm{V}}^2(\mathcal{M}_1 \mid \mathcal{M}_0) = 1 - \frac{\nu_1^*}{\nu_0^*} = 1 - \frac{\sigma_1^2}{\sigma_0^2}. \tag{9}$$

where $\sigma_j^2$ is the noise variance of model $\mathcal{M}_j$.

It is important to distinguish the population partial effect size $\rho_{\mathrm{V}}^2$ from the out-of-sample predictive partial effect size, denoted by $\rho_{\mathrm{VP}}^2$. While the population $\rho_{\mathrm{V}}^2$ is evaluated at the theoretical optimum $\boldsymbol{\theta}^*$, the predictive $\rho_{\mathrm{VP}}^2$ captures the out-of-sample predictive performance of a model fitting procedure. Plugging an estimated parameter $\hat{\boldsymbol{\theta}}_j$ into the population cross-entropy function gives the out-of-sample cross-entropy $h_j^*(\hat{\boldsymbol{\theta}}_j)$. Taking the expectation over the training dataset (or the estimator $\hat{\boldsymbol{\theta}}_j$), we obtain the expected predictive cross-entropy $\mathbb{E}[h_j^*(\hat{\boldsymbol{\theta}}_j)]$. Exponentiating this expected cross-entropy yields the expected predictive EV, $\exp(2\mathbb{E}[h_j^*(\hat{\boldsymbol{\theta}}_j)])$, from which we obtain the **expected out-of-sample predictive effect size** $\rho_{\mathrm{VP}}^2$:

$$\rho_{\mathrm{VP}}^2(\widehat{\mathcal{M}}_1 \mid \widehat{\mathcal{M}}_0) = 1 - \frac{\exp\Big(2\mathbb{E}\big[h_1^*(\hat{\boldsymbol{\theta}}_1)\big]\Big)}{\exp\Big(2\mathbb{E}\big[h_0^*(\hat{\boldsymbol{\theta}}_0)\big]\Big)}. \tag{10}$$

## 2.2 Interpretation via Equally Likely States

For a discrete response $Y$, entropic variance, $\nu(Y) = \exp(2h(Y))$, represents the squared *effective support size* (Hill, 1973; Jelinek et al., 1977; Jurafsky & Martin, 2009; Shannon, 1948). The base quantity, $N(Y) = \exp(h(Y))$, is the equivalent number of equally likely states in a uniform distribution that produce the same entropy. For instance, $v = 4$ yields $N(Y) = 2$, indicating uncertainty exactly equivalent to two equally likely outcomes. By letting $N_j(Y) = \exp(h(Y \mid \mathcal{M}_j))$ denote this effective number of states under model $\mathcal{M}_j$, the EV-based $\rho^2_{\mathrm{V}}$ can be expressed directly as:

$$\rho^2_{\mathrm{V}} = 1 - \frac{N_1^2(Y)}{N_0^2(Y)}. \tag{11}$$

This formulation demonstrates that $\rho^2_{\mathrm{V}}$ represents the proportional reduction in the *squared* effective numbers of states required to describe the response uncertainty after updating from the null model ($\mathcal{M}_0$) to the candidate model ($\mathcal{M}_1$). However, continuous differential entropy is scale-dependent. For continuous variables, $\nu(Y)$ does not count discrete categories but instead measures the *effective volume* of the probability density. We will show that within a location-scale regular family, $\rho^2_{\mathrm{V}}$ can be interpreted directly via the proportional reduction of squared kernel volumes and familiar regular variances.

## 2.3 Interpretation via Geometric Means of Kernel Volumes

The abstraction of EV $\nu_j^*(\mathbf{x}_i)$ finds a concrete geometric interpretation in the predictive density space with the squared volume of the kernel of the predictive density. Let $f_j(y \mid \mathbf{x}_i, \boldsymbol{\theta}_j^*)$ be an **absolutely continuous** predictive density from a regular location-scale family, and let $\mathrm{Vol}_j^{\mathrm{ker}}(\mathbf{x}_i, \boldsymbol{\theta}_j^*) = 1/\max_y f_j(y \mid \mathbf{x}_i, \boldsymbol{\theta}_j^*)$ denote the effective spatial volume of its standardized density kernel. Theorem S1 (Sec. A.1) shows that the true population EV $\nu_j^* = \exp(2h_j^*(\boldsymbol{\theta}_j^*))$ factors exactly into the squared geometric mean of the single-observation kernel volumes, scaled by a family-specific shape constant:

$$\nu_j^* = \mathrm{GM}^2\left( [\mathrm{Vol}_j^{\mathrm{ker}}(\mathbf{x}_i, \boldsymbol{\theta}_j^*)]_{i=1}^n \right) \cdot \mathrm{Vol}^2_{\mathrm{family}}, \tag{12}$$

where $\mathrm{Vol}_{\mathrm{family}}$ is completely invariant to the parameter vector $\boldsymbol{\theta}_j^*$ and $\mathrm{GM}^2(\mathbf{x})$ is the squared geometric mean of $\mathbf{x}$. With the constant $\mathrm{Vol}_{\mathrm{family}}$ cancelled out when comparing nested models of the same location-scale family, the population $\rho^2_{\mathrm{V}}$ simplifies to one minus the squared ratio of the geometric means of individual kernel volumes:

$$\rho_{\mathrm{V}}^2 = 1 - \frac{G_n^2\left(\,[\mathrm{Vol}_j^{\mathrm{ker}}(\mathbf{x}_i, \boldsymbol{\theta}_1^*)]_{i=1}^n\,\right)}{G_n^2\left(\,[\mathrm{Vol}_j^{\mathrm{ker}}(\mathbf{x}_i, \boldsymbol{\theta}_0^*)]_{i=1}^n\,\right)}. \tag{13}$$

## 2.4 Interpretation via Geometric Means of Regular Variances

For continuous response variables modeled within location-scale families, the entropic variance can translate directly into the coordinate space of classical regular variance. As formally proven by Theorem S2 in Sec. A.2, the individual EV factors exactly into the product of a family signature constant and the conditional regular variance, $\sigma_{j,i}^2 = \mathrm{Var}\,(Y_i \mid \mathbf{x}_i, \boldsymbol{\theta}_j^*)$: $v_{j,i}^*(\boldsymbol{\theta}_j^*; \mathbf{x}_i) = v_{\mathrm{family}} \cdot \sigma_{j,i}^2$. Substituting this factorization into the population EV definition (Eq. (6)), the constant scalar $v_{\mathrm{family}}$ factors out of the geometric mean:

$$\nu_j^* = v_{\mathrm{family}} \cdot \mathrm{GM}\left(\,[\sigma_{j,i}^2]_{i=1}^n\,\right). \tag{14}$$

When comparing two nested models $\mathcal{M}_0 \subset \mathcal{M}_1$ of the same location-scale family, the universal scaling effect $v_{\mathrm{family}}$ perfectly cancels out, hence, $\rho_{\mathrm{V}}^2$ simplifies to:

$$\rho_{\mathrm{V}}^2(\mathcal{M}_1 \mid \mathcal{M}_0) = 1 - \frac{\mathrm{GM}\left(\,[\sigma_{1,i}^2]_{i=1}^n\,\right)}{\mathrm{GM}\left(\,[\sigma_{0,i}^2]_{i=1}^n\,\right)}. \tag{15}$$

Eq. (15) establishes that **$\boldsymbol{\rho_{\mathrm{V}}^2}$ exactly measures the proportional reduction in the geometric mean of conditional regular variances**. Notably, for OLS, the geometric mean of $\sigma_{j,i}^2$ reduces to the constant noise variance $\sigma_j^2$ of each $y_i$.

## 2.5 Interpretation via Likelihood Mahalanobis Distance (MD)

In regular likelihood families, our EV-based metrics bridge cleanly with classical statistical geometry. As formally established in Theorem S4, when evaluating nested models where the restricted model $\mathcal{M}_0$ is a subset of the full model $\mathcal{M}_1$, the partial effect size is fundamentally driven by the geometric distance between their parameter spaces. Consider partitioning the full parameter vector as $\boldsymbol{\theta}_1 = (\boldsymbol{\psi}, \boldsymbol{\phi})$, where $\boldsymbol{\psi}$ represents the parameters of interest and $\boldsymbol{\phi}$ denotes the nuisance parameters. If the restricted model $\mathcal{M}_0$ is defined by the null constraint $\boldsymbol{\psi} = \mathbf{0}$, the true parameter vector under the full model takes the form $\boldsymbol{\theta}_1^* = (\boldsymbol{\psi}^*, \boldsymbol{\phi}^*)$. By leveraging the asymptotic covariance of $\hat{\boldsymbol{\theta}}_1$, the EV-based $\rho_{\mathrm{V}}^2$ simplifies to a function of the squared Mahalanobis distance within the restricted parameter space:

$$\rho_{\mathrm{V}}^2(\mathcal{M}_1 \mid \mathcal{M}_0) \approx 1 - \exp\Big(-(\boldsymbol{\psi}^*)^\top \Sigma_{\psi\psi}^{-1} \boldsymbol{\psi}^*\Big), \tag{16}$$

where $\Sigma_{\psi\psi}$ is the marginal asymptotic covariance of $\hat{\boldsymbol{\psi}}$, which is derived via the Schur complement of the inverse Fisher Information matrix. In classical likelihood theory, the scaled quadratic distance $n(\boldsymbol{\psi}^*)^\top \Sigma_{\psi\psi}^{-1} \boldsymbol{\psi}^*$ is the non-centrality parameter ($\lambda_{\mathrm{D}}$) of the likelihood ratio deviance. This equivalence is conceptually crucial, as it formally defines the specific relationship between the EV-based $\rho_{\mathrm{V}}^2$ and the non-centrality of LRT under contiguous local alternatives (Cox & Hinkley, 1974; Vaart, 1998). This interpretation enables the statistical test and confidence interval construction for $\rho_{\mathrm{V}}^2$ in large sample.

## 2.6 Discrete Standardization and $\rho_{\mathrm{SV}}^2$

For continuous outcomes, the EV is bounded below by zero. For discrete outcomes, a point-mass distribution achieves zero entropy ($h^* = 0$), yielding a minimum attainable EV of $e^0 = 1$. *Considering the minimum attainable EV is crucial in the situations where $v_{\mathrm{null}}$ of the null model is small, for example, the Bernolli model with extreme overall prevelance; see Sec. F for detailed discussion.*

We mathematically formalize this fundamental lower bound as $\nu_{\min}^*$, where $\nu_{\min}^* = 0$ for continuous outcomes and $\nu_{\min}^* = 1$ for discrete outcomes. This constant establishes the theoretical floor for any model's EV. This lower bound implies that, when comparing a candidate model $\mathcal{M}_1$ to a reference model $\mathcal{M}_0$ with EV $\nu_0^*$, the maximum attainable EV reduction is $\nu_0^* - \nu_{\min}^*$ rather than $\nu_0^*$. Consequently, the unadjusted $\rho_{\mathrm{V}}^2$ can never reach 1 for discrete outcomes. To restore the scale $[0, 1]$ we need to multiply a **discrete adjustment factor** anchored at the simpler $\mathcal{M}_0$:

$$c_0 = \frac{\nu_0^*}{\nu_0^* - \nu_{\min}^*} = \begin{cases} 1 & \text{if } Y \text{ is continuous} \\ \frac{\nu_0^*}{\nu_0^*-1} & \text{if } Y \text{ is discrete} \end{cases}. \tag{17}$$

The **standardized partial predictive effect size $\rho^2_{\mathrm{SV}}$** is written as:

$$\rho^2_{\mathrm{SV}}(\mathcal{M}_1 \mid \mathcal{M}_0) = c_0 \times \rho^2_{\mathrm{V}}(\mathcal{M}_1 \mid \mathcal{M}_0) = \frac{\nu_0^*}{\nu_0^* - \nu_{\min}^*} \cdot \left(1 - \frac{\nu_1^*}{\nu_0^*}\right) = \frac{\nu_0^* - \nu_1^*}{\nu_0^* - \nu_{\min}^*}. \tag{18}$$

## 2.7 Global $\rho^2_{\mathrm{SV}}$ and Multiplicative Property

For a sequence of nested models $\mathcal{M}_{\mathrm{null}} = \mathcal{M}_0 \subset \mathcal{M}_1 \subset \ldots \subset \mathcal{M}_J$, evaluating each model against the intercept-only null defines the **global predictive effect size**:

$$\rho^2_{\mathrm{V}}(\mathcal{M}_j) \equiv \rho^2_{\mathrm{V}}(\mathcal{M}_j \mid \mathcal{M}_{\mathrm{null}}) = \frac{\nu_{\mathrm{null}}^* - \nu_j^*}{\nu_{\mathrm{null}}^*}. \tag{19}$$

The discrete adjustment needs to be applied to the global $\rho^2_{\mathrm{V}}$:

$$\rho^2_{\mathrm{SV}}(\mathcal{M}_j) \equiv c_{\mathrm{null}} \times \rho^2_{\mathrm{V}}(\mathcal{M}_j) = \frac{\nu_{\mathrm{null}}^* - \nu_j^*}{\nu_{\mathrm{null}}^* - \nu_{\min}^*}. \tag{20}$$

The discrete adjustment to $\rho^2_{\mathrm{V}}(\mathcal{M}_j)$ results in the modification of the Cox-Snell $R^2$ by Nagelkerke (1991). In this paper, we extend this discrete normalization to partial $\rho^2_{\mathrm{SV}}$. With $\rho^2_{\mathrm{SV}}$, the proportion of unexplained standardized EV decomposes multiplicatively across the sequence regardless of whether the outcome is continuous or discrete:

$$1 - \rho^2_{\mathrm{SV}}(\mathcal{M}_J) = \prod_{j=1}^{J} \left[\, 1 - \rho^2_{\mathrm{SV}}(\mathcal{M}_j \mid \mathcal{M}_{j-1}) \,\right]. \tag{21}$$

This follows directly from the telescoping product: for discrete outcomes, $1 - \rho^2_{\mathrm{SV}}(\mathcal{M}_j \mid \mathcal{M}_{j-1}) = (\nu_j^* - 1)/(\nu_{j-1}^* - 1)$, so the product collapses to $(\nu_J^* - 1)/(\nu_{\mathrm{null}}^* - 1) = 1 - \rho^2_{\mathrm{SV}}(\mathcal{M}_J)$; for continuous outcomes the analogous telescoping holds with $\nu_{\min}^* = 0$.

## 2.8 Invariance to Monotone Transformations and Relabelling

Consider a strictly monotonic transformation $W = g(Y)$. The Jacobian shifts the cross-entropy by an additive constant $c = \mathbb{E}_{f^*}[\log|\, g'(Y)\, |]$, yielding $h_j^*(W) = h_j^*(Y) - c$. The EV absorbs this shift as a multiplicative scaling factor: $\nu_j^*(W) = \exp(-2c)\nu_j^*(Y)$. Because $\rho^2_{\mathrm{V}}$ is a ratio of EVs, this constant cancels perfectly:

$$\rho_{\mathrm{V}}^2(W) = 1 - \frac{\exp(-2c)\nu_1^*(Y)}{\exp(-2c)\nu_0^*(Y)} = \rho_{\mathrm{V}}^2(Y). \tag{22}$$

Note that the geometric mean interpretation (Eq. (15)) cannot explain this invariance, as non-linear transformations alter the parametric family. Crucially, neither entropy- nor variance-based metrics share this invariance:

- **Entropy-based $\boldsymbol{\rho_{\mathrm{H}}^2}$** fails because the shift alters the baseline denominator:

$$\rho_{\mathrm{H}}^2(W) = \frac{\Delta h^*(Y)}{h_0^*(Y) - c} \neq \rho_{\mathrm{H}}^2(Y). \tag{23}$$

- **Variance-based $\boldsymbol{\rho_{\text{variance}}^2 = 1 - \frac{\sum \sigma_{1,i}^2}{\sum \sigma_{0,i}^2}}$** fails because non-linear transformations induce observation-specific factors in the arithmetic sum (see Sec. E.3).

For discrete outcomes, EV is strictly invariant to category relabelling. Let $W = g(Y)$ be any bijection on the support of $Y$. Because discrete log-likelihoods depend only on assigned probability masses rather than numeric labels, no Jacobian arises. Thus, $h_j^*(W) = h_j^*(Y)$ and $\nu_j^*(W) = \nu_j^*(Y)$, ensuring $\rho_{\mathrm{SV}}^2(W) = \rho_{\mathrm{SV}}^2(Y)$.

# 3 Estimators $R_{\mathrm{SV}}^2$ and $R_{\mathrm{SVP}}^2$ and Statistical Inferences

## 3.1 Empirical Entropy and Entropic Variance

Consider $n$ independent observations $(y_i, \mathbf{x}_i)$ and a sequence of nested parametric models $\mathcal{M}_0 \subset \ldots \subset \mathcal{M}_J$. Each $\mathcal{M}_j$ specifies a conditional predictive density $f_j(y_i \mid \mathbf{x}_i, \boldsymbol{\theta}_j)$. For a given parameter vector $\boldsymbol{\theta}_j$, the **empirical total entropy**, $\hat{H}_j(\boldsymbol{\theta}_j)$, is the negative log-likelihood, and the **average empirical entropy**, $\hat{h}_j(\boldsymbol{\theta}_j)$, is its per-observation mean:

$$\hat{H}_j(\boldsymbol{\theta}_j) = -\sum_{i=1}^{n} \log f_j(y_i \mid \mathbf{x}_i, \boldsymbol{\theta}_j), \quad \hat{h}_j(\boldsymbol{\theta}_j) = \frac{\hat{H}_j(\boldsymbol{\theta}_j)}{n}. \tag{24}$$

Our EV-based predictive effect size rely on the **empirical EV per observation**, $\hat{v}_j(\boldsymbol{\theta}_j)$, and **the empirical total EV** of $n$ observations, $\hat{V}_j(\boldsymbol{\theta}_j)$, which are defined as follows:

$$\hat{v}_j(\boldsymbol{\theta}_j) = \exp\Big(2\hat{h}_j(\boldsymbol{\theta}_j)\Big), \quad \hat{V}_j(\boldsymbol{\theta}_j) = n \exp\left(\frac{2\hat{H}_j(\boldsymbol{\theta}_j)}{n}\right). \tag{25}$$

For notational convenience, when the parameter argument is omitted, the empirical EV is implicitly evaluated at the MLE estimates; that is, $\hat{v}_j \equiv \hat{v}_j(\hat{\boldsymbol{\theta}}_j)$.

The empirical total EV of regular likelihood families exactly generalizes the RSS of OLS. Table 1 outlines various versions of EV and $R^2$ alongside their couterparts in OLS, aiming to elucidate the formal definitions and theorems in the subsequent text.

Table 1: Correspondance of EVs of Regular Families and $\sigma^2$ of OLS

| | **Pop. EV** | **Emp. EV** | **Emp. SNR** | **Adj. EV** | $1-R^2_{\text{V}}$ | **Test EV** | **Est. Test EV** | $1-R^2_{\text{VP}}$ |
|---|---|---|---|---|---|---|---|---|
| **Sym.** | $\nu^*$ | $\hat{v}$ | $\widehat{\text{SNR}}_{\text{V}}$ | $\hat{v}^c$ | $\frac{\hat{v}_1^c}{\hat{v}_0^c}$ | $v^{\text{test}}$ | $\hat{v}^{\text{test}}$ | $\frac{\hat{v}_1^{\text{test}}}{\hat{v}_0^{\text{test}}}$ |
| **OLS** | $\sigma^2$ | $\hat{\sigma}^2 = \frac{\text{RSS}}{n}$ | $\frac{\text{RSS}_0 - \text{RSS}_1}{\text{RSS}_1}$ | $\frac{n\hat{\sigma}^2}{n-p}$ | $\frac{n-p_0}{n-p_1}\frac{\hat{\sigma}_1^2}{\hat{\sigma}_0^2}$ | MSPE | $\frac{n^2\hat{\sigma}^2}{(n-p)^2}$ | $\frac{(n-p_0)^2}{(n-p_1)^2}\frac{\hat{\sigma}_1^2}{\hat{\sigma}_0^2}$ |
| **RF** | $\exp(2h^*)$ | $\exp(2\hat{h})$ | $\frac{\hat{v}_0 - \hat{v}_1}{\hat{v}_1}$ | $\frac{n\hat{v}}{n-p}$ | $\frac{n-p_0}{n-p_1}\frac{\hat{v}_1}{\hat{v}_0}$ | $\exp(2\mathbb{E}(h^*(\hat{\boldsymbol{\theta}})))$ | $\frac{n^2\hat{v}}{(n-p)^2}$ | $\frac{(n-p_0)^2}{(n-p_1)^2}\frac{\hat{v}_1}{\hat{v}_0}$ |

Fig. 1 illustrates EV using a Bernoulli intercept-only model ($\theta = 0.4$). Panel (a) compares the population EV curve, $\nu^*(\theta)$ (blue), against 10 empirical training curves, $\hat{v}(\theta)$ (dashed).

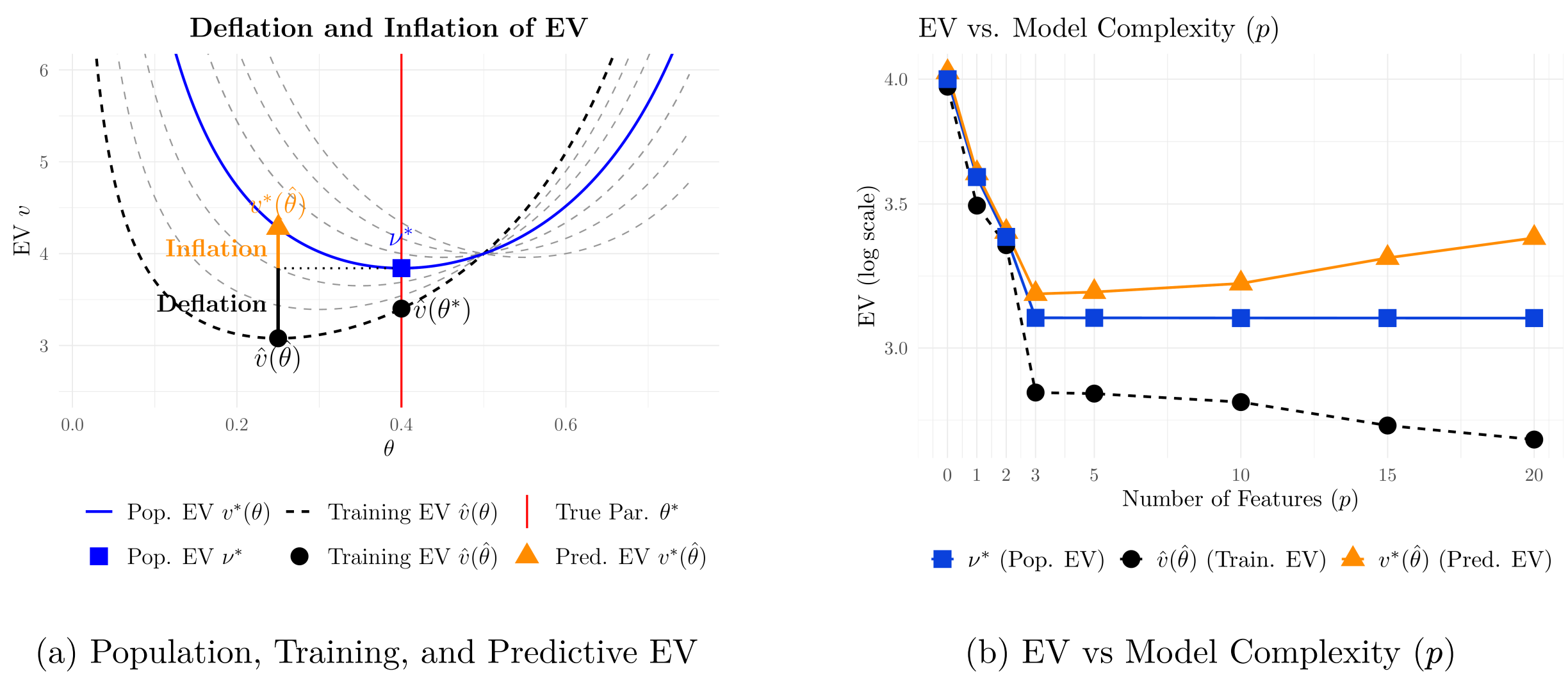

(a) Population, Training, and Predictive EV

(b) EV vs Model Complexity $(p)$

Figure 1: Illustration of the Population, Training, and Predictive EV.

An extreme sample (black) highlights the deflation of training EV and corresponding inflation of the out-of-sample predictive EV, $v^*(\hat{\boldsymbol{\theta}})$. Panel (b) tracks a single dataset ($n = 200$) as 20 covariates (only three informative) are sequentially added, demonstrating how superfluous predictors artificially deflate training variance while severely inflating predictive variance above the population minimum, $\nu^*$.

### 3.2 The Central EV-SNR Distribution

A fundamental challenge in extending exact finite-sample inference beyond OLS is that the precise $\chi^2$ distribution of the residual sum of squares (RSS) does not straightforwardly generalize to regular likelihood inference. As is widely documented in the generalized linear models (GLM) literature (McCullagh & Nelder, 1989), the finite-sample distributions of empirical entropy (negative log-likelihood) are typically analytically intractable. This paper addresses this limitation by demonstrating that the distributional properties of the classical **signal-to-noise ratio** in OLS can be broadly extended to the geometry of generalized variances (EVs) across regular likelihood models.

The empirical EV-SNR, denoted as $\widehat{\text{SNR}}_{\text{V}}$, serves as a critical statistic for comparing a restricted model $\mathcal{M}_0$ to a nested, fuller model $\mathcal{M}_1$. Generalizing the OLS signal-to-noise ratio, it quantifies the relative gain in information volume. The empirical EV-SNR is also an exponential function of the familiar **deviance**, $D_{1\setminus 0}$, the reduction in total empirical entropy between a restricted model $\mathcal{M}_0$ and a nested model $\mathcal{M}_1$:

$$D_{1\setminus 0} = 2\Big(\hat{H}_0(\hat{\boldsymbol{\theta}}_0) - \hat{H}_1(\hat{\boldsymbol{\theta}}_1)\Big) = 2n\Delta\hat{h} = -2\log(\Lambda), \tag{26}$$

where $\Lambda$ is the traditional LRT statistic. $\widehat{\text{SNR}}_{\text{V}}$ can be equivalently expressed as follows:

$$\widehat{\text{SNR}}_{\text{V}} = \frac{\hat{v}_0 - \hat{v}_1}{\hat{v}_1} = \exp\Big(2\Delta\hat{h}\Big) - 1 = \exp\left(\frac{D_{1\setminus 0}}{n}\right) - 1. \tag{27}$$

To establish the statistical significance of the fuller model, the finite-sample behavior of this empirical SNR under the null hypothesis must be characterized. Leveraging the aforementioned geometric correspondence, we extend the exact finite-sample mechanics of the OLS $F$-test to all regular likelihood families. Let $k = p_1 - p_0$ denote the difference in the number of estimated parameters between $\mathcal{M}_1$ and $\mathcal{M}_0$. Under the null hypothesis, which posits that a true population signal of $\phi_{\mathrm{V}}^2 = 0$, Theorem S6 (Sec. B.3) proves that the empirical SNR admits the following asymptotic approximation:

$$\widehat{\mathrm{SNR}}_{\mathrm{V}} \quad \dot{\sim} \quad \frac{\chi_k^2}{\chi_{n-p_1}^2}. \tag{28}$$

Scaling $\widehat{\mathrm{SNR}}_{\mathrm{V}}$ by the appropriate degrees of freedom yields an equivalent central $F$-distribution for the EV-based $F$-statistic:

$$F_{\mathrm{V}} = \left(\frac{n-p_1}{k}\right)\widehat{\mathrm{SNR}}_{\mathrm{V}} \quad \dot{\sim} \quad F_{k,n-p_1}. \tag{29}$$

As $n \to \infty$, applying a first-order Taylor expansion demonstrates that $n\widehat{\mathrm{SNR}}_{\mathrm{V}} = n\left[\exp\left(\frac{D_{1\setminus 0}}{n}\right) - 1\right] \approx D_{1\setminus 0}$. Because the $\chi_{n-p_1}^2/n \xrightarrow{p} 1$, the approximating distribution of $n\widehat{\mathrm{SNR}}_{\mathrm{V}}$ transferred from Eq. (28) converges in distribution to $\chi_k^2$. This naturally recovers the classical asymptotic limit established by Wilks's theorem for the deviance.

### 3.3 The $R_{\mathrm{V}}^2$ and $R_{\mathrm{SV}}^2$ Statistic and $F_{\mathrm{V}}$ Test for $\rho_{\mathrm{V}}^2$

Analogous to the downward bias of the maximum likelihood estimator (MLE) for error variance in OLS, the empirical EV is biased relative to the population variance, $\nu^*$. As detailed in Theorem S7 (Sec. B.4), the expectation of the empirical EV is attenuated due to evaluating the model on the data used for parameter estimation: $\mathbb{E}[\hat{v}(\hat{\boldsymbol{\theta}})] \approx \nu^*\left(1 - \frac{p}{n}\right)$.

In OLS, substituting $n\hat{v} = 2\pi e$ RSS and $\nu^* = 2\pi e\sigma^2$ recovers the distribution RSS $/\sigma^2 \sim \chi_{n-p}^2$. Taking the expectation yields the finite-sample bias of the variance MLE: $\mathbb{E}[\hat{\sigma}^2] = \sigma^2(1 - p/n)$. Extending this logic to regular likelihood models, we mitigate this bias by introducing a **bias-corrected EV**, $\hat{v}^c$. This applies a degrees-of-freedom penalty to the total empirical EV, $\hat{V}_j = n\hat{v}_j$. For a model $\mathcal{M}_j$ with $p_j$ parameters, the corrected variance is:

$$\hat{v}_j^c = \frac{\hat{V}_j}{n - p_j} = \frac{\hat{v}_j}{1 - p_j/n}. \tag{30}$$

The **adjusted EV coefficient of determination**, $R_{\mathrm{V}}^2$, quantifies the proportional reduction in this bias-corrected EV. It serves as an unbiased estimator for $\rho_{\mathrm{V}}^2$:

$$R_{\mathrm{V}}^2 = 1 - \frac{\hat{v}_1^c}{\hat{v}_0^c} = 1 - \left(\frac{n - p_0}{n - p_1}\right) \exp(-2\Delta\hat{h}). \tag{31}$$

For discrete response, $R_{\mathrm{V}}^2$ is scaled by a discrete adjustment factor $\hat{c}_0 = \hat{v}_0^c/(\hat{v}_0^c - 1)$ (where $\hat{c}_0 = 1$ for continuous outcomes). The **empirical standardized EV coefficient of determination**, $R_{\mathrm{SV}}^2$, is thus formulated as:

$$R_{\mathrm{SV}}^2 = \hat{c}_0 R_{\mathrm{V}}^2. \tag{32}$$

To test the null hypothesis $H_0 : \rho_{\mathrm{V}}^2 = 0$, we utilize the central distribution of the empirical SNR established in Eq. (29). The EV-based $F$-statistic, $F_{\mathrm{V}}^{(\mathrm{obs})}$, is formulated as: $F_{\mathrm{V}}^{(\mathrm{obs})} = \frac{n-p_1}{k}\left[\exp(2\Delta\hat{h}) - 1\right]$. The significance of the covariates' predictive contribution is evaluated directly against the central $F$-distribution, with the $p$-value given by $P(F_{k,n-p_1} > F_{\mathrm{V}}^{(\mathrm{obs})})$. This provides a finite-sample correction to the likelihood ratio test (LRT). By incorporating the variance of the full model into its denominator, the $F_{\mathrm{V}}$ statistic is better-calibrated, particularly when the sample size $n$ is small or the parameter dimension $p$ is large.

### 3.4 CIs for $\rho_{\mathrm{V}}^2$ and $\rho_{\mathrm{SV}}^2$

Under the alternative hypothesis (i.e., when the restricted model is misspecified), the population EV-SNR is strictly positive: $\phi_{\mathrm{V}}^2 = \frac{\rho_{\mathrm{V}}^2}{1-\rho_{\mathrm{V}}^2} > 0$. Let the non-centrality parameter denoted by $\lambda_{\mathrm{V}} = n\phi_{\mathrm{V}}^2$. As established in Theorem S9, asymptotically the empirical EV SNR follows a non-central distribution, hence, $F_{\mathrm{V}}$ follows the non-central $F$ distribution:

$$F_{\mathrm{V}} \quad \dot{\sim} \quad F_{k,n-p_1}(n\phi_{\mathrm{V}}^2). \tag{33}$$

This structural formulation directly generalizes classical OLS theory. Substituting the OLS variance components into $\lambda_{\mathrm{V}}$ exactly recovers the standard ANOVA non-centrality parameter, $\lambda = n(\sigma_0^2 - \sigma_1^2)/\sigma_1^2$. Beyond hypothesis testing, a $100(1-\alpha)\%$ CI for $\rho_{\mathrm{V}}^2$ can be constructed by numerically inverting this sampling distribution. Because an increase in $\phi_{\mathrm{V}}^2$ shifts the probability mass of the non-central distribution to the right, the cumulative tail probability—evaluated at a fixed observed statistic $F_{\mathrm{V}}^{(\mathrm{obs})}$—is strictly monotonically decreasing with respect to $\phi_{\mathrm{V}}^2$. This monotonicity guarantees the existence of unique and stable roots during numerical inversion. The inversion procedure first determines the valid bounds $[\hat{\phi}_{\mathrm{L}}^2, \hat{\phi}_{\mathrm{U}}^2]$ for the population SNR $\phi_{\mathrm{V}}^2$:

$$\begin{aligned} \hat{\phi}_{\mathrm{U}}^2 &= \mathrm{argmax}_{\phi_{\mathrm{V}}^2} \left\{ \phi_{\mathrm{V}}^2 : P\Big(F_{k,n-p_1}(n\phi_{\mathrm{V}}^2) \leq F_{\mathrm{V}}^{(\mathrm{obs})}\Big) \geq \frac{\alpha}{2} \right\}, \\ \hat{\phi}_{\mathrm{L}}^2 &= \mathrm{argmin}_{\phi_{\mathrm{V}}^2 \geq 0} \left\{ \phi_{\mathrm{V}}^2 : P\Big(F_{k,n-p_1}(n\phi_{\mathrm{V}}^2) \geq F_{\mathrm{V}}^{(\mathrm{obs})}\Big) \geq \frac{\alpha}{2} \right\}. \end{aligned} \tag{34}$$

The bounds in Eq. (34) are monotonically transformed to the CI for $\rho_{\mathrm{V}}^2$:

$$\mathrm{CI}\ (\rho_{\mathrm{V}}^2) = \left[ \frac{\hat{\phi}_{\mathrm{L}}^2}{1+\hat{\phi}_{\mathrm{L}}^2}, \frac{\hat{\phi}_{\mathrm{U}}^2}{1+\hat{\phi}_{\mathrm{U}}^2} \right]. \tag{35}$$

For discrete outcomes, the CI for $\rho_{\mathrm{SV}}^2$ follows immediately by applying the estimated discrete adjustment factor $\hat{c}_0 = \hat{v}_0/(\hat{v}_0 - 1)$ to each bound:

$$\mathrm{CI}\ (\rho_{\mathrm{SV}}^2) = \left[ \hat{c}_0 \cdot \frac{\hat{\phi}_{\mathrm{L}}^2}{1+\hat{\phi}_{\mathrm{L}}^2}, \min\left(1, \hat{c}_0 \cdot \frac{\hat{\phi}_{\mathrm{U}}^2}{1+\hat{\phi}_{\mathrm{U}}^2}\right) \right]. \tag{36}$$

### 3.5 $R_{\mathrm{SVP}}^2$ for Estimating Out-of-sample Predictive Effect Size

While $R_{\mathrm{V}}^2$ targets the population effect size $\rho_{\mathrm{V}}^2$, practitioners deploying models are fundamentally concerned with true out-of-sample prediction. A model's genuine performance on unseen data faces a penalty corresponding to the estimation bias. To establish a target for bias correction, we evaluate the fitted model over an independent test sample $Y^{\mathrm{test}}$ generated from the exact same underlying process $f^*$. Let $\hat{\boldsymbol{\theta}}$ be the vector of parameters estimated from the training data. The **predictive cross-entropy** $h^*(\hat{\boldsymbol{\theta}})$ (Eq. (4)) evaluates the fitted model's expected entropy over the unobserved test data $Y^{\mathrm{test}}$; since it is itself random through $\hat{\boldsymbol{\theta}}$, we define the **expected predictive test EV** by exponentiating its *average* over the sampling distribution of the training estimates: $v^{\text{test}} = \exp\left(\, 2\,\mathbb{E}_{\hat{\boldsymbol{\theta}}}[h^*(\hat{\boldsymbol{\theta}})]\,\right)$. Theorem S8 establishes that, for regular likelihood families, this predictive test EV suffers from a finite-sample inflation — a squared inflation factor relative to the expected empirical training EV: $v^{\text{test}} \approx \frac{\mathbb{E}[\hat{v}(\hat{\boldsymbol{\theta}})]}{(1-p/n)^2}$. The squared exponent reflects that the predictive target must correct for two compounding effects: the downward bias of the training fit and the upward inflation of out-of-sample prediction. Applying this squared penalty to the empirical training variance yields an estimator for the predictive test EV:

$$\hat{v}_j^{\text{test}} = \frac{\hat{v}_j}{(1 - p_j/n)^2}. \tag{37}$$

We define $R^2_{\text{VP}}$ as the proportional reduction in the estimated predictive test EV:

$$R^2_{\text{VP}} = 1 - \frac{\hat{v}_1^{\text{test}}}{\hat{v}_0^{\text{test}}} = 1 - \left(\frac{n - p_0}{n - p_1}\right)^2 \exp(-2\Delta\hat{h}). \tag{38}$$

Because $R^2_{\text{VP}}$ accounts for both estimation optimism and deployment uncertainty, it imposes a stricter complexity penalty than $R^2_{\text{V}}$. As with $R^2_{\text{SV}}$, for a discrete response we apply the discreteness adjustment $\hat{c}_0^{\text{test}} = \hat{v}_0^{\text{test}}/(\hat{v}_0^{\text{test}} - 1)$ (with $\hat{c}_0^{\text{test}} = 1$ for a continuous response) to define the standardized predictive EV $R^2_{\text{SVP}}$:

$$R^2_{\text{SVP}} = \hat{c}_0^{\text{test}} R^2_{\text{VP}}. \tag{39}$$

# 4 Simulation Studies

## 4.1 Logistic Regression with No Pre-selection

To evaluate the EV-based $R^2_{\text{SV}}$ metrics, we use a logistic regression data generating process (DGP), in which binary responses $Y_i \in \{0, 1\}$ are generated from standard normal covariates via the logit link:

$$\text{logit}\ (\pi_i) = \beta_0 + \sum_{j=1}^{3} \beta_j^x x_{ij} + \sum_{j=1}^{10} \beta_j^w w_{ij} + \sum_{j=1}^{50} \beta_j^z z_{ij}. \tag{40}$$

The covariates are partitioned into three distinct blocks: 3 strong individual signals ($x$, $\beta_j^x = 2$), 10 weak background signals ($w$, $\beta_j^w = 0.2$), and 50 pure noise variables ($z$, $\beta_j^z = 0$) grouped into five blocks of 10. The intercept $\beta_0$ is numerically calibrated for a target marginal prevalence. To establish ground truth, we generate a superpopulation of 100,000 observations from this DGP to calculate the true population partial and global $\rho^2_{\mathrm{SV}}$.

We fit the full logistic regression model utilizing all 63 available predictors. The variables are evaluated sequentially by inherent signal strength: 3 strong individual signals ($x_1 - x_3$), a group ($w$) of 10 weak background signals, and five pure noise groups ($z_1 - z_5$). For each added group, we calculate both the global and partial standardized EV coefficients of determination, $R^2_{\mathrm{SV}}$, and their expected predictive $R^2_{\mathrm{SVP}}$.

Fig. 2 visualizes the results from a single dataset. The left panel demonstrates the estimation accuracy of the empirical $R^2_{\mathrm{SV}}$, which tightly tracks the superpopulation ground truth $\rho^2_{\mathrm{SV}}$ with all confidence intervals successfully covering the true global and partial parameters. To validate the predictive metric directly, we also overlay the exact leave-one-out cross-validation (LOOCV) predictive $R^2$, computed by refitting each model with every case removed in turn; the analytic $R^2_{\text{SVP}}$ matches this brute-force LOOCV benchmark, confirming that the squared inflation factor reproduces genuine out-of-sample prediction error without resampling. The right panel illustrates the sensitivity of the expected predictive metric, $R^2_{\text{SVP}}$, in capturing the out-of-sample generalizability. While $R^2_{\text{SVP}}$ appropriately registers performance gains from informative predictors ($x$ and $w$), it imposes a strict penalty immediately upon the introduction of uninformative variables ($z_1 - z_5$). By sharply identifying the exact threshold where model complexity transitions into overfitting, these results validate the theoretical performance and utility of the proposed EV-based predictive effect size statistics $R^2_{\text{SV}}$ and $R^2_{\text{SVP}}$.

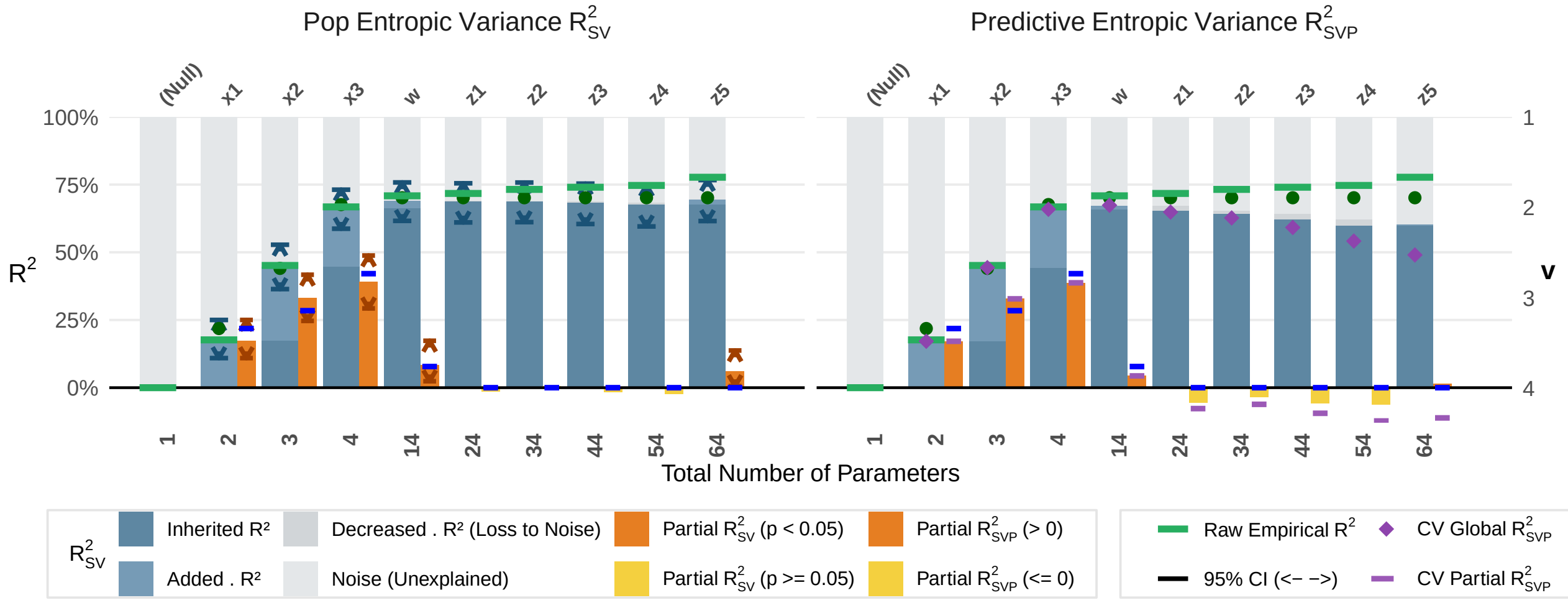

Figure 2: Comparison of estimated vs. true partial and global $\rho^2_{\mathrm{SV}}$ for a single simulated dataset.

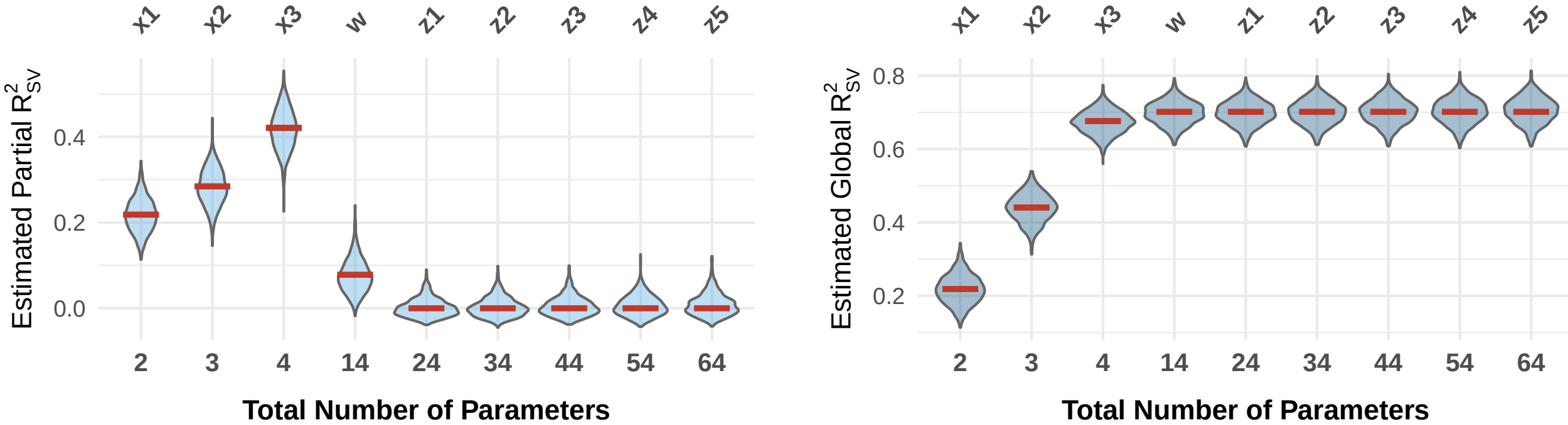

Figure 3: Distributions of estimated partial and global $R^2_{\text{SV}}$. Red bars indicate the true values.

The aggregated results across the 1,000 simulated datasets are shown in Fig. 3 and summarized in Table 2. These results confirm the robust finite-sample properties of the proposed estimators. Median $R^2_{\text{SV}}$ estimates tightly track theoretical true values (indicated by red bars in Fig. 3), with confidence interval coverage meeting the nominal 95% target for partial effects and proving slightly conservative (approaching 98%) for global effects.

Evaluating the pure noise groups ($z_1 - z_5$) demonstrates the distinct superiority of the $F_{\text{V}}$ test, which yields substantially lower Type I error rates compared to the traditional $\chi^2$ and $F_{\text{MD}}$ tests. Beyond hypothesis testing, applying a very loose effect size threshold ($R^2_{\text{SV}} \geq$ 0.05 or $R^2_{\text{SVP}} \geq 0.05$) provides near-perfect elimination of these uninformative predictors.

Table 2: Summary of inferences for $N_{\text{sim}} = 1000$ datasets simulated from a logistic regression model with prevalence $= 0.5$. **Variable Selection** reports selection rates using five different methods. To summarize the performance of partial and global $R^2_{\text{SV}}$, the table reports their true value (True), median (Med), interquartile range (IQR), and coverage rate (CR) of their CIs.

| | | Variable Selection | | | | | Estimate of Partial $\rho^2_{\text{SV}}$ | | | | Estimate of Global $\rho^2_{\text{SV}}$ | | | |
|---|---|---|---|---|---|---|---|---|---|---|---|---|---|---|
| Term | $p$ | $F$ | $\chi^2$ | $F_{\text{MD}}$ | $R^2_{\text{SV}}$ | $R^2_{\text{SVP}}$ | True | Med | IQR | CR | True | Med | IQR | CR |
| x1 | 1 | 1.000 | 1.000 | 1.000 | 1.000 | 1.000 | 0.218 | 0.216 | 0.055 | 0.947 | 0.218 | 0.216 | 0.055 | 0.947 |
| x2 | 1 | 1.000 | 1.000 | 1.000 | 1.000 | 1.000 | 0.284 | 0.281 | 0.060 | 0.961 | 0.441 | 0.441 | 0.052 | 0.975 |
| x3 | 1 | 1.000 | 1.000 | 1.000 | 1.000 | 1.000 | 0.421 | 0.417 | 0.065 | 0.967 | 0.676 | 0.673 | 0.043 | 0.981 |
| w | 10 | 0.870 | 0.879 | 0.980 | 0.733 | 0.327 | 0.078 | 0.072 | 0.053 | 0.948 | 0.701 | 0.699 | 0.042 | 0.977 |
| z1 | 10 | 0.069 | 0.081 | 0.379 | 0.033 | 0.000 | 0.000 | −0.002 | 0.027 | 0.955 | 0.701 | 0.699 | 0.043 | 0.978 |
| z2 | 10 | 0.077 | 0.098 | 0.446 | 0.034 | 0.001 | 0.000 | −0.001 | 0.029 | 0.955 | 0.701 | 0.701 | 0.045 | 0.977 |
| z3 | 10 | 0.089 | 0.120 | 0.498 | 0.046 | 0.000 | 0.000 | 0.001 | 0.030 | 0.950 | 0.701 | 0.702 | 0.045 | 0.976 |
| z4 | 10 | 0.101 | 0.161 | 0.566 | 0.044 | 0.004 | 0.000 | 0.004 | 0.033 | 0.955 | 0.701 | 0.703 | 0.047 | 0.978 |
| z5 | 10 | 0.120 | 0.190 | 0.619 | 0.073 | 0.006 | 0.000 | 0.005 | 0.032 | 0.932 | 0.701 | 0.706 | 0.048 | 0.973 |

Crucially, the weak signal block ($w$) highlights the stark distinction between statistical significance and practical predictive utility. Traditional $p$-value tests detect $w$ with high statistical power (87%–98%). As expected from theory, the selection rate of such weak signals via statistical significance will inevitably approach 1 as the sample size $n$ increases, driven by the corresponding growth of the non-centrality parameter. A vanishingly small $p$-value in large datasets merely confirms that an effect is strictly non-zero, offering no insight into its practical magnitude. In contrast, as $n$ increases, the empirical estimates of the $R^2$ metrics become increasingly accurate. This precision provides researchers with a highly reliable, standardized guideline to dictate variable inclusion based on actual information gain rather than mere statistical detectability. Consequently, the expected predictive threshold ($R^2_{\text{SVP}} \geq 0.05$) selects the $w$ block in only 32.7% of the simulations, which indeed drops to 6% if using $R^2_{\text{SVP}} \geq 0.1$. This correctly reflects that the conditional predictive contribution of these 10 variables (true partial $\rho^2_{\text{SV}} = 0.078$) is so small that they are heavily offset by the penalty of increased dimensionality, discouraging their inclusion for out-of-sample prediction or further investigation. In summary, EV-based $R^2_{\text{SV}}$ and $R^2_{\text{SVP}}$ provide more nuanced metrics for variable selection than $p$-value alone.

## 4.2 Logistic Regression with Data-Splitting LASSO Ordering

High-dimensional regression requires variable selection. While LASSO effectively identifies sparse active sets, a cross-validation optimal $\lambda$ frequently over-selects noise. In this section, we apply the proposed entropic variance based $R^2$ to logistic regression to demonstrate how evaluating predictive effect sizes can further purify LASSO-selected features. However, using the same dataset for both variable ordering and model evaluation induces a severe optimistic bias in statistical inference and learning. This represents an enduring challenge in modern statistics (see (Ambroise & McLachlan, 2002; Li, 2012; Li et al., 2008) in the context of estimating predictivity). To counteract artificially inflated false discovery rates (FDR) and optimistic estimates of $\rho^2$, we propose to use data-splitting methods to strictly separate the LASSO variable ordering from the evaluation of the EV-$R^2$.

To evaluate this data-splitting strategy, we simulate independent binary responses $Y_i \in \{0, 1\}$ for $n = 500$ observations from a logistic model comprising $p = 400$ standard normal covariates. These predictors are composed of strong and weak signals, as well as pure noise:

$$\text{logit }(\pi_i) = \beta_0 + \sum_{j=1}^{10} \beta_j^{\text{S}} S_{ij} + \sum_{j=1}^{10} \beta_j^{\text{W}} W_{ij} + \sum_{j=1}^{380} \beta_j^{\text{N}} N_{ij}. \tag{41}$$

The non-zero coefficients are assigned random signs with magnitudes $\mid \beta_j^{\text{S}} \mid \sim U(1, 2)$ and $\mid \beta_j^{\text{W}} \mid \sim U(0.1, 0.5)$, while the noise coefficients are fixed at zero ($\beta_j^{\text{N}} = 0$). An independent superpopulation of $N = 10^5$ observations is generated to establish the exact ground truth for the partial and global EV-$R^2$ across all evaluated variable subsets.

Executing both LASSO ordering and EV-$R^2$ evaluation on the same dataset—termed internal LASSO ordering—induces an optimistic selection bias. Because LASSO prioritizes artificially predictive noise variables, this double-use of data causes the empirical global $R^2_{\mathrm{SV}}$ trajectory to artificially inflate as noise is sequentially added (Fig. S15). By re-evaluating spurious correlations exploited during the initial LASSO path, the estimator overestimates the true $\rho^2_{\mathrm{SV}}$, causing confidence intervals to fail to cover the truth at later steps.

To eliminate optimistic bias, we implement a data-splitting procedure that partitions the data into an ordering set and an evaluation set. Using the **ordering set**, we determine the LASSO entry path of candidate predictors. For comparison with internal ordering, we evaluate the EV-$R^2$ across the same number of predictors. We cumulatively add these variables to unpenalized logistic regression models, which is then fitted on the **evaluation set**, computing the empirical partial and global $R^2_{\mathrm{SV}}$ alongside their predictive counterparts, $R^2_{\mathrm{SVP}}$. For each step, the theoretical ground truth ($\rho^2_{\mathrm{SV}}$) of the subset is computed by fitting the model with the $10^5$ observations of the superpopulation. Using a single dataset with $n = 500$ cases, Fig. 4 visualizes the trajectories of the global population $R^2_{\mathrm{SV}}$ and the predictive $R^2_{\mathrm{SVP}}$ on the evaluation set, with partial EV-$R^2$ values shown as vertical lines. Applying a 0.05 threshold for both evaluation criteria ($R^2 > 0.05$ and $p$ -value $< 0.05$), the plot reveals that all strong signals are retained by both the EV-$R^2$ and the $F_{\mathrm{V}}$ test. While the $F_{\mathrm{V}}$ test admits weak signals and noise variables, their partial $R^2_{\mathrm{SVP}}$ values remain below the 0.05 cutoff, filtering them out. Furthermore, the predictive $R^2_{\mathrm{SVP}}$ penalizes model complexity more than the population estimator $R^2_{\mathrm{SV}}$, eliminating weak signals and noise variables. The confidence intervals for the partial $R^2_{\mathrm{SV}}$ cover the population truth. The global $R^2_{\mathrm{SV}}$ values stay close to their population truth, with CIs covering the true values, until the unpenalized model overfits to raw $R^2_{\mathrm{SV}} = 1$ after adding noise variables.

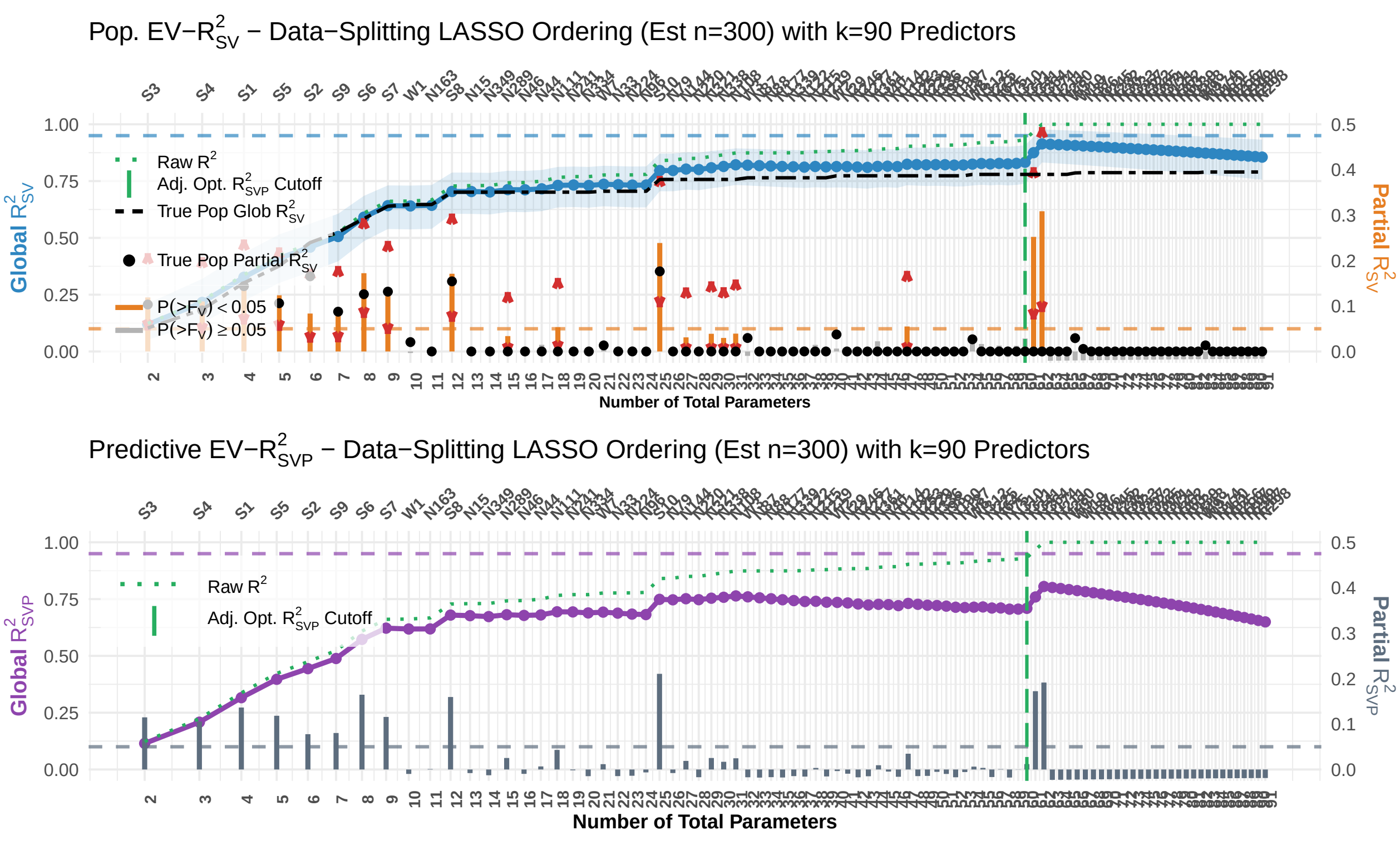

Figure 4: Estimation trajectories of global and partial $R^2_{\mathrm{SV}}$, and predictive $R^2_{\mathrm{SVP}}$, for a single simulated dataset. Predictors enter sequentially via their LASSO path. Curves compare cumulative global estimates against the population truth; vertical bars denote individual partial EV-$R^2$ contributions. Red error bars and blue bands indicate 95% confidence intervals for the partial and global $\rho^2_{\mathrm{SV}}$, respectively.

As seen in Fig. 4, due to evaluating on unpenalized MLE, the maximum global $R^2_{\mathrm{SVP}}$ typically occurs exactly at the point of overfitting. Therefore, the selection procedure identifies the step with maximum global $R^2_{\mathrm{SVP}}$ and, in the case that the corresponding raw $R^2_{\mathrm{SV}}$ is near 1 (e.g., $\geq 0.95$), we move the step backward to prune predictors with spuriously high partial $R^2_{\mathrm{SVP}}$ that immediately precede the peak. The final step is denoted by "adjusted optimal $R^2_{\mathrm{SVP}}$ cutoff" in Fig. 4. Using the predictors selected by this **adjusted optimal cutoff**, we further apply four different selection criteria ($F_{\mathrm{V}}$, $\chi^2$, and partial $R^2_{\mathrm{SV}}$ and $R^2_{\mathrm{SVP}}$) to determine the final sets of selected predictors.

Using the above selection methods, we conduct an experiment with 100 simulated datasets, each with $n = 500$ cases. To implement the sample partitioning, each dataset is divided into an ordering set with $n_{\text{ord}} = 200$ cases and an evaluation set with $n_{\text{eval}} = 300$ cases. For each dataset, we evaluate the performance of different selection criteria and compare the results of internal ordering against data-splitting. The averages of these metrics over the 100 datasets are reported in Table 3. For each criterion, the table reports the total number of selected variables, the counts of true strong and weak signals correctly identified (out of 10 each), and the resulting FDR, along with the precision and recall rates for both signal groups.The results in Table 3 demonstrate that the EV-$R^2$ criteria (partial $R^2_{\text{SV}}$ and $R^2_{\text{SVP}}$) yield a substantially lower FDR compared to the alternative methods. Notably, applying these EV-$R^2$ criteria reduces the FDR from 0.807 under the original LASSO-CV to 0.056. While this pruning slightly decreases the overall recall of true signals, the retention rates remain high at 0.892 and 0.864 for partial $R^2_{\text{SV}}$ and $R^2_{\text{SVP}}$. Furthermore, the comparison between evaluation regimes confirms the necessity of the sample partitioning: variable selection based on internal ordering performs worse across these operating characteristics, driven by the previously discussed optimistic bias, whereas the data-splitting method successfully controls the FDR while isolating true signals.

Table 3: Comparison of Variable selection performances of Internal LASSO Ordering and Data-Splitting LASSO Ordering, averaged over 100 simulated datasets. For each selection criterion, the table reports the total number of variables selected (#Sel), the counts of true strong (#S) and weak (#W) signals correctly identified (out of 10 each), and the resulting FDR (%N), Precisions (%S, %W), and recalls (Rec-S, Rec-W) respectively for Strong and Weak signals.

| | Internal LASSO Ordering | | | | | | | | Data-Splitting LASSO Ordering | | | | | | | |
|---|---|---|---|---|---|---|---|---|---|---|---|---|---|---|---|---|
| **Criteria** | **#Sel** | **#S** | **#W** | **FDR** | **%S** | **%W** | **Rec-S** | **Rec-W** | **#Sel** | **#S** | **#W** | **FDR** | **%S** | **%W** | **Rec-S** | **Rec-W** |
| **LASSO-CV** | 84 | 10 | 6 | 0.807 | 0.123 | 0.070 | 1.000 | 0.575 | - | - | - | - | - | - | - | - |
| **Adj. Opt.** $R^2_{\text{SVP}}$ | 55 | 10 | 5 | 0.728 | 0.184 | 0.088 | 1.000 | 0.478 | 54 | 10 | 2 | 0.730 | 0.227 | 0.043 | 0.985 | 0.236 |
| $\chi^2$ | 32 | 10 | 3 | 0.576 | 0.316 | 0.109 | 1.000 | 0.349 | 16 | 10 | 1 | 0.309 | 0.635 | 0.056 | 0.977 | 0.097 |
| $F_{\text{V}}$ | 31 | 10 | 3 | 0.561 | 0.329 | 0.110 | 1.000 | 0.340 | 15 | 10 | 1 | 0.276 | 0.672 | 0.051 | 0.976 | 0.083 |
| **partial** $R^2_{\text{SV}}$ | 15 | 10 | 1 | 0.244 | 0.681 | 0.074 | 0.997 | 0.118 | 11 | 9 | 0 | 0.140 | 0.823 | 0.037 | 0.892 | 0.043 |
| **partial** $R^2_{\text{SVP}}$ | 13 | 10 | 1 | 0.136 | 0.804 | 0.060 | 0.997 | 0.081 | 10 | 9 | 0 | 0.056 | 0.916 | 0.028 | 0.864 | 0.029 |

Table 4: Estimation accuracy and 95% confidence interval coverage for the empirical partial and global $R^2_{\mathrm{SV}}$. Results report the Mean Absolute Error (MAE) and empirical coverage rates, averaged across evaluated steps and variables within each signal group over 100 simulated datasets.

| | Internal LASSO Ordering | | | | | Data-Splitting LASSO Ordering | | | | |
|---|---|---|---|---|---|---|---|---|---|---|
| | | Partial $\rho^2_{\mathrm{SV}}$ | | Global $\rho^2_{\mathrm{SV}}$ | | | Partial $\rho^2_{\mathrm{SV}}$ | | Global $\rho^2_{\mathrm{SV}}$ | |
| **Term** | **Steps** | **MAE** | **CR** | **MAE** | **CR** | **Steps** | **MAE** | **CR** | **MAE** | **CR** |
| **Strong** | 10 | 0.0295 | 95.6% | 0.0352 | 91.5% | 10 | 0.0374 | 95.0% | 0.0383 | 96.2% |
| **Weak** | 5 | 0.0223 | 91.3% | 0.0580 | 61.6% | 3 | 0.0238 | 93.5% | 0.0341 | 96.5% |
| **Noise** | 39 | 0.0230 | 68.5% | 0.0823 | 32.0% | 41 | 0.0119 | 92.2% | 0.0350 | 95.7% |

At the adjusted optimal cutoff, we evaluate the estimation accuracy of the empirical partial and global $R^2_{\mathrm{SV}}$ against the population truth. Table 4 summarizes the Mean Absolute Error (MAE) and 95% confidence interval (CI) coverage, averaged across the specific variables (strong, weak, and noise) that enter the active set over 100 simulated datasets. While MAEs remain uniformly small ($\approx 0.02$) across all signal classifications, CI coverage diverges substantially by evaluation method. Data-splitting achieves robust coverage exceeding 90% for all variable groups. In contrast, internal ordering exhibits severe undercoverage. As visualized in Fig. S15, evaluating on the same data used to select predictors induces an upward bias in the empirical $R^2_{\mathrm{SV}}$ estimates, systematically shifting the resulting confidence intervals above the population truth.

Fig. 5 plots the global EV-$R^2$ metrics and estimation errors at the adjusted optimal $R^2_{\mathrm{SVP}}$ cutoff. Internal LASSO ordering exhibits optimistic bias, overestimating global EV-$R^2$ by approximately 0.15 on average; consequently, its estimated predictive $R^2_{\mathrm{SVP}}$ artificially exceeds the true population $R^2_{\mathrm{SV}}$. In contrast, data-splitting achieves comparable true population $R^2_{\mathrm{SV}}$ at the optimal step, preserving predictive performance. While data-splitting retains a minor upward bias in empirical $R^2_{\mathrm{SV}}$ (median error $\approx 0.08$) due to near-separation, its predictive estimator ($R^2_{\mathrm{SVP}}$) correctly applies the complexity penalty, distributing predominantly below the mean true $R^2_{\mathrm{SV}}$.

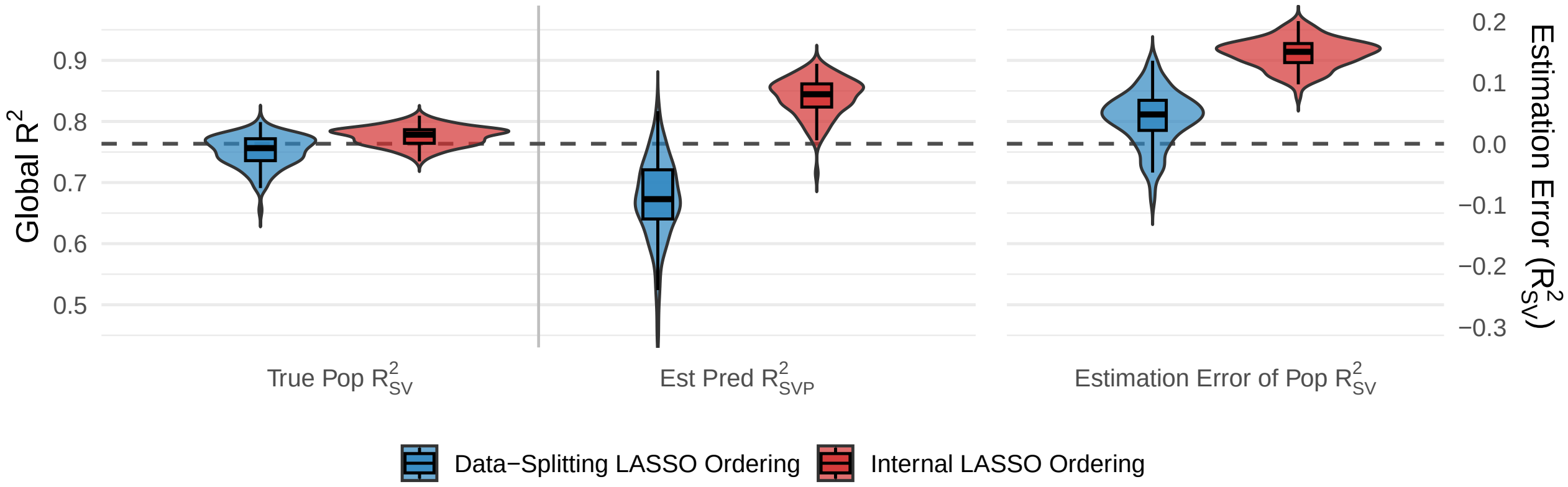

Figure 5: Distribution of global EV-$R^2$ metrics and estimation error at the optimal $R^2_{\mathrm{SVP}}$ cutoff step. The left panel compares the true population EV-based $R^2_{\mathrm{SV}}$ and the estimated predictive EV-based $R^2_{\mathrm{SVP}}$ for both Internal and Data-Splitting LASSO ordering methods. The right panel isolates the estimation error of $R^2_{\mathrm{SV}}$, with the dashed reference line indicating zero error to evaluate methodological bias and variance.

# 5 An Application to PD Microbiome Data

To demonstrate the practical utility of the EV-$R^2$, we analyzed a human gut microbiome dataset associated with Parkinson's disease (PD) pathogenesis (Hill-Burns et al., 2017). Following a rigorous initial quality control phase that excluded specific technical replicates and samples exhibiting insufficient sequencing depth (fewer than 5,000 total reads), the final analytical cohort comprised 328 distinct subjects. To appropriately handle the compositional nature and inherent overdispersion of 16S rRNA sequencing data, the raw microbial counts were normalized by total sample read depth and subsequently subjected to an arcsine square root variance-stabilizing transformation. These transformed microbial features were then merged with patient age, binary biological sex, and their interaction term. This comprehensive data integration yielded a final predictive matrix encompassing a total of 378 variables. To facilitate reproducibility, this fully cleaned and transformed high-dimensional dataset is accessible within the accompanying `ANOEN` R package.

To mitigate optimistic selection bias, the dataset was randomly partitioned into an ordering set ($n = 128$) and an independent evaluation set ($n = 200$). First, cross-validated LASSO logistic regression was utilized to determine the total number of predictors to evaluate. Next, a subsequent LASSO model was fitted exclusively on the ordering set to establish an initial variable order based on their LASSO entry order. This sequence was then re-arranged: predictors yielding a strong partial predictive $R^2$ ($R^2_{\text{SV}} > 0.05$) within the ordering set were front-loaded, while preserving the original LASSO order for improving their degree-freedom. Finally, logistic regression models were sequentially fitted on the hold-out evaluation set according to this refined order, allowing us to compute rigorous, unbiased EV-$R^2$ metrics for each successive model.

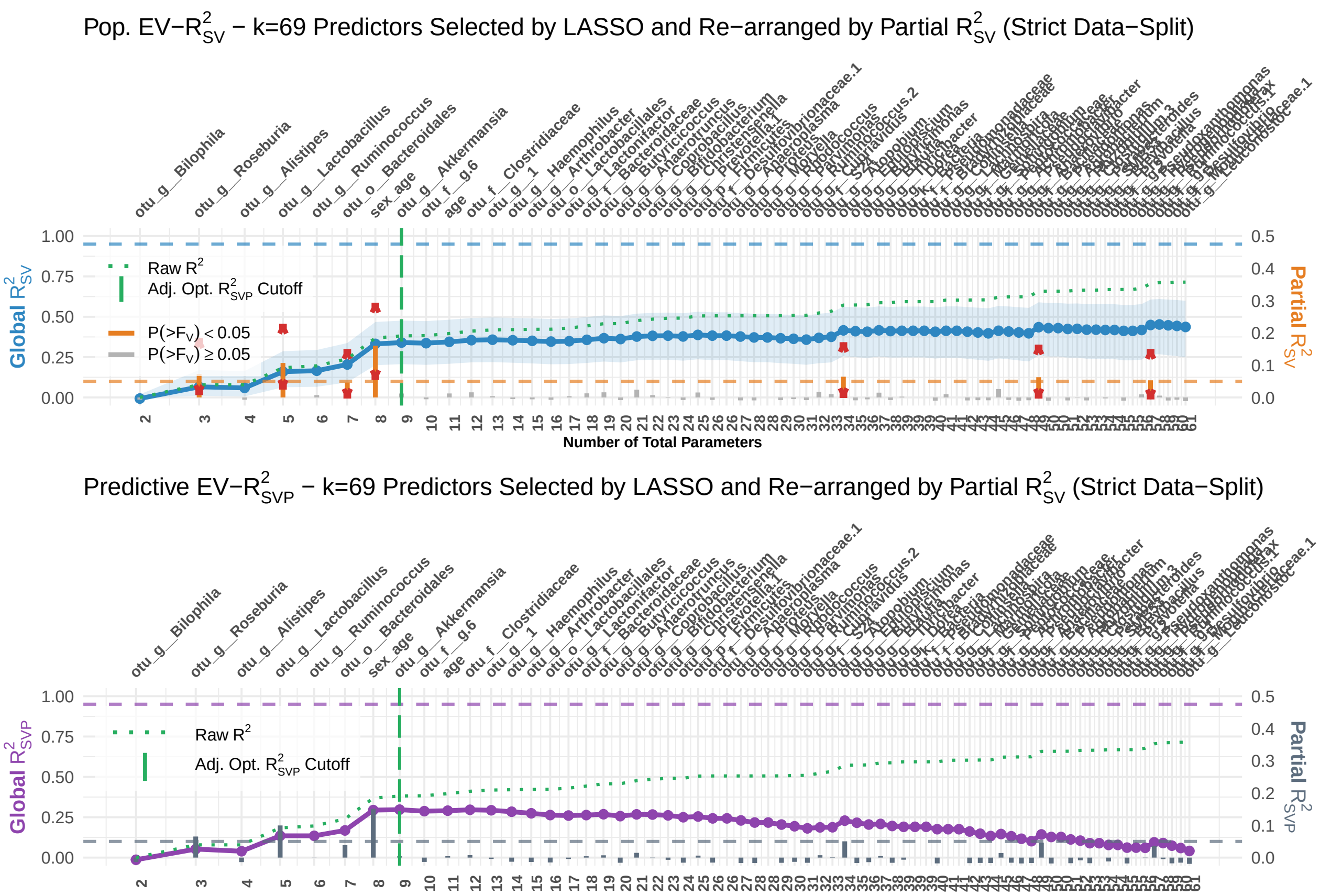

Figure 6: Out-of-sample EV-$R^2$ trajectories for the Parkinson microbiome dataset, evaluated along a predictor sequence derived from an independent ordering set.

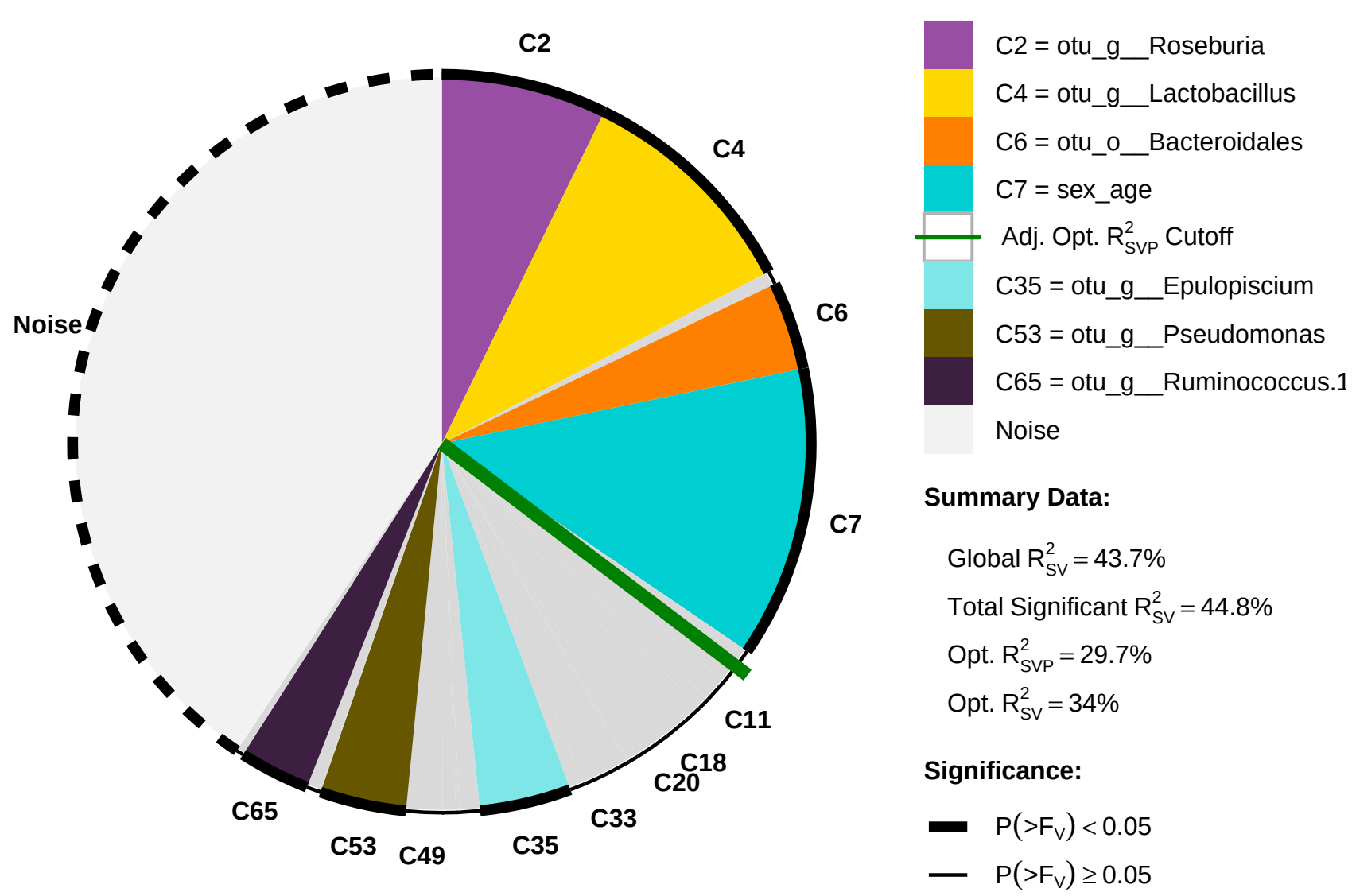

Figure 7: Out-of-sample $R^2_{\mathrm{SV}}$ contributors in a Parkinson microbiome data.

The trajectory of the EV-$R^2$ metrics across the pre-selected variable sequence is illustrated in Fig. 6. Evaluated under the strict data-splitting framework, the peak out-of-sample predictive performance, denoted by the optimal $R^2_{\mathrm{SVP}}$, is achieved at a distinct model size encompassing only 4 predictors with statistical significance based on the $F_{\mathrm{V}}$ test ($p < 0.05$), yielding an optimal $R^2_{\mathrm{SV}}$ of 34%. To further deconstruct this unbiased predictive power, the pie chart (Fig. 7) visualizes the proportional contributions of each term to the total $R^2_{\mathrm{SV}}$, with the optimal predictive cutoff explicitly marked by the heavy dashed green line. The specific taxonomic and demographic variables that achieve statistical significance along the evaluation sequence are detailed in the legend of Fig. 7. Notably, this realistic 34% estimate contrasts starkly with findings from internal LASSO ordering, which identified approximately 20 statistically significant predictors and an inflated $R^2_{\mathrm{SV}}$ of 81% as shown in Sec. G.3. This dramatic higher estimate is likely caused by the double-use of the data.

The four predictors in the optimal $R^2_{\mathrm{SVP}}$ model link gut dysbiosis and host demographics to PD, aligning with established literature. The largest component, *Roseburia*, produces short-chain fatty acids like butyrate. Its depletion compromises gut barrier integrity and drives systemic inflammation—key early events in PD pathogenesis (Aho et al., 2019; Wallen et al., 2020). *Lactobacillus* and the order *Bacteroidales* follow. *Lactobacillus* over-abundance, a consistent PD signature involved in neurotransmitter metabolism, expands due to chronic constipation and altered gut motility (Barichella et al., 2019; Hill-Burns et al., 2017). Shifts in *Bacteroidales* reflect structural disruptions that alter gut-immune homeostasis and exacerbate peripheral immune activation (Bedarf et al., 2017; Keshavarzian et al., 2015). The model optimally terminates with the age-sex interaction, a recognized PD risk factor. Beyond this boundary, *Epulopiscium*, *Pseudomonas*, and *Ruminococcus* maintain statistical significance ($p < 0.05$). *Ruminococcus* was found to associate with the gut-brain axis via mucus degradation (Romano et al., 2021). Nevertheless, inclusion of these lower-ranked taxa degrades the out-of-sample prediction. These taxa may capture redundant variance already explained by the top predictors. Consequently, their informational value is outweighed by the penalty of increased model complexity.

# 6 Conclusion and Future Work

In this paper, we introduced entropic variance as the rigorous mathematical generalization of the OLS residual variance for regular parametric families. By establishing that exponentiated empirical cross-entropy—rather than raw entropy—preserves the geometric properties of residual variance, we resolved long-standing paradoxes in defining a universally applicable predictive effect size. Our proposed estimators, $R^2_{\mathrm{SV}}$ and $R^2_{\mathrm{SVP}}$, project unbounded log-likelihoods onto a standardized, interpretable $[0, 1]$ scale while intrinsically correcting for finite-sample training optimism and model complexity.

Crucially, we established a unified inferential machinery based on a novel characterization of the empirical EV Signal-to-Noise Ratio. We demonstrated that the ratio of empirical EVs maps to a non-central $F$-distribution ($F_{\mathrm{V}}$), enabling refined $p$-values and continuous confidence intervals without requiring intractable, model-specific Fisher information matrices. Extensive simulation studies and real-world applications verify these finite-sample distributions and confirm the practical superiority of conducting variable selection via predictive effect size $R^2_{\text{SV}}$ and $R^2_{\text{SVP}}$. Notably, our simulation studies demonstrate that applying our proposed EV-$R^2$ cutoff to a data-splitting LASSO path drastically reduces the false discovery rate from an original 80% down to 6%, while preserving true signal recall. Furthermore, applying this rigorous data-splitting framework to a Parkinson's disease microbiome dataset identified a biologically interpretable model of just four predictors achieving a realistic out-of-sample $R^2_{\text{SV}}$ of 34%, successfully correcting the artificially inflated 81% estimate produced by naive internal evaluation.

Looking forward, several critical avenues for future research remain. A primary direction is generalizing the EV-$R^2$ framework to Bayesian inference and penalized regression models characterized by singular likelihoods, as well as addressing the complex optimism induced by internal variable pre-selection. In these regimes, adjusting for finite-sample bias using simple parameter counts ($p$) becomes fundamentally inadequate. Defining a mathematically rigorous measure of effective model complexity beyond basic parameter counting presents a profound challenge. Future theoretical extensions could explore integrating Watanabe's real log canonical threshold ($\lambda$) derived from Singular Learning Theory (Watanabe, 2009), and adapting the topological penalties of Extended Bayesian Information Criterion (Chen & Chen, 2008), to formally quantify effective complexity in singular and/or post-selected models. Finally, in highly complex settings where such theoretical approximations are unavailable, the proposed predictive effect size $\rho^2_{\text{SVP}}$ can still be robustly computed through computationally intensive cross-validation, ensuring the practical utility of the EV-$R^2$ metrics across modern statistical learning paradigms.

## Data and Software Availability

The proposed methodologies are implemented in the open-source R package **ANOEN**, which will be released on GitHub and CRAN upon publication. It provides efficient functions to estimate and visualize predictive effect sizes ($R^2_{\text{SV}}$ and $R^2_{\text{SVP}}$) for generalized and standard linear models (`glm`, `lm`). For full reproducibility, the package includes the cleaned Parkinson's disease microbiome dataset and Quarto scripts to replicate all simulations and empirical analyses. For peer review, the preliminary package and replication materials are available via Figshare at [https://figshare.com/s/65d8bd7067124de6187d].

## Supplementary Materials

The Supplementary Materials provide comprehensive geometric interpretations, rigorous asymptotic proofs for the proposed distributions, theoretical recovery of classical OLS theory, and extensive comparative analyses with alternative pseudo-$R^2$ measures, alongside extended simulation and empirical results.

## Acknowledgments

We used Google Gemini Pro 3.1 to refine the manuscript's grammar and phrasing, conduct literature searches, and assist with R scripting. Despite this assistance, all scientific ideas, analytical methods, and final code are the original work of the authors. All AI-generated content was thoroughly reviewed and validated to ensure scientific accuracy and integrity.

# References

Aho, V. T., Pereira, P. A., Voutilainen, V., Paulin, L., Pekkonen, E., Auvinen, P., & Scheperjans, F. (2019). Gut microbiota in Parkinson's disease: Temporal stability and relations to disease progression. *Ebiomedicine*, *44*, 691–707.

Akaike, H. (1973). Information theory and an extension of the maximum likelihood principle. In B. N. Petrov & F. Csáki (Eds.), *Second International Symposium on Information Theory: Second International Symposium on Information Theory.*

Ambroise, C., & McLachlan, G. J. (2002). Selection Bias in Gene Extraction on the Basis of Microarray Gene-expression Data. *PNAS*, *99*(10), 6562–6566.

Barichella, M., Severgnini, M., Cilia, R., Cassani, E., Bolliri, C., Caronni, S., Ferri, V., Francetti, R., Pusani, E., Celia, C., & others. (2019). Unraveling gut microbiota in Parkinson's disease and atypical parkinsonism. *Movement Disorders*, *34*(3), 396–405.

Bartlett, M. S. (1937). Properties of sufficiency and statistical tests. *Proceedings of the Royal Society of London. Series a-Mathematical and Physical Sciences*, *160*(901), 268–282.

Bedarf, J. R., Hildebrand, F., Coelho, L. P., Sunagawa, S., Bahram, M., Goeser, F., Bork, P., & Wüllner, U. (2017). Metabolic implications of gut microbiota alterations in Parkinson's disease. *Cell Reports*, *19*(4), 845–851.

Billheimer, D. (2019). Predictive inference and scientific reproducibility. *The American Statistician*, *73*(sup1), 291–295.

Burnham, K. P., & Anderson, D. R. (2002). *Model Selection and Multimodel Inference: A Practical*

*Information-Theoretic Approach* (2nd ed.). Springer-Verlag.

Cameron, A. C., & Windmeijer, F. A. G. (1996). R-squared measures for count data regression models with applications to health-care utilization. *Journal of Business & Economic Statistics*, *14*(2), 209–220.

Chatterjee, S. (2021). A new coefficient of correlation. *Journal of the American Statistical Association*, *116*(536), 2009–2022.

Chen, J., & Chen, Z. (2008). Extended Bayesian information criteria for model selection with large model spaces. *Biometrika*, *95*(3), 759–771.

Chernozhukov, V., Wüthrich, K., & Zhu, Y. (2023). Exact and Robust Conformal Inference Methods for Predictive Machine Learning With Dependent Data. *Journal of the American Statistical Association*, *118*(541), 173–191. https://doi.org/10.1080/01621459.2021.1920957

Clements, M. P., & Hendry, D. F. (1993). On the limitations of comparing mean square forecast errors. *Journal of Forecasting*, *12*(8), 617–637.

Cordeiro, G. M. (1983). Improved likelihood ratio statistics for generalized linear models. *Journal of the Royal Statistical Society: Series B (Methodological)*, *45*(3), 404–413.

Cox, D. R., & Hinkley, D. V. (1974). *Theoretical Statistics*. Chapman, Hall.

Cox, D. R., & Snell, E. J. (1989). *Analysis of Binary Data* (2nd ed.). Chapman, Hall/CRC.

Deuber, D., Li, J., Engelke, S., & Maathuis, M. H. (2024). Estimation and Inference of Extremal Quantile Treatment Effects for Heavy-Tailed Distributions. *Journal of the American Statistical Association*. https://doi.org/10.1080/01621459.2023.2252141

Ding, P., Feller, A., & Miratrix, L. (2019). Decomposing treatment effect variation. *Journal of the American Statistical Association*, *114*(525), 304–317.

Domencich, T. A., & McFadden, D. (1975). *Urban Travel Demand: A Behavioral Analysis*. North-Holland Publishing Company.

Efron, B. (1978). Regression and ANOVA with Zero-One Data: Measures of Residual Variation. *Journal of the American Statistical Association*, *73*(361), 113–121.

Gelman, A., Goodrich, B., Gabry, J., & Vehtari, A. (2019). R-squared for Bayesian regression models. *The American Statistician*, *73*(3), 307–309.

Guo, F. R., & Shah, R. D. (2025). Rank-transformed subsampling: inference for multiple data splitting and exchangeable p-values. *Journal of the Royal Statistical Society Series B: Statistical Methodology*, *87*(1), 256–286.

Hayakawa, T. (1977). The likelihood ratio criterion and the asymptotic expansion of its distribution. *Annals of the Institute of Statistical Mathematics*, *29*(1), 359–378.

Hensher, D. A., Rose, J. M., & Greene, W. H. (2005). *Applied Choice Analysis: A Primer*. Cambridge University Press.

Hill, M. O. (1973). Diversity and evenness: a unifying notation and its consequences. *Ecology*, *54*(2), 427–432.

Hill-Burns, E. M., Debelius, J. W., Morton, J. T., Wisnivesky, W. T., Vazquez-Baeza, Y.,

McFadden, E., Knight, R., & Payami, H. (2017). Parkinson's disease and Parkinson's disease medications have distinct signatures of the gut microbiome. *Movement Disorders*, *32*(5), 739–749.

Jaeger, B. C., Edwards, L. J., Das, K., & Sen, P. K. (2017). An $R^2$ statistic for fixed effects in the generalized linear mixed model. *Journal of Applied Statistics*, *44*(6), 1086–1105.

Jelinek, F., Mercer, R. L., Bahl, L. R., & Baker, J. K. (1977). Perplexity—a measure of the difficulty of speech recognition tasks. *The Journal of the Acoustical Society of America*, *62*(S1), S63–S63.

Jurafsky, D., & Martin, J. H. (2009). *Speech and Language Processing: An Introduction to Natural Language Processing, Computational Linguistics, and Speech Recognition* (2nd ed.). Pearson Prentice Hall.

Keshavarzian, A., Green, S. J., Engen, P. A., Voigt, R. M., Naqib, A., Forsyth, C. B., Mutlu, E., & Shannon, K. M. (2015). Colonic bacterial composition in Parkinson's disease. *Movement Disorders*, *30*(10), 1351–1360.

Lawley, D. N. (1956). A general method for approximating to the distribution of likelihood ratio criteria. *Biometrika*, *43*(3/4), 295–303.

Li, L. (2012). Bias-Corrected Hierarchical Bayesian Classification With a Selected Subset of High-Dimensional Features. *Journal of the American Statistical Association*, *107*(497), 120–134.

Li, L., Zhang, J., & Neal, R. (2008). A Method for Avoiding Bias from Feature Selection with Application to Naive Bayes Classification Models. *Bayesian Analysis*, *3*(1), 171–196.

Lin, M., Lucas, H. C., Jr., & Shmueli, G. (2013). Research Commentary—Too Big to Fail: Large Samples and the *p*-Value Problem. *Information Systems Research*, *24*(4), 906–917.

McCullagh, P., & Nelder, J. A. (1989). *Generalized Linear Models* (2nd ed.). Chapman, Hall.

McFadden, D. (1974). Conditional Logit Analysis of Qualitative Choice Behavior. In P. Zarembka (Ed.), *Frontiers in Econometrics: Frontiers in Econometrics* (pp. 105–142). Academic Press.

Nagelkerke, N. J. D. (1991). A note on a general definition of the coefficient of determination. *Biometrika*, *78*(3), 691–692.

Nakagawa, S., Johnson, P. C. D., & Schielzeth, H. (2017). The coefficient of determination $R^2$ and intra-class correlation coefficient from generalized linear mixed-effects models revisited and expanded. *Journal of the Royal Society Interface*, *14*(134), 20170213.

Pepe, M. S., Janes, H., Longton, G., Leisenring, W., & Newcomb, P. (2004). Limitations of the odds ratio in gauging the performance of a diagnostic, prognostic, or screening marker. *American Journal of Epidemiology*, *159*(9), 882–890.

Rights, J. D., & Sterba, S. K. (2019). Quantifying explained variance in multilevel models: An integrative framework for defining R-squared measures. *Psychological Methods*, *24*(3), 302–319.

Romano, S., Savva, G. M., Bedarf, J. R., Charles, I. G., Hildebrand, F., & Narbad, A. (2021). Meta-analysis of the Parkinson's disease gut microbiome suggests alterations linked to intestinal inflammation. *Npj Parkinson's Disease*, *7*(1), 27.

Shannon, C. E. (1948). A mathematical theory of communication. *The Bell System Technical*

*Journal*, *27*(3), 379–423.

Sur, P., Chen, Y., & Candès, E. J. (2019). The likelihood ratio test in high-dimensional logistic regression is asymptotically a rescaled chi-square. *Probability Theory and Related Fields*, *175*(1–2), 487–558.

Tjur, T. (2009). Coefficients of determination in logistic regression models—A new proposal: The coefficient of discrimination. *The American Statistician*, *63*(4), 366–372.

Vaart, A. W. van der. (1998). *Asymptotic Statistics*. Cambridge University Press.

Vehtari, A., Gelman, A., & Gabry, J. (2017). Practical Bayesian model evaluation using leave-one-out cross-validation and WAIC. *Statistics and Computing*, *27*(5), 1413–1432.

Wallen, Z. D., Appah, M., Dean, M. N., Sesler, C. L., Factor, S. A., Molho, E., Zabetian, C. P., Standaert, D. G., & Payami, H. (2020). Characterizing dysbiosis of gut microbiome in PD: evidence for overabundance of opportunistic pathogens. *Npj Parkinson's Disease*, *6*(1), 11.

Wasserstein, R. L., & Lazar, N. A. (2016). The ASA statement on p-values: context, process, and purpose. *The American Statistician*, *70*(2), 129–133.

Wasserstein, R. L., Schirm, A. L., & Lazar, N. A. (2019). Moving to a world beyond "p<0.05". *The American Statistician*, *73*(sup1), 1–19.

Watanabe, S. (2009). *Algebraic Geometry and Statistical Learning Theory* (Vol. 25). Cambridge University Press.

White, H. (1982). Maximum likelihood estimation of misspecified models. *Econometrica*, *50*(1), 1–25.

Zhang, D. (2017). A coefficient of determination for generalized linear models. *The American Statistician*, *71*(4), 310–316.

Zhang, D. (2022). Coefficients of Determination for Mixed-Effects Models. *Journal of Agricultural, Biological and Environmental Statistics*, *27*(4), 674–689.

# Supplementary Materials for "An Entropy-based Coefficient of Determination with Adjustment of Optimization Bias"

Longhai Li
2026-08-05

**Abstract**: This supplementary document provides comprehensive theoretical extensions, detailed proofs, and comparative analyses supporting the entropic variance (EV) framework. We begin by establishing a geometric perspective via a kernel volumetric interpretation of EV and its relationship to the geometric mean of regular variances (Sec. A.1, Sec. A.2). The core of this supplement is dedicated to the rigorous asymptotic theory of empirical entropic variances (Sec. B). We present complete mathematical proofs—primarily utilizing moment-generating functions—for the absolute deflation theorem (Sec. B.2), the central and non-central distributions of the relative EV-SNR (Sec. B.3, Sec. B.6), and the asymptotic $\chi^2$ decomposition of the EV drop (Sec. B.7.1). Within this theory, we formalize the mechanisms governing the deflation of training EVs (Sec. B.4) and the inflation of expected predictive EV. Next, we demonstrate the framework's backward compatibility by detailing how EV theory perfectly recovers classical OLS predictive effect sizes and inferential statistics (Sec. C). Following these theoretical foundations, we analyze the structural relationships between EV-based metrics and classical information criteria (Sec. D). We then provide extensive comparative analyses against other pseudo-$R^2$ measures (Sec. E), explicitly demonstrating the invariance of $\rho^2_{\mathrm{V}}$ under monotonic transformations (Sec. E.3) and evaluating its performance in Bernoulli models (Sec. F). Finally, we supply supplementary tables and figures expanding upon the main manuscript's simulation studies and real-world microbiome application (Sec. G.1, Sec. G.2, Sec. G.3).

# A Geometric Interpretation of EV and $\rho^2_{\mathrm{V}}$

## A.1 Kernel volumetric Interpretation of Entropic $\rho^2_{\mathrm{V}}$

**Theorem S1 (Kernel volumetric Interpretation of Entropic $\rho^2$).** *Let $f_j(y \mid \mathbf{x}_i, \boldsymbol{\theta}^*_j)$ be a predictive density from a regular location-scale family, and assume the true DGP is contained within this family (i.e., $f^* \in \mathcal{M}_j$). For each observation $i$, define the maximum achievable density as $M_j(\mathbf{x}_i) = \max_y f_j(y \mid \mathbf{x}_i, \boldsymbol{\theta}^*_j)$. The effective spatial volume of its standardized density kernel is given by $\mathrm{Vol}^{\mathrm{ker}}_j(\mathbf{x}_i, \boldsymbol{\theta}^*_j) = 1/M_j(\mathbf{x}_i)$.*

*Under this assumption, the individual EV, $\nu^*_j(\mathbf{x}_i)$, factors exactly into the squared kernel volume multiplied by a family-specific constant:*

$$\nu^*_j(\mathbf{x}_i) = \mathrm{Vol}^{\mathrm{ker}}_j(\mathbf{x}_i, \boldsymbol{\theta}^*_j)^2 \cdot \mathrm{Vol}_{\mathrm{family}}, \tag{S1}$$

*where $\mathrm{Vol}_{\mathrm{family}}$ is invariant to the parameter vector $\boldsymbol{\theta}^*_j$. Consequently, the population EV $\nu^*_j$ is the squared geometric mean of these observation-specific kernel volumes:*

$$\nu^*_j = \mathrm{GM}\left( [\mathrm{Vol}^{\mathrm{ker}}_j(\mathbf{x}_i, \boldsymbol{\theta}^*_j)]^n_{i=1}\right)^2 \cdot \mathrm{Vol}_{\mathrm{family}}. \tag{S2}$$

*Proof.*

1. **Kernel Standardization:** For a given observation $\mathbf{x}_i$, the standardized density kernel is defined by scaling the predictive density by its peak, yielding $k_j(y \mid \mathbf{x}_i) = f_j(y \mid \mathbf{x}_i, \boldsymbol{\theta}_j^*)/M_j(\mathbf{x}_i)$. By this definition, the maximum of $k_j(y \mid \mathbf{x}_i)$ is exactly 1, and its geometric volume is $\mathrm{Vol}_j^{\mathrm{ker}}(\mathbf{x}_i, \boldsymbol{\theta}_j^*) = \int k_j(y \mid \mathbf{x}_i)dy = 1/M_j(\mathbf{x}_i)$.

2. **Cross-Entropy Decomposition:** We can express the original density as the ratio $f_j(y \mid \mathbf{x}_i, \boldsymbol{\theta}_j^*) = k_j(y \mid \mathbf{x}_i)/\mathrm{Vol}_j^{\mathrm{ker}}(\mathbf{x}_i, \boldsymbol{\theta}_j^*)$. Substituting this relationship into the expected conditional cross-entropy $h_j^*(\boldsymbol{\theta}_j^*) = -\frac{1}{n}\sum_{i=1}^n \mathbb{E}_{f^*}[\log f_j(Y_i \mid \mathbf{x}_i, \boldsymbol{\theta}_j^*)]$ separates the scalar volume from the functional kernel shape:

$$h_j^*(\boldsymbol{\theta}_j^*) = \frac{1}{n}\sum_{i=1}^n \log \mathrm{Vol}_j^{\mathrm{ker}}(\mathbf{x}_i, \boldsymbol{\theta}_j^*) - \frac{1}{n}\sum_{i=1}^n \mathbb{E}_{f^*}[\log k_j(Y_i \mid \mathbf{x}_i)]. \tag{S3}$$

3. **Logarithmic Arithmetic to Geometric Mean:** Utilizing the property that the arithmetic mean of logarithms is strictly equivalent to the logarithm of the geometric mean, we can rewrite the first term to obtain:

$$h_j^*(\boldsymbol{\theta}_j^*) = \log\Big(\mathrm{GM}\ \big(\ [\mathrm{Vol}_j^{\mathrm{ker}}(\mathbf{x}_i, \boldsymbol{\theta}_j^*)]_{i=1}^n\ \big)\Big) - \frac{1}{n}\sum_{i=1}^n \mathbb{E}_{f^*}[\log k_j(Y_i \mid \mathbf{x}_i)]. \tag{S4}$$

4. **Exponentiation to EV:** The EV is defined as $\nu_j^* = \exp(2h_j^*(\boldsymbol{\theta}_j^*))$. Exponentiating the decomposed cross-entropy expression eliminates the logarithm on the geometric mean term:

$$\nu_j^* = \mathrm{GM}\ \big(\ [\mathrm{Vol}_j^{\mathrm{ker}}(\mathbf{x}_i, \boldsymbol{\theta}_j^*)]_{i=1}^n\big)^2 \cdot \exp\left(-\frac{2}{n}\sum_{i=1}^n \mathbb{E}_{f^*}[\log k_j(Y_i \mid \mathbf{x}_i)]\right). \tag{S5}$$

5. **Scale Invariance:** For a location-scale family, the density takes the standard form,

$$f_j(y \mid \mathbf{x}_i, \boldsymbol{\theta}_j^*) = \frac{1}{\sigma_i} g\ \big(\ \frac{y-\mu_i}{\sigma_i}\ \big), \tag{S6}$$

where $g(z)$ is the standard base density and $M_g = \max_z g(z)$. It follows that $M_j(\mathbf{x}_i) = M_g/\sigma_i$, which simplifies the standardized kernel to $k_j(y \mid \mathbf{x}_i) = g\ \big(\ \frac{y-\mu_i}{\sigma_i}\ \big)\ /M_g$. Under the transformation $Z_i = (Y_i - \mu_i)/\sigma_i$, the expected log-kernel isolates the base distribution:

$$\mathbb{E}_{f^*}[\log k_j(Y_i \mid \mathbf{x}_i)] = \mathbb{E}_{Z\sim g}[\log g(Z) - \log M_g] = \int g(z) \log\left(\frac{g(z)}{M_g}\right) dz. \tag{S7}$$

This integral depends exclusively on the functional form of $g$. Thus, the family constant $\mathrm{Vol}_{\mathrm{family}} = \exp\big(-2\int g(z)\log(g(z)/M_g)dz\ \big)$ is fully invariant to $\boldsymbol{\theta}_j^*$.

**Examples of $\mathrm{Vol}_{\mathrm{family}}$**

(1) **Gaussian:** The standard base density is $g(z) = \frac{1}{\sqrt{2\pi}}e^{-z^2/2}$ with a peak maximum $M_g = 1/\sqrt{2\pi}$. The standardized kernel is $k_g(z) = e^{-z^2/2}$. Evaluating the expecta-tion yields $\mathbb{E}[\log k_g(Z)] = \mathbb{E}[-Z^2/2] = -1/2$. This results in a family constant of $\text{Vol}_{\text{Gaussian}} = \exp(-2 \cdot -1/2) = e$.

(2) **Uniform:** The standard base density is $g(z) = 1$ for $z \in [0, 1]$ with a peak maximum $M_g = 1$. The standardized kernel is $k_g(z) = 1$. The expected log-kernel is exactly $\mathbb{E}[\log k_g(Z)] = 0$, resulting in a family constant of $\text{Vol}_{\text{Uniform}} = \exp(-2 \cdot 0) = 1$.

(3) **Laplace:** The standard base density is $g(z) = \frac{1}{2}e^{-\mid z \mid}$ with a peak maximum $M_g = 1/2$. The standardized kernel is $k_g(z) = e^{-\mid z \mid}$. Evaluating the expectation yields $\mathbb{E}[\log k_g(Z)] = \mathbb{E}[- \mid Z \mid] = -1$. This results in a family constant of $\text{Vol}_{\text{Laplace}} = \exp(-2 \cdot -1) = e^2$.

## A.2 Using Geometric Mean of Regular Variances

This section provides the formal information-theoretic proof that the population entropic variance of a continuous location-scale family resolves into the geometric mean of regular variances, leading to the simplified variance-reduction ratio for $\rho^2_{\text{V}}$.

### A.2.1 Factorization of Entropic Variance with Regular Variance

**Theorem S2 (Factorization of Entropic Variance with Regular Variance).** *Let $Y$ be a continuous random variable with a probability density function $f(y)$ belonging to a location-scale family. Let $\sigma^2_Y$ denote the regular variance* $\text{Var}\ (Y)$*. The entropic variance of $Y$, defined as $v(Y) = \exp(2h(Y))$, factors into:*

$$v(Y) = v_{\text{family}} \cdot \sigma^2_Y, \tag{S8}$$

*where $v_{\text{family}} = \exp(2h(Z))$ is the family base EV, derived from a standardized reference variable $Z$ of the same family such that* $\text{Var}\ (Z) = 1$*.*

*Proof.* By the definition of a location-scale family, the random variable $Y$ can be expressed as an affine transformation of the standardized random variable $Z$:

$$Y = \mu + \sigma_Y Z, \tag{S9}$$

where $\mu = \mathbb{E}[Y]$ and $\sigma_Y = \sqrt{\text{Var}\ (Y)}$.

By the change of variables formula, the probability density function of $Y$ is defined by scaling the density of $Z$ with the Jacobian of the transformation ($\frac{1}{\sigma_Y}$):

$$f_Y(y) = \frac{1}{\sigma_Y} f_Z\left(\frac{y - \mu}{\sigma_Y}\right). \tag{S10}$$

We start with the foundational definition of the differential entropy of $Y$ as the expected negative log-density:

$$h(Y) = \mathbb{E}_Y[-\log f_Y(Y)]. \tag{S11}$$

Substituting the density of $Y$ into this expectation yields:

$$h(Y) = \mathbb{E}_Y\left[-\log\left(\frac{1}{\sigma_Y} f_Z\left(\frac{Y-\mu}{\sigma_Y}\right)\right)\right]. \tag{S12}$$

Using the properties of logarithms, we can separate the scale constant from the density function:

$$h(Y) = \mathbb{E}_Y\left[-\log\left(\frac{1}{\sigma_Y}\right) - \log f_Z\left(\frac{Y-\mu}{\sigma_Y}\right)\right] \tag{S13}$$

$$h(Y) = \mathbb{E}_Y[\log(\sigma_Y)] + \mathbb{E}_Y\left[-\log f_Z\left(\frac{Y-\mu}{\sigma_Y}\right)\right]. \tag{S14}$$

Because $\sigma_Y$ is a deterministic constant, its expectation is simply itself: $\mathbb{E}_Y[\log(\sigma_Y)] = \log(\sigma_Y)$. For the second term, we recognize that the argument inside the function is exactly the standardized variable $Z = \frac{Y-\mu}{\sigma_Y}$. Therefore, taking the expectation over $Y$ is probabilistically equivalent to taking the expectation over $Z$:

$$\mathbb{E}_Y[-\log f_Z(Z)] = \mathbb{E}_Z[-\log f_Z(Z)]. \tag{S15}$$

This expectation is exactly the definition of the differential entropy of $Z$. Thus, the expression cleanly resolves to:

$$h(Y) = \log(\sigma_Y) + h(Z). \tag{S16}$$

We define the entropic variance of $Y$ as the exponentiated twice differential entropy:

$$v(Y) = \exp(2h(Y)) = \exp(2[h(Z) + \log(\sigma_Y)]). \tag{S17}$$

Distributing the exponentiation yields:

$$v(Y) = \exp(2h(Z)) \cdot \exp(2\log(\sigma_Y)) = \exp(2h(Z)) \cdot \sigma_Y^2. \tag{S18}$$

By defining the constant family base EV as $v_{\text{family}} \equiv \exp(2h(Z))$, we recover the exact factorization:

$$v(Y) = v_{\text{family}} \cdot \sigma_Y^2. \tag{S19}$$

### A.2.2 Examples of Family Base EV ($v_{\text{family}}$)

To make the concept of the family base EV concrete, we can evaluate $v_{\text{family}} = \frac{\exp(2h(Y))}{\text{Var }(Y)}$ for several common continuous distributions. Because $v_{\text{family}}$ measures the "entropic bloat" or informational volume of a distribution relative to its Euclidean variance, it serves as a shape-specific geometric constant.

#### The Normal Distribution (Maximum Entropy)

For a Gaussian random variable $Y \sim \mathcal{N}(\mu, \sigma_Y^2)$, the differential entropy is $h(Y) = \frac{1}{2}\log(2\pi e \sigma_Y^2)$. Exponentiating twice and dividing by the variance ($\sigma_Y^2$) yields:

$$v_{\text{Normal}} = 2\pi e \approx 17.079. \tag{S20}$$

By Shannon's Maximum Entropy Principle, the Normal distribution contains the maximum possible uncertainty for a given Euclidean variance. Therefore, $2\pi e$ acts as the theoretical upper bound for $v_{\text{family}}$ among all continuous distributions defined on the real line.

**The Laplace Distribution (Heavy Tails, Sharp Peak)**

For a Laplace random variable $Y \sim \text{Laplace}\ (\mu, b)$, the variance is $\sigma_Y^2 = 2b^2$ and the differential entropy is $h(Y) = \log(2be)$. Substituting $b = \sigma_Y/\sqrt{2}$ and evaluating the signature gives:

$$v_{\text{Laplace}} = 2e^2 \approx 14.778. \tag{S21}$$

Because the Laplace distribution is more sharply peaked around the mean than the Gaussian, it occupies less informational volume for the same amount of Euclidean variance, resulting in a strictly smaller $v_{\text{family}}$.

**The Gamma Distribution (Shape-Dependent Signature)**

The Gamma distribution, $Y \sim \text{Gamma}\ (\alpha, \beta)$, provides a crucial counter-example because it is a *shape-scale* family, not a location-scale family. Its differential entropy depends heavily on the shape parameter $\alpha$, while the rate parameter $\beta$ merely scales it. The variance is $\text{Var}\ (Y) = \alpha/\beta^2$. Calculating the signature yields:

$$v_{\text{Gamma}}(\alpha) = \frac{e^{2\alpha}\Gamma(\alpha)^2 e^{2(1-\alpha)\psi(\alpha)}}{\alpha}, \tag{S22}$$

where $\Gamma(\cdot)$ is the Gamma function and $\psi(\cdot)$ is the digamma function.

Unlike the Normal and Laplace distributions, $v_{\text{Gamma}}$ is not a single universal constant; it is a strictly increasing function of the shape parameter $\alpha$. The following table illustrates this shape-dependency across several values of $\alpha$.

| Shape ($\alpha$) | $v_{\text{Gamma}}(\alpha)$ | Distribution Equivalent |
|---|---:|---|
| **0.5** | 2.397 | Heavily right-skewed (e.g., $\chi^2$ with 1 df) |
| **1.0** | 7.389 | Exponential Distribution |
| **2.0** | 11.720 | |

This table beautifully illustrates the mechanics of Information Geometry. As the shape parameter $\alpha$ increases, the Gamma distribution becomes increasingly symmetric. Consequently, its entropic volume ($v_{\text{Gamma}}$) steadily climbs toward the theoretical maximum bounds of the Gaussian distribution ($2\pi e \approx 17.079$). The entropic variance naturally absorbs not just the scale of the data, but the specific geometric reality dictated by its shape.

### A.2.3 $\rho_V^2$ as proportional reduction of Geometric Mean of Variances

**Theorem S3 (Geometric Mean Variance Reduction).** *For a dataset of $n$ independent observations and nested predictive models $\mathcal{M}_0 \subset \mathcal{M}_1$ sharing the same continuous* *location-scale family, the population partial effect size $\rho^2_{\mathrm{V}}$ is strictly equal to the complement of the ratio of the geometric means of their conditional regular variances.*

*Proof.* Let the conditional regular variance for the $i$-th observation under model $\mathcal{M}_j$ be $\sigma^2_{j,i} = \mathrm{Var}\ (Y_i \mid \mathbf{x}_i, \boldsymbol{\theta}^*_j)$. Applying Theorem S2 to the individual EV defined in the main text gives:

$$v^*_{j,i}(\boldsymbol{\theta}^*_j; \mathbf{x}_i) = v_{\text{family}} \cdot \sigma^2_{j,i}. \tag{S23}$$

The population EV is defined as the geometric mean of the individual EVs across the $n$ observations:

$$\nu^*_j = \mathrm{GM}\ \left(\ [v^*_{j,i}(\boldsymbol{\theta}^*_j; \mathbf{x}_i)]^n_{i=1}\ \right) = \left(\prod_{i=1}^{n} v_{\text{family}} \cdot \sigma^2_{j,i}\right)^{1/n}. \tag{S24}$$

Because $\mathrm{v}_{\text{family}}$ is a global constant with respect to the covariates $\mathbf{x}_i$ and the parameters $\boldsymbol{\theta}^*_j$, it factors entirely out of the geometric mean:

$$\nu^*_j = v_{\text{family}} \left(\prod_{i=1}^{n} \sigma^2_{j,i}\right)^{1/n} = v_{\text{family}} \cdot \mathrm{GM}\ \left(\ [\sigma^2_{j,i}]^n_{i=1}\ \right). \tag{S25}$$

The entropic variance $\rho^2$ is defined as the relative reduction in population EV. Substituting the factored forms for $\mathcal{M}_0$ and $\mathcal{M}_1$ yields:

$$\rho^2_{\mathrm{V}}(\mathcal{M}_1 \mid \mathcal{M}_0) = 1 - \frac{\nu^*_1}{\nu^*_0} = 1 - \frac{v_{\text{family}} \cdot \mathrm{GM}\ \left(\ [\sigma^2_{1,i}]^n_{i=1}\ \right)}{v_{\text{family}} \cdot \mathrm{GM}\ \left(\ [\sigma^2_{0,i}]^n_{i=1}\ \right)}. \tag{S26}$$

The family signature $\mathrm{v}_{\text{family}}$ cancels out, proving the corollary:

$$\rho^2_{\mathrm{V}}(\mathcal{M}_1 \mid \mathcal{M}_0) = 1 - \frac{\mathrm{GM}\ \left(\ [\sigma^2_{1,i}]^n_{i=1}\ \right)}{\mathrm{GM}\ \left(\ [\sigma^2_{0,i}]^n_{i=1}\ \right)}. \tag{S27}$$

**Remark (Entropic Variance for Discrete Variables)**

Theorem S2 applies exclusively to continuous random variables. This factorization does not extend to discrete variables because discrete Shannon entropy is a dimensionless quantity derived purely from probability masses, making it inherently scale-invariant. Unlike continuous differential entropy, which captures changes in Euclidean spread via the Jacobian of a transformation, discrete entropy requires no such scaling factor.

Because discrete entropy is dimensionless, separating a geometric signature from a Euclidean scale is unnecessary. Instead, the Entropic Variance for a discrete response $Y$, $v(Y) = \exp(2h(Y))$, directly represents the squared perplexity or effective support size. The underlying quantity $\exp(h(Y))$ yields the equivalent number of equally likely states (Shannon, 1948), a concept formalized as the "effective number of types" in ecology (Hill, 1973) and "perplexity" in information theory (Jelinek et al., 1977; Jurafsky & Martin, 2009). Consequently, $v(Y)$ provides a standalone, non-Euclidean metric of categorical uncertainty by measuring the squared effective number of categories.

## A.3 Using Mahalanobis Distance on Log-likelihood

**Theorem S4 (Asymptotic Equivalence of EV Reduction and Mahalanobis Distance).** *Assuming the true data generating process is contained within the full model ($f^* \in \mathcal{M}_1$) under regular likelihood conditions, the population partial effect size $\rho^2_{\mathrm{V}}$ is asymptotically equivalent to a transformation of the squared Mahalanobis distance between the true and restricted parameters.*

*Proof.* By Bartlett's identities (Bartlett, 1937; White, 1982), the expected negative Hessian of the cross-entropy function is equivalent to the expected Fisher Information matrix evaluated under the true density $f_1^* = f_1(Y_i \mid \mathbf{x}_i, \boldsymbol{\theta}_1^*)$:

$$\overline{\mathcal{I}}_1(\boldsymbol{\theta}_1^*) = -\frac{1}{n}\sum_{i=1}^{n} \mathbb{E}_{f_1^*}\left[\nabla^2_{\boldsymbol{\theta}_1} \log f_1(Y_i \mid \mathbf{x}_i, \boldsymbol{\theta}_1)\Big|_{\boldsymbol{\theta}_1=\boldsymbol{\theta}_1^*}\right]. \tag{S28}$$

Because the restricted model is nested ($\mathcal{M}_0 \subset \mathcal{M}_1$), both models can be evaluated within the full parameter space, implying $h_0^*(\boldsymbol{\theta}_0^*) = h_1^*(\boldsymbol{\theta}_0^*)$. We approximate the cross-entropy difference $\Delta h^*$ using a second-order Taylor expansion of $h_1^*(\boldsymbol{\theta})$ around its minimum $\boldsymbol{\theta}_1^*$. This natural approximation yields a quadratic form:

$$\Delta h^* \approx \frac{1}{2}(\boldsymbol{\theta}_0^* - \boldsymbol{\theta}_1^*)^\top \overline{\mathcal{I}}_1(\boldsymbol{\theta}_1^*)(\boldsymbol{\theta}_0^* - \boldsymbol{\theta}_1^*) = \frac{1}{2}D^2_{\mathrm{MD}}\left(\ \boldsymbol{\theta}_0^*, \boldsymbol{\theta}_1^*; \overline{\mathcal{I}}_1(\boldsymbol{\theta}_1^*)^{-1}\ \right), \tag{S29}$$

where $D^2_{\mathrm{MD}}(a, b; \Sigma) = (a-b)^\top \Sigma^{-1}(a-b)$ denotes the squared Mahalanobis distance. Substituting this approximation directly into the definition of $\rho^2_{\mathrm{V}}$ gives:

$$\rho^2_{\mathrm{V}}(\mathcal{M}_1 \mid \mathcal{M}_0) \approx 1 - \exp\Big(-D^2_{\mathrm{MD}}\left(\ \boldsymbol{\theta}_0^*, \boldsymbol{\theta}_1^*; \overline{\mathcal{I}}_1(\boldsymbol{\theta}_1^*)^{-1}\ \right)\Big). \tag{S30}$$

To isolate the parameters of interest, consider a partition of the full parameter vector $\boldsymbol{\theta}_1 = (\boldsymbol{\psi}, \boldsymbol{\phi})$, where $\boldsymbol{\psi}$ represents the parameters of interest and $\boldsymbol{\phi}$ denotes the nuisance parameters. If the restricted model $\mathcal{M}_0$ is defined by the constraint $\boldsymbol{\psi} = \mathbf{0}$, the true parameter vector under $\mathcal{M}_1$ is $\boldsymbol{\theta}_1^* = (\boldsymbol{\psi}^*, \boldsymbol{\phi}^*)$.

Correspondingly, we partition the inverse Fisher Information matrix into blocks:

$$\overline{\mathcal{I}}_1(\boldsymbol{\theta}_1^*)^{-1} = \begin{pmatrix} \Sigma_{\psi\psi} & \Sigma_{\psi\phi} \\ \Sigma_{\phi\psi} & \Sigma_{\phi\phi} \end{pmatrix}, \tag{S31}$$

where $\Sigma_{\psi\psi}$ is the marginal asymptotic covariance of $\boldsymbol{\psi}$. By applying the Schur complement, the Mahalanobis distance in Eq. (S29) simplifies entirely to the restricted parameter space, completing the proof:

$$D^2_{\mathrm{MD}}\left(\ \boldsymbol{\theta}_0^*, \boldsymbol{\theta}_1^*; \overline{\mathcal{I}}_1(\boldsymbol{\theta}_1^*)^{-1}\ \right) = (\boldsymbol{\psi}^*)^\top \Sigma^{-1}_{\psi\psi} \boldsymbol{\psi}^*. \tag{S32}$$

# B Asymptotic Theory of Empirical Entropic Variances

## B.1 Notations and Definitions

Our predictive metrics are based on the cross-entropy between the true data-generating process (DGP), denoted $f^*$, and the parametric model family $\mathcal{M}_j$ with parameter $\boldsymbol{\theta}_j$. **To formally define these metrics, we adopt a consistent notational convention.** Throughout this paper, the superscript $*$ (e.g., $h^*(\cdot)$ or $v^*(\cdot)$) designates quantities evaluated in expectation with respect to the true DGP, $f^*$, whereas a circumflex $(\hat{\cdot})$ indicates their empirical counterparts calculated from the training data. The parameter vector $\boldsymbol{\theta}$ of the **evaluated predictive density** $f_j(y_i \mid \mathbf{x}_i, \boldsymbol{\theta})$ is included as the argument of these metric functions to indicate it is the metric of the evaluted model with parameter vector $\boldsymbol{\theta}$.

Consider $n$ independent observations $(y_i, \mathbf{x}_i)$ and a sequence of nested parametric models $\mathcal{M}_0 \subset ... \subset \mathcal{M}_J$. Each $\mathcal{M}_j$ specifies a conditional predictive density $f_j(y_i \mid \mathbf{x}_i, \boldsymbol{\theta}_j)$. We use the $\hat{A}$ for empirical quantity $A$ calculated on the sample, and use $A^*$ for the corresponding population quantity taking expectations over $f^*$.

**Population EVs and Effect Sizes**

- **Expected Entropy and Population EV:** For a given parameter vector $\boldsymbol{\theta}$, the **expected mean-cross-entropy** $h_j^*(\boldsymbol{\theta})$ is the arithmetic mean of the individual expected entropies under the true DGP. The **population EV function** $v_j^*(\boldsymbol{\theta})$ is the corresponding geometric mean of the individual EVs:

$$\begin{aligned} h_j^*(\boldsymbol{\theta}) &= \frac{1}{n}\sum_{i=1}^{n} \mathbb{E}_{Y_i \sim f^*}\left[ -\log f_j(Y_i \mid \mathbf{x}_i, \boldsymbol{\theta}) \right], \\ v_j^*(\boldsymbol{\theta}) &= \exp(2h_j^*(\boldsymbol{\theta})). \end{aligned} \tag{S33}$$

- **The True Parameter and Optimum EV:** The true parameter $\boldsymbol{\theta}_j^*$ is the projection that minimizes the total expected cross-entropy: $\boldsymbol{\theta}_j^* = \arg\min_{\boldsymbol{\theta}} h_j^*(\boldsymbol{\theta})$. The **population EV** $\nu_j^*$ is defined as the EV function evaluated at this theoretical optimum:

$$\nu_j^* = v_j^*(\boldsymbol{\theta}_j^*) = \exp(2h_j^*(\boldsymbol{\theta}_j^*)). \tag{S34}$$

- **Population EV SNR:** When comparing two nested models $\mathcal{M}_0 \subset \mathcal{M}_1$, the **population SNR** represents the Signal-to-Noise Ratio of the information gain, defined as the relative inflation of the restricted model's optimum EV over the fuller model's optimum EV:

$$\phi_{\mathrm{V}}^2 = \frac{\nu_0^* - \nu_1^*}{\nu_1^*} = \frac{\nu_0^*}{\nu_1^*} - 1 = \exp\left( 2(h_0^*(\boldsymbol{\theta}_0^*) - h_1^*(\boldsymbol{\theta}_1^*)) \right) - 1. \tag{S35}$$

- **Entropy-Based Partial Effect Sizes:** Analogous to the proportion of variance explained in OLS, the **population partial effect size** $\rho_{\mathrm{V}}^2$ measures the intrinsic theoretical predictivity gained by the fuller model:

$$\rho^2_{\mathrm{V}}(\mathcal{M}_1 \mid \mathcal{M}_0) = \frac{\phi^2_{\mathrm{V}}}{1+\phi^2_{\mathrm{V}}} = 1 - \frac{\nu^*_1}{\nu^*_0}. \tag{S36}$$

Conversely, the **predictive partial effect size** $\rho^2_{\mathrm{VP}}$ captures the actual deployment performance by evaluating the *estimated* parameters on the population space:

$$\rho^2_{\mathrm{VP}}(\widehat{\mathcal{M}}_1 \mid \widehat{\mathcal{M}}_0) = 1 - \frac{\exp(2\mathbb{E}\big(h^*_1(\hat{\boldsymbol{\theta}}_1)\big))}{\exp(2\mathbb{E}\big(h^*_0(\hat{\boldsymbol{\theta}}_0)\big))}. \tag{S37}$$

**Empirical Entropy and EV**

- **Empirical Entropy:** The **total empirical entropy**, $\hat{H}_j(\boldsymbol{\theta}_j)$, is the negative log-likelihood evaluated at a specific parameter, and the **average empirical entropy**, $\hat{h}_j(\boldsymbol{\theta}_j)$, is its per-observation mean:

$$\begin{aligned} \hat{H}_j(\boldsymbol{\theta}_j) &= -\sum_{i=1}^{n} \log f_j(y_i \mid \mathbf{x}_i, \boldsymbol{\theta}_j), \\ \hat{h}_j(\boldsymbol{\theta}_j) &= \frac{\hat{H}_j(\boldsymbol{\theta}_j)}{n}. \end{aligned} \tag{S38}$$

- **Empirical Deviance:** Instead of anchoring to a hypothetical saturated model, we measure deviance relative to a model's own theoretical optimum. Let $\hat{H}_j(\boldsymbol{\theta}^*_j)$ denote the empirical entropy evaluated at the true KL projection parameter $\boldsymbol{\theta}^*_j$ for model $\mathcal{M}_j$. The **absolute deviance** of a model measures the empirical reduction in entropy achieved by its maximum likelihood estimate $\hat{\boldsymbol{\theta}}_j$ relative to its own theoretical topline:

$$D^{\mathrm{abs}}_j = 2\Big[\hat{H}_j(\boldsymbol{\theta}^*_j) - \hat{H}_j(\hat{\boldsymbol{\theta}}_j)\Big]. \tag{S39}$$

  For two nested models $\mathcal{M}_0 \subset \mathcal{M}_1$, the **relative deviance** directly measures the incremental reduction in entropy between the two empirical fits:

$$D_{1\setminus 0} = 2\Big[\hat{H}_0(\hat{\boldsymbol{\theta}}_0) - \hat{H}_1(\hat{\boldsymbol{\theta}}_1)\Big]. \tag{S40}$$

- **Relationship between $\boldsymbol{D_{1\setminus 0}}$ and $\boldsymbol{D^{\mathrm{abs}}_j}$:**
  - Because absolute deviances are anchored to model-specific toplines, $D_{1\setminus 0}$ is not generally equal to $D^{\mathrm{abs}}_1 - D^{\mathrm{abs}}_0$. If the fuller model captures true signal ($\rho^2_{\mathrm{V}} > 0$), its theoretical minimum entropy is lower, meaning $\hat{H}_0(\boldsymbol{\theta}^*_0) \neq \hat{H}_1(\boldsymbol{\theta}^*_1)$.
  - However, under the null hypothesis ($\rho^2_{\mathrm{V}} = 0$), the true KL projection onto $\mathcal{M}_1$ collapses perfectly into the restricted subspace $\mathcal{M}_0$ (i.e., $\boldsymbol{\theta}^*_1 = \boldsymbol{\theta}^*_0$). Thus, both models yield the exact same probability distribution, $f_1(y \mid \mathbf{x}, \boldsymbol{\theta}^*_1) = f_0(y \mid \mathbf{x}, \boldsymbol{\theta}^*_0)$. Consequently, they share the exact same theoretical topline ($\hat{H}_0(\boldsymbol{\theta}^*_0) = \hat{H}_1(\boldsymbol{\theta}^*_1)$), and only in this specific scenario does the classical identity hold: $D_{1\setminus 0} = D^{\mathrm{abs}}_1 - D^{\mathrm{abs}}_0$.
- **Empirical EV:** We define the **empirical EV** per-observation, $\hat{v}_j(\boldsymbol{\theta}_j)$, and the **total**

**empirical EV**, $\hat{V}_j(\boldsymbol{\theta}_j)$, as EV summaries of the model fit:

$$\begin{aligned} \hat{v}_j(\boldsymbol{\theta}_j) &= \exp\Big(2\hat{h}_j(\boldsymbol{\theta}_j)\Big), \\ \hat{V}_j(\boldsymbol{\theta}_j) &= n\hat{v}_j(\boldsymbol{\theta}_j) = n\exp\left(\frac{2\hat{H}_j(\boldsymbol{\theta}_j)}{n}\right). \end{aligned} \tag{S41}$$

## B.2 The Absolute Deflation Theorem for EVs

**Theorem S5 (The Deflation of Training EV).** *Let $\mathcal{M}$ be a regular parametric model with dimension $p$. Using the EV notations established in Sec. B.1, let $\hat{v}(\hat{\boldsymbol{\theta}})$ denote the empirical EV evaluated at the MLE, and $\hat{v}(\boldsymbol{\theta}^*)$ denote the topline EV evaluated at the true parameter $\boldsymbol{\theta}^*$.*

*We define the **Absolute EV SNR** ($\widehat{\mathrm{SNR}}_{\mathrm{V}}^{\mathrm{abs}}$) as the relative inflation of the truth topline over the training fit:*

$$\widehat{\mathrm{SNR}}_{\mathrm{V}}^{\mathrm{abs}} = \frac{\hat{v}(\boldsymbol{\theta}^*) - \hat{v}(\hat{\boldsymbol{\theta}})}{\hat{v}(\hat{\boldsymbol{\theta}})}. \tag{S42}$$

*Under the null hypothesis that the model class $\mathcal{M}$ contains the true DGP, the finite-sample geometry of the variance space is approximated by the ratio distribution $X_{\mathrm{ratio}} = \chi^2_p/\chi^2_{n-p}$. This yields three equivalent stochastic representations for the absolute deflation:*

1. ***Ratio of Independent Chi-Squares:***

$$\widehat{\mathrm{SNR}}_{\mathrm{V}}^{\mathrm{abs}} \quad \dot{\sim} \quad \frac{\chi^2_p}{\chi^2_{n-p}}. \tag{S43}$$

2. ***Beta for the Proportion of Residual EV:***

$$\widehat{\mathrm{ER}}^{\mathrm{abs}} = \frac{1}{1+\widehat{\mathrm{SNR}}_{\mathrm{V}}^{\mathrm{abs}}} = \frac{\hat{v}(\hat{\boldsymbol{\theta}})}{\hat{v}(\boldsymbol{\theta}^*)} \quad \dot{\sim} \quad \mathrm{Beta}\left(\frac{n-p}{2}, \frac{p}{2}\right). \tag{S44}$$

3. ***The central $F$ distribution for $\mathbf{F}_{\mathbf{V}}$:.***

$$F_{\mathrm{V}}^{\mathrm{abs}} = \left(\frac{n-p}{p}\right)\widehat{\mathrm{SNR}}_{\mathrm{V}}^{\mathrm{abs}} \quad \dot{\sim} \quad F_{p,n-p}. \tag{S45}$$

*This approximation is **first-order accurate** at the $O(n^{-1})$ scale. At the $O(n^{-2})$ scale, the MGFs exhibit a discrepancy governed by the Bartlett correction $c$. The residual error of the ratio approximation expands as:*

$$M_{\widehat{\mathrm{SNR}}_{\mathrm{V}}^{\mathrm{abs}}}(t) - M_{X_{\mathrm{ratio}}}(t) = \frac{t}{n^2}\left[c - \frac{p(p+2)}{2}\right] + O(n^{-3}). \tag{S46}$$

*Proof* (Proof via MGFs).

**Step 1: Asymptotic Expansion of the Deviance**

The EV is defined by the empirical average entropy, $\hat{v} = \exp(2\hat{H}/n)$. Let $D = 2[\hat{H}(\boldsymbol{\theta}^*) - \hat{H}(\hat{\boldsymbol{\theta}})]$ be the absolute deviance of the fitted model to the the true model. The absolute deflation ratio can be expressed with the SNR:

$$\widehat{\mathrm{SNR}}_{\mathrm{V}}^{\mathrm{abs}} = \exp\left(\frac{D}{n}\right) - 1, \tag{S47}$$

where From likelihood theory, the deviance admits a higher-order MGF expansion yielding the Bartlett-corrected moments:

$$\mathbb{E}[D] = p + \frac{c}{n} + O(n^{-2}), \qquad \mathbb{E}[D^2] = p(p+2) + O(n^{-1}), \tag{S48}$$

where $c$ encapsulates the statistical curvature of the specific model manifold.

**Step 2: MGF of the Estimator $\widehat{\mathrm{SNR}}_{\mathrm{V}}^{\mathrm{abs}}$**

We establish the MGF of the Estimator by applying a Taylor expansion to the exponential function around $D/n = 0$:

$$\widehat{\mathrm{SNR}}_{\mathrm{V}}^{\mathrm{abs}} = \frac{D}{n} + \frac{D^2}{2n^2} + O_p(n^{-3}). \tag{S49}$$

For a fixed real $t$, the exponential mapping $e^{t\widehat{\mathrm{SNR}}_{\mathrm{V}}^{\mathrm{abs}}}$ expands as:

$$\exp\left(t\widehat{\mathrm{SNR}}_{\mathrm{V}}^{\mathrm{abs}}\right) = 1 + \frac{tD}{n} + \frac{tD^2}{2n^2} + \frac{t^2 D^2}{2n^2} + O_p(n^{-3}). \tag{S50}$$

Taking the expectation and substituting the Bartlett-corrected moments yields:

$$M_{\widehat{\mathrm{SNR}}_{\mathrm{V}}^{\mathrm{abs}}}(t) = 1 + \frac{t}{n}\left(p + \frac{c}{n}\right) + \frac{t + t^2}{2n^2} p(p+2) + O(n^{-3}). \tag{S51}$$

Distributing the terms and strictly truncating at the $O(n^{-2})$ scale provides the exact operational MGF:

$$\begin{aligned} M_{\widehat{\mathrm{SNR}}_{\mathrm{V}}^{\mathrm{abs}}}(t) = 1 + \frac{tp}{n} + \frac{t}{n^2}\left[c + \frac{p(p+2)}{2}\right] \\ + \frac{t^2 p(p+2)}{2n^2} + O(n^{-3}). \end{aligned} \tag{S52}$$

**Step 3: MGF of the Approximating Distribution $X_{\mathrm{ratio}}$**

Consider the ratio approximation $X_{\mathrm{ratio}} = U/V$, where $U \sim \chi^2_p$ and $V \sim \chi^2_{n-p}$ are independent. Expanding its inverse moments asymptotically in $n$:

$$\begin{aligned} \mathbb{E}[U/V] &= \frac{p}{n-p-2} = \frac{p}{n} + \frac{p(p+2)}{n^2} + O(n^{-3}), \\ \mathbb{E}[U^2/V^2] &= \frac{p(p+2)}{n^2} + O(n^{-3}). \end{aligned} \tag{S53}$$

Substituting these into the MGF expansion $M_{X_{\text{ratio}}}(t) = 1 + t\mathbb{E}[U/V] + \frac{t^2}{2}\mathbb{E}[U^2/V^2]$ yields:

$$M_{X_{\text{ratio}}}(t) = 1 + \frac{tp}{n} + \frac{tp(p+2)}{n^2} + \frac{t^2 p(p+2)}{2n^2} + O(n^{-3}). \tag{S54}$$

**Step 4: Derivation of the MGF Discrepancy**

We subtract the MGF of the ratio approximation from the Estimator MGF to isolate the residual error:

$$M_{\widehat{\text{SNR}}_{\text{V}}^{\text{abs}}}(t) - M_{X_{\text{ratio}}}(t) = \frac{t}{n^2}\left[c - \frac{p(p+2)}{2}\right] + O(n^{-3}). \tag{S55}$$

This exact difference confirms first-order accuracy while establishing the geometric bounds of the approximation.

**Remarks on the Approximating Discrepancy by Comparing to Wilks' Theorem**

The advantage of $X_{\text{ratio}}$ becomes starkly apparent when contrasted with classical likelihood theory. The classical limit relies on $D \dot{\sim} \chi^2_p$. Mapped into the variance space, this implies the approximating distribution $X_{\text{Wilks}} = \exp(\chi^2_p/n) - 1$ for $\widehat{\text{SNR}}_{\text{V}}^{\text{abs}}$. Expanding $X_{\text{Wilks}}$ to match the estimator's form gives:

$$X_{\text{Wilks}} = \frac{\chi^2_p}{n} + \frac{(\chi^2_p)^2}{2n^2} + O_p(n^{-3}). \tag{S56}$$

Using the standard moments of the $\chi^2_p$ distribution ($\mathbb{E}[\chi^2_p] = p$ and $\mathbb{E}[(\chi^2_p)^2] = p(p+2)$), the MGF of this classical representation expands as:

$$\begin{aligned} M_{X_{\text{Wilks}}}(t) &= 1 + t\mathbb{E}[X_{\text{Wilks}}] + \frac{t^2}{2}\mathbb{E}[X^2_{\text{Wilks}}] \\ &= 1 + \frac{tp}{n} + \frac{tp(p+2)}{2n^2} + \frac{t^2 p(p+2)}{2n^2} + O(n^{-3}). \end{aligned} \tag{S57}$$

Subtracting this from the Estimator MGF reveals the classical residual error:

$$M_{\widehat{\text{SNR}}_{\text{V}}^{\text{abs}}}(t) - M_{X_{\text{Wilks}}}(t) = \frac{tc}{n^2} + O(n^{-3}). \tag{S58}$$

By evaluating this discrepancy, we uncover a profound geometric reality. The classical exponential mapping successfully handles the scaling of the numerator, but because it completely lacks the $\chi^2_{n-p}$ denominator, it fails to account for the dimensional degradation of the residual space. In the canonical setting of exact Gaussian linear models, the Bartlett correction is explicitly $c = p(p+2)/2$. In this scenario, the residual error for the ratio approximation $X_{\text{ratio}}$ collapses perfectly to zero ($M_{\widehat{\text{SNR}}_{\text{V}}^{\text{abs}}} - M_{X_{\text{ratio}}} = 0$), making the ratio $F$-test exact for all $n$. In stark contrast, the classical $X_{\text{Wilks}}$ approximation retains an unmitigated error of $t[p(p+2)/2]/n^2$ even in exact linear models. Thus, $X_{\text{ratio}}$ naturally absorbs the fundamental Euclidean geometry of the model manifold, rendering it significantly more accurate in finite samples.

## B.3 The Central Distribution for Relative EV-SNR

### Central EV-SNR Distribution

**Theorem S6 (Central EV-SNR Distribution).** *Let $\mathcal{M}_0 \subset \mathcal{M}_1$ be two nested regular parametric models with dimensions $p_0$ and $p_1$, where $k = p_1 - p_0$. Using the notation established in Sec. B.1, we define the* ***Relative EV-SNR*** *($\widehat{\text{SNR}}_{\text{V}}$) as the relative improvement in EV:*

$$\widehat{\text{SNR}}_{\text{V}} = \frac{\hat{v}_0(\hat{\boldsymbol{\theta}}_0) - \hat{v}_1(\hat{\boldsymbol{\theta}}_1)}{\hat{v}_1(\hat{\boldsymbol{\theta}}_1)}. \tag{S59}$$

*Under the null hypothesis where the fuller model provides no intrinsic predictive gain ($\rho_{\text{V}}^2 = 0$), $\widehat{\text{SNR}}_{\text{V}}$ is approximated by the ratio distribution $X_{\text{ratio}} = \chi^2_k/\chi^2_{n-p_1}$. This provides three equivalent distributional forms:*

1. ***Independent Central Chi-Squares Ratio for EV-SNR:***

$$\widehat{\text{SNR}}_{\text{V}} \quad \dot{\sim} \quad \frac{\chi^2_k}{\chi^2_{n-p_1}}. \tag{S60}$$

2. ***Beta Distribution for the Proportion of Residual EV:***

$$\widehat{\text{ER}} = \frac{1}{1+\widehat{\text{SNR}}_{\text{V}}} = \frac{\hat{v}_1(\hat{\boldsymbol{\theta}}_1)}{\hat{v}_0(\hat{\boldsymbol{\theta}}_0)} \quad \dot{\sim} \quad \text{Beta}\left(\frac{n-p_1}{2}, \frac{k}{2}\right). \tag{S61}$$

3. ***Central $F$-distribution for $F_{\text{V}}$:***

$$F_{\text{V}}^{\text{rel}} \equiv \left(\frac{n-p_1}{k}\right)\widehat{\text{SNR}}_{\text{V}} \quad \dot{\sim} \quad F_{k,n-p_1}. \tag{S62}$$

*This approximation is accurate to $O(n^{-1})$. At the $O(n^{-2})$ scale, the deviation between the Estimator and the ratio approximation is quantified by:*

$$M_{\widehat{\text{SNR}}_{\text{V}}}(t) - M_{X_{\text{ratio}}}(t) = \frac{t}{n^2}\left[c_{01} + \frac{k(k+2)}{2} - k(p_1+2)\right] + O(n^{-3}), \tag{S63}$$

*where $c_{01}$ represents the relative Bartlett curvature between the nested manifolds.*

*Proof.*

**Step 1: Multiplicative Identity and Asymptotic Independence**

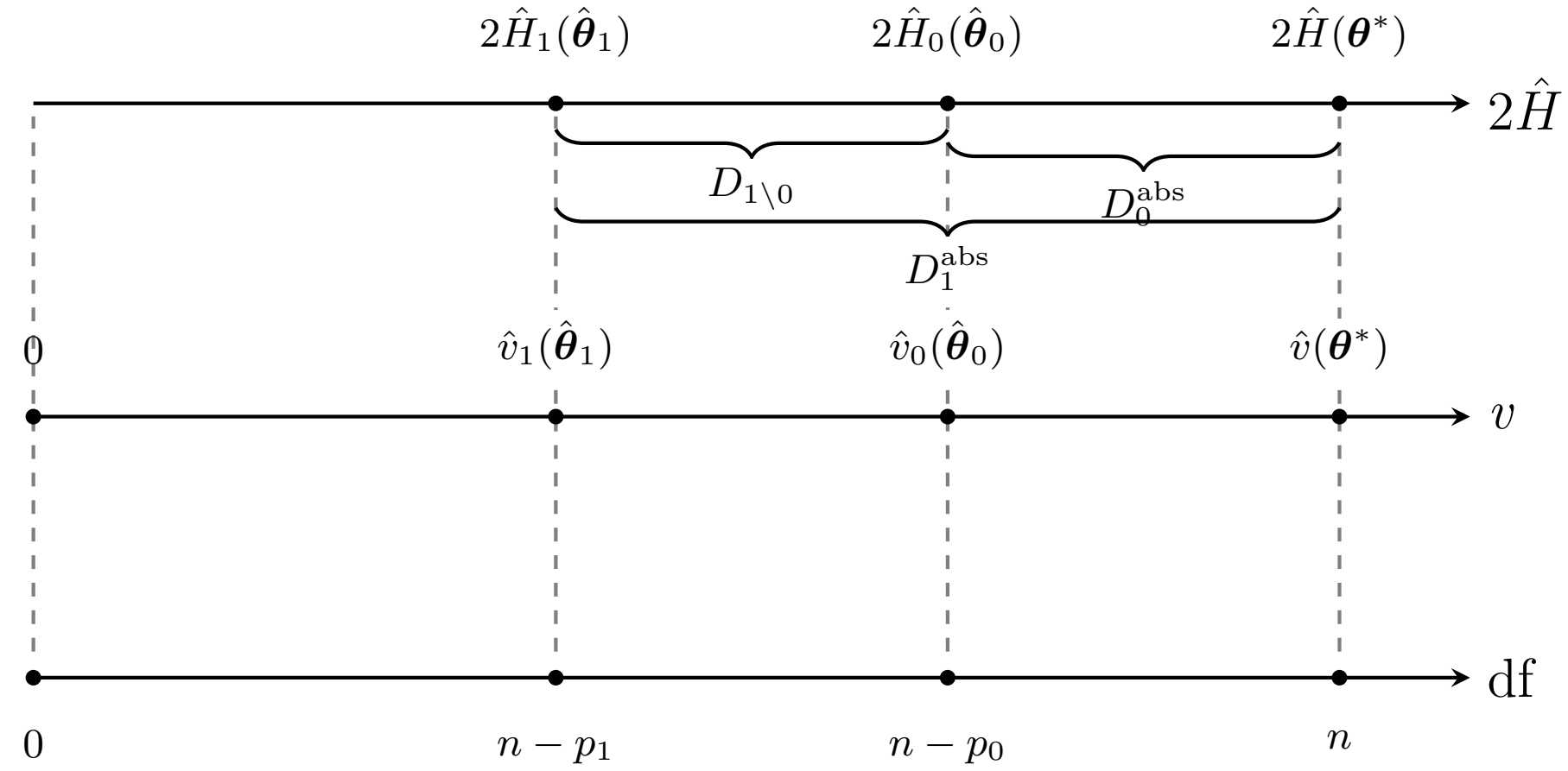

Figure S8: Illustration of the Proof of Theorem S6: empirical entropy ($2\hat{H}$), deviance, EV ($v$), residual degrees of freedom (df).

We begin directly with the definitions of the empirical EV proportions. The total absolute error rate of the fuller model ($\widehat{\text{ER}}_1^{\text{abs}}$), the absolute error rate of the restricted model ($\widehat{\text{ER}}_0^{\text{abs}}$), and the relative retention rate between them ($\widehat{\text{ER}}_{1\backslash 0}$), which corresponds to the Proportion of Residual EV, $\widehat{\text{ER}}$, in the theorem statement) are defined as:

$$\widehat{\text{ER}}_1^{\text{abs}} = \frac{\hat{v}_1(\hat{\boldsymbol{\theta}}_1)}{\hat{v}(\boldsymbol{\theta}^*)}, \quad \widehat{\text{ER}}_0^{\text{abs}} = \frac{\hat{v}_0(\hat{\boldsymbol{\theta}}_0)}{\hat{v}(\boldsymbol{\theta}^*)}, \quad \widehat{\text{ER}}_{1\backslash 0} = \frac{\hat{v}_1(\hat{\boldsymbol{\theta}}_1)}{\hat{v}_0(\hat{\boldsymbol{\theta}}_0)}. \tag{S64}$$

By basic algebraic substitution, the total absolute deflation trivially factors into a multiplicative identity of the nested components:

$$\widehat{\text{ER}}_1^{\text{abs}} = \left(\frac{\hat{v}_0(\hat{\boldsymbol{\theta}}_0)}{\hat{v}(\boldsymbol{\theta}^*)}\right)\left(\frac{\hat{v}_1(\hat{\boldsymbol{\theta}}_1)}{\hat{v}_0(\hat{\boldsymbol{\theta}}_0)}\right) = \widehat{\text{ER}}_0^{\text{abs}} \cdot \widehat{\text{ER}}_{1\backslash 0}. \tag{S65}$$

To establish the stochastic properties of these components, we map them to their corresponding deviances. Because EV is the exponential mapping of cross-entropy, the variance proportions correspond exactly to the classical deviance statistics via $\widehat{\text{ER}} = \exp(-D^{\text{abs}}/n)$. The multiplicative variance identity is therefore structurally equivalent to the additive deviance decomposition:

$$D_1^{\text{abs}} = D_0^{\text{abs}} + D_{1\backslash 0}. \tag{S66}$$

Under classical likelihood geometry, the parameter space of the restricted model $\mathcal{M}_0$ is a nested manifold within $\mathcal{M}_1$. The orthogonal projection of the data onto these nested manifolds dictates that the restricted absolute deviance ($D_0^{\text{abs}}$) and the relative deviance increment ($D_{1\backslash 0}$) converge to asymptotically independent chi-square random variables.

Because $\widehat{\text{ER}}_0^{\text{abs}}$ and $\widehat{\text{ER}}_{1\backslash 0}$ are monotonic transformations of asymptotically independent random variables ($D_0^{\text{abs}} \perp D_{1\backslash 0}$), they themselves must be asymptotically independent:

$$\widehat{\mathrm{ER}}_0^{\mathrm{abs}} \perp \widehat{\mathrm{ER}}_{1\backslash 0}. \tag{S67}$$

This asymptotic independence naturally permits the factorization of their asymptotic fractional moments for any power $s$:

$$\mathbb{E}[(\widehat{\mathrm{ER}}_1^{\mathrm{abs}})^s] \approx \mathbb{E}[(\widehat{\mathrm{ER}}_0^{\mathrm{abs}})^s] \cdot \mathbb{E}[\widehat{\mathrm{ER}}_{1\backslash 0}^s]. \tag{S68}$$

**Step 2: Identification via Fractional Moments**

By the theorem of absolute deflation, the absolute error rates converge to Beta distributions: $\widehat{\mathrm{ER}}_1^{\mathrm{abs}} \dot\sim \text{Beta}\,(\frac{n-p_1}{2}, \frac{p_1}{2})$ and $\widehat{\mathrm{ER}}_0^{\mathrm{abs}} \dot\sim \text{Beta}\,(\frac{n-p_0}{2}, \frac{p_0}{2})$. For a Beta distribution with parameters $a$ and $b$, the $s$-th fractional moment is $\mathbb{E}[R^s] = \frac{\Gamma(a+s)\Gamma(a+b)}{\Gamma(a)\Gamma(a+b+s)}$.

Notice that for both $\widehat{\mathrm{ER}}_1^{\mathrm{abs}}$ and $\widehat{\mathrm{ER}}_0^{\mathrm{abs}}$, the sum of their parameters is identically $a + b = \frac{n}{2}$. Expanding their respective moments yields:

$$\mathbb{E}[(\widehat{\mathrm{ER}}_1^{\mathrm{abs}})^s] = \frac{\Gamma\big(\frac{n-p_1}{2} + s\big)\Gamma\big(\frac{n}{2}\big)}{\Gamma\big(\frac{n-p_1}{2}\big)\Gamma\big(\frac{n}{2} + s\big)}, \quad \mathbb{E}[(\widehat{\mathrm{ER}}_0^{\mathrm{abs}})^s] = \frac{\Gamma\big(\frac{n-p_0}{2} + s\big)\Gamma\big(\frac{n}{2}\big)}{\Gamma\big(\frac{n-p_0}{2}\big)\Gamma\big(\frac{n}{2} + s\big)}. \tag{S69}$$

Because $\mathbb{E}[(\widehat{\mathrm{ER}}_1^{\mathrm{abs}})^s] \approx \mathbb{E}[(\widehat{\mathrm{ER}}_0^{\mathrm{abs}})^s] \cdot \mathbb{E}[\widehat{\mathrm{ER}}_{1\backslash 0}^s]$, we isolate the moments of the relative gain by dividing them. The shared $\Gamma\big(\frac{n}{2}\big)$ and $\Gamma\big(\frac{n}{2} + s\big)$ terms perfectly cancel:

$$\mathbb{E}[\widehat{\mathrm{ER}}_{1\backslash 0}^s] = \frac{\mathbb{E}[(\widehat{\mathrm{ER}}_1^{\mathrm{abs}})^s]}{\mathbb{E}[(\widehat{\mathrm{ER}}_0^{\mathrm{abs}})^s]} = \frac{\Gamma\big(\frac{n-p_1}{2} + s\big)\Gamma\big(\frac{n-p_0}{2}\big)}{\Gamma\big(\frac{n-p_1}{2}\big)\Gamma\big(\frac{n-p_0}{2} + s\big)}. \tag{S70}$$

Substituting the nested dimensional relationship $p_0 = p_1 - k$, the $n - p_0$ terms become $(n - p_1) + k$. Rewriting the equation with this substitution directly reveals the functional form of a new Beta moment:

$$\mathbb{E}[\widehat{\mathrm{ER}}_{1\backslash 0}^s] = \frac{\Gamma\big(\frac{n-p_1}{2} + s\big)\Gamma\big(\frac{n-p_1}{2} + \frac{k}{2}\big)}{\Gamma\big(\frac{n-p_1}{2}\big)\Gamma\big(\frac{n-p_1}{2} + \frac{k}{2} + s\big)}. \tag{S71}$$

This matches the exact structure $\frac{\Gamma(a+s)\Gamma(a+b)}{\Gamma(a)\Gamma(a+b+s)}$ where $a = \frac{n-p_1}{2}$ and $b = \frac{k}{2}$. This sequence uniquely identifies the relative retention rate $\widehat{\mathrm{ER}}_{1\backslash 0}$ (defined as $\widehat{\mathrm{ER}}$) as a Beta $\big(\frac{n-p_1}{2}, \frac{k}{2}\big)$ random variable, establishing the **Proportion of Residual EV** representation.

**Step 3: Algebraic Mapping to SNR and EV-F**

By the definition of the Beta distribution, $\widehat{\mathrm{ER}} \equiv \widehat{\mathrm{ER}}_{1\backslash 0} \stackrel{d}{=} V/(V + U)$ where $V \sim \chi^2_{n-p_1}$ and $U \sim \chi^2_k$ are independent. Inverting $\widehat{\mathrm{ER}}_{1\backslash 0}$ yields the **Ratio of Chi-Squares**:

$$\widehat{\mathrm{SNR}}_{\mathrm{V}} = \widehat{\mathrm{ER}}_{1\backslash 0}^{-1} - 1 \quad \dot\sim \quad \left(1 + \frac{\chi^2_k}{\chi^2_{n-p_1}}\right) - 1 = \frac{\chi^2_k}{\chi^2_{n-p_1}}. \tag{S72}$$

Scaling this ratio by $\frac{n-p_1}{k}$ directly recovers the **Relative EV-F Statistic**. Expanding the MGF of this ratio representation to the second order yields the $O(n^{-2})$ residual discrepancy against the true Bartlett-corrected relative deviance, confirming its alignment.

### Approximating Discrepancy on Central EV-SNR Distribution

As in the absolute deflation case, we must contrast $X_{\text{ratio}}$ with the classical limit. Classical likelihood theory relies on the deviance increment $D_{1\setminus 0} \dot{\sim} \chi^2_k$. Mapped into the variance space, this suggests the Wilks-implied approximation $X_{\text{Wilks}} = \exp(\chi^2_k/n) - 1$. Expanding its MGF gives:

$$M_{X_{\text{Wilks}}}(t) = 1 + \frac{tk}{n} + \frac{t^2 k(k+2)}{2n^2} + O(n^{-3}). \tag{S73}$$

The residual error for $X_{\text{Wilks}}$ is:

$$M_{\widehat{\text{SNR}}_{\text{V}}}(t) - M_{X_{\text{Wilks}}}(t) = \frac{tc_{01}}{n^2} + O(n^{-3}). \tag{S74}$$

The EV ratio approximation $X_{\text{ratio}}$ intentionally leverages the $\chi^2_{n-p_1}$ denominator to construct an intrinsic penalty for model complexity, injecting the $-k(p_1+2)$ shift into the residual error. The classical $X_{\text{Wilks}}$ mapping lacks a denominator entirely, trapping it with an unmitigated geometric error of $k(k+2)/2$. In the canonical setting of exact linear models, $c_{01} = k(p_1+2) - k(k+2)/2$, allowing the $X_{\text{ratio}}$ discrepancy to vanish perfectly. Thus, $X_{\text{ratio}}$ consistently offers a robust finite-sample surrogate that honors the dimensional degradation of the variance space far better than standard likelihood approximations.

## B.4 Deflation of Training EVs Against Population EV

**Theorem S7 (Deflation of Training Variances against their True Values).** *Let $\nu^* = \exp(2h^*(\boldsymbol{\theta}^*))$ denote the true population EV. The empirical variance evaluated at the maximum likelihood estimates, $\hat{v}(\hat{\boldsymbol{\theta}})$, is strictly bounded by the empirical topline at the true parameters, $\hat{v}(\boldsymbol{\theta}^*)$. By Theorem S5, this residual proportion is approximated by a Beta distribution:*

$$\frac{\hat{v}(\hat{\boldsymbol{\theta}})}{\hat{v}(\boldsymbol{\theta}^*)} \quad \dot{\sim} \quad \text{Beta}\left(\frac{n-p}{2}, \frac{p}{2}\right). \tag{S75}$$

*Leveraging the topline's expectation $\mathbb{E}[\hat{v}(\boldsymbol{\theta}^*)] \approx \nu^*$, we establish an exact finite-sample downward bias that mathematically formalizes training optimism:*

$$\mathbb{E}[\hat{v}(\hat{\boldsymbol{\theta}})] \approx \nu^* \left(1 - \frac{p}{n}\right). \tag{S76}$$

*Proof.* By the definition of absolute deviance, we can decompose the empirical topline variance as the product of the minimized training variance and the exponential of the deviance:

$$\hat{v}(\boldsymbol{\theta}^*) = \hat{v}(\hat{\boldsymbol{\theta}}) \cdot \exp\left(\frac{D}{n}\right). \tag{S77}$$

In regular families, the minimized empirical entropy and the absolute deviance are asymp-totically independent. Taking the expectation of both sides, the product factors:

$$\mathbb{E}[\hat{v}(\boldsymbol{\theta}^*)] \approx \mathbb{E}[\hat{v}(\hat{\boldsymbol{\theta}})] \cdot \mathbb{E}\left[\exp\left(\frac{D}{n}\right)\right]. \tag{S78}$$

From Theorem S5, $\exp(D/n) - 1 \dot{\sim} \chi^2_p/\chi^2_{n-p}$. Thus, $\exp(D/n) \dot{\sim} 1 + \chi^2_p/\chi^2_{n-p}$, which is the reciprocal of a Beta $(\frac{n-p}{2}, \frac{p}{2})$ variable. The expectation of this reciprocal Beta distribution is:

$$\mathbb{E}\left[\exp\left(\frac{D}{n}\right)\right] \approx \frac{\frac{n-p}{2} + \frac{p}{2} - 1}{\frac{n-p}{2} - 1} = \frac{n-2}{n-p-2}. \tag{S79}$$

Substituting the topline expectation $\mathbb{E}[\hat{v}(\boldsymbol{\theta}^*)] \approx \nu^*$ and solving for our target:

$$\mathbb{E}[\hat{v}(\hat{\boldsymbol{\theta}})] \approx \nu^* \left(\frac{n-p-2}{n-2}\right) \approx \nu^* \left(1 - \frac{p}{n}\right). \tag{S80}$$

The final approximation holds by ignoring the small constant offsets in the denominator for large $n$.

## B.5 Inflation of Expected Predictive EV

Let $\hat{\boldsymbol{\theta}}$ be a vector of parameters estimated from a training set of size $n$, and let $Y^{\text{test}}$ be an independent test sample generated from the exact same DGP and design $\mathbf{X}$.

Let us recall the **expected predictive cross-entropy** of the fitted model by evaluating the model's expected negative log-likelihood over the unobserved test data:

$$h^*(\hat{\boldsymbol{\theta}}) = \mathbb{E}_{Y^{\text{test}} \sim f^*}\left[-\frac{1}{n}\sum_{i=1}^{n} \log f(Y_i^{\text{test}} \mid \mathbf{x}_i, \hat{\boldsymbol{\theta}})\right]. \tag{S81}$$

The quantity $h^*(\hat{\boldsymbol{\theta}})$ is itself random through $\hat{\boldsymbol{\theta}}$. Averaging it over the sampling distribution of the training estimates gives a deterministic population target, and the **predictive test EV** is defined by exponentiating this *averaged* predictive cross-entropy:

$$v^{\text{test}} = \exp\left(\, 2\, \mathbb{E}_{\hat{\boldsymbol{\theta}}}[h^*(\hat{\boldsymbol{\theta}})] \,\right). \tag{S82}$$

We exponentiate the expectation rather than averaging the exponential because $v^{\text{test}}$ is intended as a fixed deployment-risk target, not a random variable; this also makes it the structural counterpart of the population EV $v^*(\boldsymbol{\theta})$ (Eq. (5), index $j$ dropped) evaluated at the *expected* fitted entropy.

**Theorem S8 (Inflation of Expected Predictive EV).** *For regular likelihood families, the predictive test EV obtained by exponentiating the expected predictive cross-entropy exhibits a finite-sample upward inflation relative to the true minimum $\nu^*$, and a corresponding squared inflation relative to the expected empirical training variance $\mathbb{E}[\hat{v}(\hat{\boldsymbol{\theta}})]$:*

$$v^{\text{test}} \approx \nu^* \, e^{p/n} \approx \frac{\nu^*}{1-p/n} \approx \frac{\mathbb{E}[\hat{v}(\hat{\boldsymbol{\theta}})]}{(1-p/n)^2}, \tag{S83}$$

*where $p$ is the dimension of $\boldsymbol{\theta}$.*

*Proof.* Let $Y$ denote the training sample of size $n$ yielding the maximum likelihood estimate $\hat{\boldsymbol{\theta}}$. By Eq. (S81), the predictive cross-entropy $h^*(\hat{\boldsymbol{\theta}})$ is already averaged over the unobserved $Y^{\text{test}}$, so it is a deterministic function of $\hat{\boldsymbol{\theta}}$ evaluated on the population surface.

We perform a second-order Taylor expansion of $h^*(\hat{\boldsymbol{\theta}})$ around the true population parameter $\boldsymbol{\theta}^*$. Since $\boldsymbol{\theta}^*$ minimizes the population cross-entropy, the gradient vanishes ($\nabla h^*(\boldsymbol{\theta}^*) = 0$) and the Hessian is the Fisher Information $\mathcal{I}(\boldsymbol{\theta}^*)$:

$$2h^*(\hat{\boldsymbol{\theta}}) \approx 2h^*(\boldsymbol{\theta}^*) + (\hat{\boldsymbol{\theta}} - \boldsymbol{\theta}^*)^\top \mathcal{I}(\boldsymbol{\theta}^*)(\hat{\boldsymbol{\theta}} - \boldsymbol{\theta}^*) = 2h^*(\boldsymbol{\theta}^*) + \frac{Q}{n}, \tag{S84}$$

where $Q = n(\hat{\boldsymbol{\theta}} - \boldsymbol{\theta}^*)^\top \mathcal{I}(\boldsymbol{\theta}^*)(\hat{\boldsymbol{\theta}} - \boldsymbol{\theta}^*)$ is the Wald statistic on the training data.

We now take the expectation over the training sampling distribution **before** exponentiating, so that only the first moment of $Q$ is required. To leading order the MLE is asymptotically normal, $\sqrt{n}(\hat{\boldsymbol{\theta}} - \boldsymbol{\theta}^*) \underset{d}{\to} \mathcal{N}(0, \mathcal{I}(\boldsymbol{\theta}^*)^{-1})$, giving $Q \underset{d}{\to} \chi^2_p$ and $\mathbb{E}[Q] \to p$. At finite $n$, however, the mean of the Wald statistic departs from its nominal degrees of freedom by a **Bartlett-type correction** (Bartlett, 1937; Lawley, 1956):

$$\mathbb{E}[Q] = p + c, \qquad c = c(\boldsymbol{\theta}^*, n) = O\left(\frac{p^2}{n}\right) \geq 0, \tag{S85}$$

where the correction $c$ is non-negative for the regular families considered, reflecting the systematic over-dispersion of the quadratic form under estimated curvature. Substituting into Eq. (S84):

$$2\,\mathbb{E}[h^*(\hat{\boldsymbol{\theta}})] \approx 2h^*(\boldsymbol{\theta}^*) + \frac{\mathbb{E}[Q]}{n} = 2h^*(\boldsymbol{\theta}^*) + \frac{p+c}{n}. \tag{S86}$$

Exponentiating Eq. (S86) and noting $\exp(2h^*(\boldsymbol{\theta}^*)) = \nu^*$ maps the averaged cross-entropy back to EV space. Substituting the finite-sample correction $c/n = \frac{1}{2}(p/n)^2 + O((p/n)^3)$ — whose coefficient $\frac{1}{2}$ is exact in the Gaussian case, where the $\chi^2_p$ moment route yields it with no higher-cumulant remainder, while for a general family $c$ depends on the third- and fourth-order cumulants of the score and admits no universal closed form — the predictive EV becomes:

$$v^{\text{test}} = \nu^* \exp\left(\frac{p}{n} + \frac{1}{2}\left(\frac{p}{n}\right)^2 + O\left( (p/n)^3 \right)\right). \tag{S87}$$

The exponent in Eq. (S87) is precisely the logarithm of the rational penalty, since $(1 - p/n)^{-1} = \exp\left( p/n + \frac{1}{2}(p/n)^2 + O((p/n)^3) \right)$. The two series therefore agree through second order and differ only at $O((p/n)^3)$, so the predictive EV collapses onto the geometric

inflation factor:

$$v^{\text{test}} = \frac{\nu^*}{1 - p/n} + O\left( (p/n)^3 \right). \tag{S88}$$

The naive first-order factor $\exp(p/n)$ retains only the leading term of the second line and omits the $\frac{1}{2}(p/n)^2$ Bartlett surplus, under-penalizing at second order; the rational form restores it, agreeing with the Bartlett-corrected target up to $O((p/n)^3)$ rather than only to $O(p/n)$.

Applying the same expansion in-sample gives the dual training deflation $\mathbb{E}[\hat{v}(\hat{\boldsymbol{\theta}})] = \nu^*(1 - p/n) + O((p/n)^2)$, so the squared penalty spans both directions to the same order:

$$\frac{\mathbb{E}[\hat{v}(\hat{\boldsymbol{\theta}})]}{(1 - p/n)^2} = \frac{\nu^*}{1 - p/n} + O\left( (p/n)^2 \right) = v^{\text{test}} + O\left( (p/n)^2 \right). \tag{S89}$$

This squared penalty is strictly heavier than the symmetric exponential correction $\exp(2p/n)$: since

$$\left(1 - \frac{p}{n}\right)^{-2} = \exp\left(\frac{2p}{n} + \left(\frac{p}{n}\right)^2 + O\left( (p/n)^3 \right)\right), \tag{S90}$$

it carries a surplus of $(p/n)^2$ in the exponent, matching the doubled correction $2c/n$ to leading order. Exponentiating the *expected* predictive entropy thus absorbs not only the mean degrees of freedom $p$ but its $O((p/n)^2)$ finite-sample inflation, formalizing predictive variance inflation as the correctly over-penalized inverse of training variance deflation.

## B.6 The Non-Central Distribution of the Relative EV-SNR

**Theorem S9 (The Non-Central Distribution of the Relative EV-SNR).** *Let $\mathcal{M}_0 \subset \mathcal{M}_1$ be two nested regular parametric models with dimensions $p_0$ and $p_1$, where $k = p_1 - p_0$. Assume the true DGP lies in $\mathcal{M}_1$ but not in $\mathcal{M}_0$, yielding a local contiguous alternative where the true population predictive gain is $\phi_{\mathrm{V}}^2 = O(n^{-1})$.*

*We define the **Relative EV SNR** ($\widehat{\mathrm{SNR}}_{\mathrm{V}}$) as the proportion of EV explained by the fuller model relative to its own baseline:*

$$\widehat{\mathrm{SNR}}_{\mathrm{V}} = \frac{\hat{v}_0(\hat{\boldsymbol{\theta}}_0) - \hat{v}_1(\hat{\boldsymbol{\theta}}_1)}{\hat{v}_1(\hat{\boldsymbol{\theta}}_1)}. \tag{S91}$$

*Under local alternatives, the finite-sample geometry of the variance space is approximated by the non-central ratio distribution $X_{\text{ratio}} = \chi^2_k(\lambda_{\mathrm{V}})/\chi^2_{n-p_1}$, where $\lambda_{\mathrm{V}} = n\phi_{\mathrm{V}}^2$. This yields three equivalent stochastic representations for the predictive signal:*

1. ***Ratio of Independent Chi-Squares:***

$$\widehat{\mathrm{SNR}}_{\mathrm{V}} \quad \dot{\sim} \quad \frac{\chi^2_k(\lambda_{\mathrm{V}})}{\chi^2_{n-p_1}}. \tag{S92}$$

2. ***Non-Central Beta for the Proportion of Residual EV:***

$$\widehat{\mathrm{ER}} = \frac{1}{1+\widehat{\mathrm{SNR}}_{\mathrm{V}}} = \frac{\hat{v}_1(\hat{\boldsymbol{\theta}}_1)}{\hat{v}_0(\hat{\boldsymbol{\theta}}_0)} \quad \dot{\sim} \quad \mathrm{Beta}\left(\frac{n-p_1}{2}, \frac{k}{2}, \lambda_{\mathrm{V}}\right). \tag{S93}$$

3. ***The Non-Central F distribution for $F_{\mathrm{V}}$:.***

$$F_{\mathrm{V}} = \left(\frac{n-p_1}{k}\right)\widehat{\mathrm{SNR}}_{\mathrm{V}} \quad \dot{\sim} \quad F_{k,n-p_1}(\lambda_{\mathrm{V}}). \tag{S94}$$

*This approximation is* ***first-order accurate*** *at the $O(n^{-1})$ scale. At the $O(n^{-2})$ scale, the MGFs exhibit a discrepancy governed by the statistical curvature $c$ and the baseline degrees of freedom $p_0$. The residual error of the ratio approximation expands as:*

$$M_{\widehat{\mathrm{SNR}}_{\mathrm{V}}}(t) - M_{X_{\mathrm{ratio}}}(t) = \frac{t}{n^2}\left[c + k\left(\frac{k}{2} - p_1 - 1\right) - \lambda_{\mathrm{V}} p_0 + \frac{\lambda_{\mathrm{V}}^2}{2}\right] + O(n^{-3}). \tag{S95}$$

*Proof* (Proof via MGFs).

**Step 1: Asymptotic Expansion of the Non-Central Deviance**

The relative EV signal can be expressed natively via the relative deviance $D_{1\backslash 0} = 2n[\hat{H}_0(\hat{\boldsymbol{\theta}}_0) - \hat{H}_1(\hat{\boldsymbol{\theta}}_1)]$:

$$\widehat{\mathrm{SNR}}_{\mathrm{V}} = \exp\left(\frac{D_{1\backslash 0}}{n}\right) - 1. \tag{S96}$$

Under $H_1$, likelihood theory dictates that the relative deviance converges to a non-central $\chi^2$ with non-centrality parameter $\lambda_D = 2n\Delta h^*$. Under local contiguous alternatives ($\rho_V^2 \to 0$), the likelihood non-centrality ($\lambda_D = -n\ln(1-\rho_V^2)$) and the variance non-centrality ($\lambda_{\mathrm{V}} = n\rho_V^2/(1-\rho_V^2)$) are asymptotically equivalent up to the first order: $\lambda_D \approx \lambda_{\mathrm{V}}$.

According to the higher-order asymptotic expansion of the likelihood ratio statistic under local alternatives (Hayakawa, 1977), the expected moments of the non-central deviance expand as:

$$\mathbb{E}[D_{1\backslash 0}] = k + \lambda_{\mathrm{V}} + \frac{c}{n} + O(n^{-2}), \tag{S97}$$

$$\mathbb{E}[D_{1\backslash 0}^2] = (k+\lambda_{\mathrm{V}})^2 + 2(k+2\lambda_{\mathrm{V}}) + O(n^{-1}), \tag{S98}$$

where $c$ encapsulates the statistical curvature of the log-likelihood manifold.

**Step 2: MGF of the Estimator $\widehat{\mathrm{SNR}}_{\mathrm{V}}$**

We establish the MGF of the true relative estimator by applying a Taylor expansion to the exponential mapping $e^{t\widehat{\mathrm{SNR}}_{\mathrm{V}}}$ around $D_{1\backslash 0}/n = 0$:

$$\exp\left(t\widehat{\mathrm{SNR}}_{\mathrm{V}}\right) = 1 + \frac{tD_{1\backslash 0}}{n} + \frac{t+t^2}{2n^2}D_{1\backslash 0}^2 + O_p(n^{-3}). \tag{S99}$$

Taking the expectation and substituting the non-central moments yields:

$$M_{\widehat{\mathrm{SNR}}_{\mathrm{V}}}(t) = 1 + \frac{t}{n}\left(k + \lambda_{\mathrm{V}} + \frac{c}{n}\right) + \frac{t+t^2}{2n^2}\left[(k+\lambda_{\mathrm{V}})^2 + 2(k+2\lambda_{\mathrm{V}})\right] + O(n^{-3}). \tag{S100}$$

Distributing the terms and strictly truncating at the $O(n^{-2})$ scale provides the operational MGF:

$$\begin{aligned} M_{\widehat{\mathrm{SNR}}_{\mathrm{V}}}(t) = 1 &+ \frac{t(k+\lambda_{\mathrm{V}})}{n} + \frac{t}{n^2}\left[c + \frac{(k+\lambda_{\mathrm{V}})^2 + 2(k+2\lambda_{\mathrm{V}})}{2}\right] \\ &+ \frac{t^2}{2n^2}\left[(k+\lambda_{\mathrm{V}})^2 + 2(k+2\lambda_{\mathrm{V}})\right] + O(n^{-3}). \end{aligned} \tag{S101}$$

**Step 3: MGF of the Approximating Distribution $X_{\text{ratio}}$**

Consider the non-central ratio approximation $X_{\text{ratio}} = U/V$, where $U \sim \chi^2_k(\lambda_{\mathrm{V}})$ and $V \sim \chi^2_{n-p_1}$ are independent. Expanding its inverse moments asymptotically in $n$:

$$\begin{aligned} \mathbb{E}[U/V] &= \frac{k+\lambda_{\mathrm{V}}}{n-p_1-2} = \frac{k+\lambda_{\mathrm{V}}}{n} + \frac{(k+\lambda_{\mathrm{V}})(p_1+2)}{n^2} + O(n^{-3}), \\ \mathbb{E}[U^2/V^2] &= \frac{(k+\lambda_{\mathrm{V}})^2 + 2(k+2\lambda_{\mathrm{V}})}{n^2} + O(n^{-3}). \end{aligned} \tag{S102}$$

Substituting these into the standard MGF expansion $M_{X_{\text{ratio}}}(t) = 1 + t\mathbb{E}[U/V] + \frac{t^2}{2}\mathbb{E}[U^2/V^2]$ yields:

$$\begin{aligned} M_{X_{\text{ratio}}}(t) = 1 &+ \frac{t(k+\lambda_{\mathrm{V}})}{n} + \frac{t(k+\lambda_{\mathrm{V}})(p_1+2)}{n^2} \\ &+ \frac{t^2}{2n^2}\left[(k+\lambda_{\mathrm{V}})^2 + 2(k+2\lambda_{\mathrm{V}})\right] + O(n^{-3}). \end{aligned} \tag{S103}$$

**Step 4: Derivation of the MGF Discrepancy**

We subtract the MGF of the ratio approximation from the Estimator MGF to isolate the residual geometric error. Notice that the quadratic $t^2$ terms cancel perfectly. Extracting the coefficients of $t/n^2$, the difference simplifies to:

$$M_{\widehat{\mathrm{SNR}}_{\mathrm{V}}}(t) - M_{X_{\text{ratio}}}(t) = \frac{t}{n^2}\left[c + \frac{(k+\lambda_{\mathrm{V}})^2 + 2(k+2\lambda_{\mathrm{V}})}{2} - (k+\lambda_{\mathrm{V}})(p_1+2)\right] + O(n^{-3}) \tag{S104}$$

By expanding the interior quadratic and grouping the remaining terms utilizing the identity $k - p_1 = -p_0$, the algebra collapses into the final discrepancy:

$$M_{\widehat{\mathrm{SNR}}_{\mathrm{V}}}(t) - M_{X_{\mathrm{ratio}}}(t) = \frac{t}{n^2}\left[c + k\left(\frac{k}{2} - p_1 - 1\right) - \lambda_{\mathrm{V}} p_0 + \frac{\lambda_{\mathrm{V}}^2}{2}\right] + O(n^{-3}). \quad \text{(S105)}$$

This confirms first-order accuracy. Notably, if the null model is trivial ($p_0 = 0 \Longrightarrow k = p_1$) and the signal is zero ($\lambda_{\mathrm{V}} = 0$), this strictly collapses to the absolute discrepancy $c - p_1(p_1 + 2)/2$, confirming structural unification.

**Remarks on the Approximating Accuracy**

The exact mapping to the classical non-central $F$-distribution fundamentally relies on the assumption of *local contiguous alternatives* (i.e., $\phi_{\mathrm{V}}^2 = O(n^{-1})$). In plain language, this requires that the true population effect size $\phi_{\mathrm{V}}^2$ cannot be excessively large relative to the sample size. If the true signal is massive (a large $\phi_{\mathrm{V}}^2$ in small sample size $n$ cases), the linear Taylor expansions used to bridge the exponential deviance and the EV ($\exp(x) - 1 \approx x$ and $\phi_{\mathrm{V}}^2 \approx \rho_V^2$) begin to systematically diverge. Therefore, this geometric approximation is exceptionally accurate for small-to-moderate signals or large sample size $n$ — precisely where power analyses and confidence bounds are most critical — but it may introduce asymptotic bias if applied to unusually large effect sizes and small sample size $n$ where the log-likelihood manifold is no longer "locally flat."

## B.7 $\chi^2$ Decomposition of EVD

To dissect the finite-sample geometry of the variance space, we define the **EV Total** for $n$ observations as $\widehat{\mathrm{EVT}}_j = n\hat{v}_j(\hat{\boldsymbol{\theta}}_j)$, which corresponds to residual sum square (RSS) of OLS. Correspondingly, the **EVD** between two nested models $\mathcal{M}_0 \subset \mathcal{M}_1$ is defined as their difference: $\widehat{\mathrm{EVD}} = \widehat{\mathrm{EVT}}_0 - \widehat{\mathrm{EVT}}_1$. EVD is the counterpart of sum squares reduction (SSR) of OLS.

### B.7.1 Asymptotic Central $\chi^2$ Distribution of EVD

**Theorem S10 (Asymptotic Central $\chi^2$ Decomposition of EVD).** *Let $\mathcal{M}_0 \subset \mathcal{M}_1$ be two nested regular parametric models with dimensions $p_0$ and $p_1$, where $k = p_1 - p_0$. To establish the information-geometric equivalent of the nested sum of squares decomposition (where $\mathrm{RSS}_0 = \mathrm{RSS}_1 + \mathrm{SSR}$), we define the **Absolute EVD** ($\widehat{\mathrm{EVD}}^{\mathrm{abs}}$) as the variance deflated during model training relative to the true population parameter:*

$$\widehat{\mathrm{EVD}}_j^{\mathrm{abs}} = \widehat{\mathrm{EVT}}^* - \widehat{\mathrm{EVT}}_j = n\left[\hat{v}(\boldsymbol{\theta}^*) - \hat{v}_j(\hat{\boldsymbol{\theta}}_j)\right]. \quad \text{(S106)}$$

*We further define the **Relative EVD** between the nested models as:*

$$\widehat{\mathrm{EVD}} = \widehat{\mathrm{EVT}}_0 - \widehat{\mathrm{EVT}}_1 = n\left[\hat{v}_0(\hat{\boldsymbol{\theta}}_0) - \hat{v}_1(\hat{\boldsymbol{\theta}}_1)\right]. \quad \text{(S107)}$$

*This permits a perfect algebraic identity in the EV space:*

$$\widehat{\mathrm{EVD}}_1^{\mathrm{abs}} = \widehat{\mathrm{EVD}}_0^{\mathrm{abs}} + \widehat{\mathrm{EVD}}. \quad \text{(S108)}$$

*Under the strict null hypothesis that the restricted model $\mathcal{M}_0$ contains the true DGP ($\rho_{\mathrm{V}}^2 =$*

*0), the geometric drops act as independent, constant-scaled $\chi^2$ partitions. Specifically:*

$$\widehat{\mathrm{EVD}}_j^{\mathrm{abs}} \xrightarrow{d} \nu^* \chi^2_{p_j} \quad \text{and} \quad \widehat{\mathrm{EVD}} \xrightarrow{d} \nu^* \chi^2_k, \tag{S109}$$

*where $\nu^* = \exp(2h^*)$ is the true population topline EV, and the baseline error $\widehat{\mathrm{EVD}}_0^{\mathrm{abs}}$ is asymptotically independent from the relative model improvement $\widehat{\mathrm{EVD}}$.*

*Proof* (Proof of Chi-square Decomposition of EVD).

**Step 1: First-Order Deviance Mapping**

We bridge the variance space to the log-likelihood space by factoring out the empirical estimators and applying the first-order Taylor expansion $\exp(x) - 1 \approx x$. For the absolute drop of model $j$:

$$\widehat{\mathrm{EVD}}_j^{\mathrm{abs}} = n\hat{v}_j(\hat{\boldsymbol{\theta}}_j)\left(\exp\left(\frac{D_j^{\mathrm{abs}}}{n}\right) - 1\right) = \hat{v}_j(\hat{\boldsymbol{\theta}}_j)\Big(D_j^{\mathrm{abs}} + O_p(n^{-1})\Big), \tag{S110}$$

where $D_j^{\mathrm{abs}} = 2n[\hat{H}(\boldsymbol{\theta}^*) - \hat{H}_j(\hat{\boldsymbol{\theta}}_j)]$ is the absolute deviance. Similarly, the relative drop between models maps to the relative deviance $D_{1\setminus 0} = 2n[\hat{H}_0(\hat{\boldsymbol{\theta}}_0) - \hat{H}_1(\hat{\boldsymbol{\theta}}_1)]$:

$$\widehat{\mathrm{EVD}} = \hat{v}_1(\hat{\boldsymbol{\theta}}_1)\Big(D_{1\setminus 0} + O_p(n^{-1})\Big). \tag{S111}$$

**Step 2: Orthogonal Score Projections and Independence**

Under the null hypothesis and local asymptotics, the absolute deviance $D_1^{\mathrm{abs}}$ is asymptotically equivalent to the quadratic form of the score vector projected onto the $p_1$-dimensional tangent space of $\mathcal{M}_1$ using the Fisher information metric.

Because the models are strictly nested ($\mathcal{M}_0 \subset \mathcal{M}_1$), this projection can be orthogonally partitioned into two components: a projection onto the $p_0$-dimensional subspace of $\mathcal{M}_0$, and a projection onto its $k$-dimensional orthogonal complement within $\mathcal{M}_1$. Algebraically, this enforces the asymptotic deviance decomposition $D_1^{\mathrm{abs}} \approx D_0^{\mathrm{abs}} + D_{1\setminus 0}$.

Because these projections are geometrically orthogonal within the Fisher information metric, Cochran's Theorem dictates that the resulting quadratic forms are asymptotically independent $\chi^2$ variables:

$$D_0^{\mathrm{abs}} \xrightarrow{d} \chi^2_{p_0} \quad \text{and} \quad D_{1\setminus 0} \xrightarrow{d} \chi^2_k, \quad \text{with} \;\; D_0^{\mathrm{abs}} \perp D_{1\setminus 0}. \tag{S112}$$

**Step 3: Synthesis via Slutsky's Theorem**

By the Law of Large Numbers, the empirical variance estimators for all correctly specified models converge in probability to the same true population variance: $\hat{v}_0(\hat{\boldsymbol{\theta}}_0) \xrightarrow{p} \nu^*$ and $\hat{v}_1(\hat{\boldsymbol{\theta}}_1) \xrightarrow{p} \nu^*$.

Substituting these limits into our first-order mappings from Step 1, and invoking Slutsky's theorem, we scale the independent likelihood ratio limits by the constant population variance:

$$\widehat{\mathrm{EVD}}_0^{\mathrm{abs}} \xrightarrow{d} \nu^* \chi^2_{p_0} \quad \text{and} \quad \widehat{\mathrm{EVD}} \xrightarrow{d} \nu^* \chi^2_k. \tag{S113}$$

Because the underlying absolute and relative deviances are asymptotically independent, multiplying them by the same asymptotic constant $\nu^*$ preserves their independence. Thus, the variance drops partition exactly like independent nested sum of squares in classical OLS.

### B.7.2 Remarks on EVD Decomposition

- **A Generalization of Wilks' Theorem on EVD**

  This algebraic equivalence demonstrates that $\widehat{\mathrm{EVD}}$ serves as the strict geometric analogue to Wilks' theorem within the EV space. Where Wilks' theorem operates on the relative deviance in the log-likelihood (entropy) space, the $\widehat{\mathrm{EVD}}$ distribution translates that behavior into the exponentiated volume space, safely scaled by the topline noise.

- **The Non-$\chi^2$ Geometry of the EV Totals (EVT)**

  Crucially, while the geometric *drops* ($\widehat{\mathrm{EVD}}^{\mathrm{abs}}$) and *reductions* ($\widehat{\mathrm{EVD}}$) follow elegant $\chi^2$ distributions, the individual EVT ($\widehat{\mathrm{EVT}}_j$) **do not**. Unlike classical OLS, where the RSS defines an isolated chi-square floor, general parametric likelihoods possess no such stable foundation. Let's observe the decomposition

  $$\widehat{\mathrm{EVT}}_j = \widehat{\mathrm{EVT}}^* - \widehat{\mathrm{EVD}}_j^{\mathrm{abs}}. \tag{S114}$$

  The foundational anchor of this term is $\widehat{\mathrm{EVT}}^* = n \exp(2\hat{H}(\boldsymbol{\theta}^*))$. Because the empirical entropy $\hat{H}(\boldsymbol{\theta}^*)$ is a simple average of independent log-likelihoods, the Central Limit Theorem dictates that it carries a dominant $O_p(n^{1/2})$ Gaussian sampling noise. Exponentiating this normally distributed error means that the total $\widehat{\mathrm{EVT}}^*$ is fundamentally a **Log-Normal** distribution, not a $\chi^2$ distribution. While the $\chi^2$ drops are bounded at $O_p(1)$, the Log-Normal anchor drifts at $O_p(n^{1/2})$. Because the totals are dominated by this Log-Normal sampling variance, no unified $\chi^2$ distribution can characterize $\widehat{\mathrm{EVT}}_j$.

- **Significance of the EV-SNR and $F_{\mathrm{V}}$ Distributions**

  This architectural limitation illustrates exactly why establishing a standardized effect size necessitates the **Relative EV-SNR**. If we attempt to evaluate model performance using absolute totals ($\widehat{\mathrm{EVT}}$), we are overwhelmed by the model-dependent, Log-Normal sampling noise of the entropy ceiling. However, by taking the *ratio* of EVs ($\widehat{\mathrm{EVD}}/\widehat{\mathrm{EVT}}_1$), the Log-Normal topline noise perfectly cancels out in the numerator, and the unobservable population scaling factor ($\nu^*$) annihilates itself. The $\chi^2$ differences are organically isolated, proving that robust, model-invariant inference must be conducted strictly within the relative SNR space.

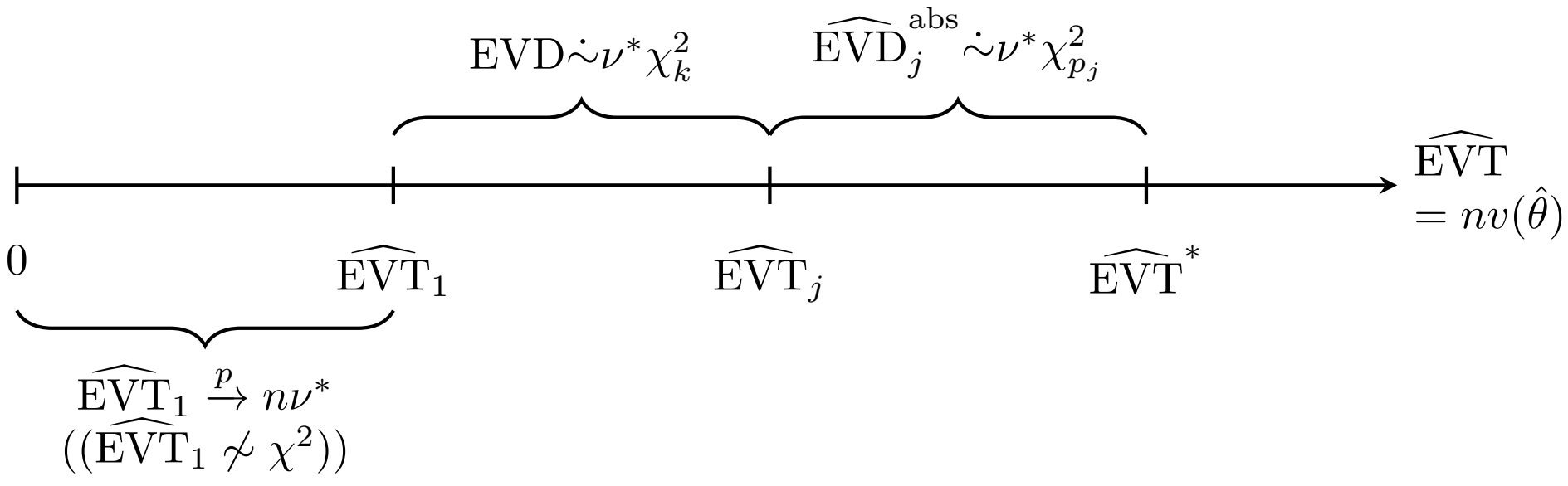

Figure S9: Decomposition of EV and Asymptotic Distributions

# C Recovery of OLS Metrics with the EV Theory

In this section, we verify that the proposed EV $R^2_{\mathrm{V}}$ flawlessly recovers the fundamental results of classical OLS when the model family is Gaussian: $Y \mid \mathbf{X} \sim \mathcal{N}(\mathbf{X}\beta, \sigma^2 I_n)$. We mirror the exact structural development of the previous section to demonstrate how each likelihood-based metric collapses into its classical counterpart.

## C.1 Recovery of Population $\rho^2$ of OLS by $\rho^2_{\mathrm{V}}$

To verify this construction, we demonstrate that $\rho^2_{\mathrm{V}}$ recovers the classical population $\rho^2$ for Gaussian linear models. Let $f^*$ be the true DGP. Consider nested models $\mathcal{M}_j : Y \mid \mathbf{X} \sim \mathcal{N}(\mathbf{X}_j \boldsymbol{\beta}^*_j, \sigma^2_j I_n)$ for $j \in \{0, 1\}$, where $\mathcal{M}_0 \subset \mathcal{M}_1$. Let $(\boldsymbol{\beta}^*_j, \sigma^2_j)$ denote the parameters minimizing the expected cross-entropy relative to $f^*$. Under the Gaussian log-likelihood (White, 1982), the evaluated cross-entropy resolves to:

$$h^*_j(\boldsymbol{\theta}^*_j) = \frac{1}{2}\log(2\pi e \sigma^2_j), \tag{S115}$$

where $\sigma^2_j = \frac{1}{n}\sum_{i=1}^n \mathbb{E}_{f^*}\left[\, (Y_i - \mathbf{x}_{ij}^\top \boldsymbol{\beta}^*_j)^2 \,\right]$. Substituting the EV $\nu^*_j = 2\pi e \sigma^2_j$ into our partial effect size definition yields:

$$\rho^2_{\mathrm{V}}(\mathcal{M}_1 \mid \mathcal{M}_0) = 1 - \frac{\nu^*_1}{\nu^*_0} = 1 - \frac{\sigma^2_1}{\sigma^2_0}. \tag{S116}$$

Here, $\sigma^2_0 = \sigma^2_1 + \frac{1}{n} \mid \boldsymbol{\mu}^*_1 - \mathbf{P}_0 \boldsymbol{\mu}^*_1 \mid^2$, with $\mathbf{P}_0$ projecting onto Col $(\mathbf{X}_0)$. The restricted variance $\sigma^2_0$ inflates because it absorbs the omitted signal. If $\mathcal{M}_0$ is the intercept-only model, $\nu^*_0$ represents the marginal variance of $Y$. See Sec. C for detailed derivations.

## C.2 Recovery of RSS with Entropic Variance

As defined in Sec. 3.1, the foundation of the framework is the empirical entropy. For a Gaussian model $\mathcal{M}_j$ with maximum likelihood estimates $\hat{\beta}_j$ and $\hat{\sigma}^2_j = \mathrm{RSS}_j/n$, the minimized average empirical cross-entropy evaluates directly to:

$$\hat{h}_j(\hat{\boldsymbol{\theta}}_j) = \frac{1}{2}\log(2\pi e \hat{\sigma}^2_j). \tag{S117}$$

Exponentiating this yields the empirical EV (Eq. (25)). We see that the EV is strictly proportional to the maximum likelihood variance estimator, and by extension, the RSS:

$$\hat{v}_j = \exp(2\hat{h}_j) = 2\pi e \hat{\sigma}_j^2 = 2\pi e \left( \frac{\text{RSS}_j}{n} \right). \tag{S118}$$

Consequently, the total empirical EV is directly proportional to the RSS: $\hat{V}_j = n\hat{v}_j = 2\pi e(\text{RSS}_j)$.

## C.3 Recovery of the Central Distribution and ANOVA $F$-Statistic

In the generalized framework, the empirical Signal-to-Noise Ratio ($\widehat{\text{SNR}}_\text{V}$) drives the central $F$-test (Theorem S6). Substituting the empirical Gaussian variances, the probability density constants perfectly cancel out, reducing the EV-SNR to the classical relative reduction in RSS:

$$\widehat{\text{SNR}}_\text{V} = \exp(2\Delta\hat{h}) - 1 = \frac{\hat{v}_0}{\hat{v}_1} - 1 = \frac{2\pi e \hat{\sigma}_0^2}{2\pi e \hat{\sigma}_1^2} - 1 = \frac{\text{RSS}_0 - \text{RSS}_1}{\text{RSS}_1}. \tag{S119}$$

Substituting this empirical ratio into the definition of our EV-based test statistic $F_\text{V}$ (Eq. (29)):

$$F_\text{V} = \frac{n - p_1}{k} \widehat{\text{SNR}}_\text{V} = \frac{(\text{RSS}_0 - \text{RSS}_1)/k}{\text{RSS}_1/(n - p_1)}. \tag{S120}$$

This is the exact formulation of the classical ANOVA $F$-test statistic. Where Theorem S6 establishes an asymptotic central $F$-distribution for general regular families under the null hypothesis, the Gaussian family yields this $F$-distribution exactly in finite samples due to the exact independence of the quadratic forms.

## C.4 Recovery of $R^2_\text{adj}$ and Deflation of Training Variances

To recover the population target $\rho^2_\text{V}$ in OLS, we derive the population EV from first principles. Let $f^*$ be an arbitrary true DGP. Consider nested Gaussian model families $\mathcal{M}_j$ : $Y \mid \mathbf{X} \sim \mathcal{N}(\mu_j, \sigma_j^2 I_n)$ for $j \in \{0, 1\}$, where $\mathcal{M}_0 \subset \mathcal{M}_1$ and $\mu_j = \mathbf{X}_j \beta_j$.

Let $\beta_j$ and $\sigma_j^2$ denote the projected population parameters that minimize the KL divergence from $f^*$ to $\mathcal{M}_j$. Because the Gaussian negative log-likelihood is quadratic, the expected predictive cross-entropy relative to $f^*$ resolves cleanly to:

$$h_j^*(\boldsymbol{\theta}_j^*) = \frac{1}{2} \log(2\pi e \sigma_j^2), \tag{S121}$$

where $\sigma_j^2 = \frac{1}{n} \sum_{i=1}^n \mathbb{E}_{f^*} \left[ (Y_i - \mathbf{x}_{ij}^\top \beta_j)^2 \right]$ represents the projected residual variance, and the constant $\frac{1}{2}\log(2\pi e)$ is the entropy of a standard normal distribution.

Transforming this into the variance space provides the exact population EV:

$$\nu_j^* = \exp(2h_j^*) = 2\pi e \sigma_j^2. \tag{S122}$$

When evaluating the population effect size $\rho^2_{\mathrm{V}}$, the probability density constant $2\pi e$ cancels perfectly:

$$\rho^2_{\mathrm{V}} = 1 - \frac{\nu_1^*}{\nu_0^*} = 1 - \frac{2\pi e \sigma_1^2}{2\pi e \sigma_0^2} = 1 - \frac{\sigma_1^2}{\sigma_0^2}. \tag{S123}$$

Structurally, the restricted variance inflates by absorbing the omitted signal: $\sigma_0^2 = \sigma_1^2 + \frac{1}{n}\|\mathbf{X}_1\beta_1 - P_0\mathbf{X}_1\beta_1\|^2$, where $P_0$ is the orthogonal projection matrix onto $\mathbf{X}_0$. When $\mathcal{M}_0$ is the intercept-only null model, $\rho^2_{\mathrm{V}}$ exactly recovers the classical population $R^2$.

To estimate this unbiasedly from training data, we must address the finite-sample deflation bias of maximum likelihood estimation. The theoretical justification for the bias-corrected variance $\hat{v}^c$ rests strictly on the expected value of the training EV, not its total distribution. Theorem S7 establishes that the expectation of the empirical EV deflates by a factor of $(1 - p/n)$. In the OLS setting, substituting $\hat{v} = 2\pi e(\mathrm{RSS}\ /n)$ and $\nu^* = 2\pi e \sigma^2$ perfectly recovers the exact finite-sample downward bias of the classical MLE variance:

$$\mathbb{E}[\hat{v}] \approx \nu^*\left(1 - \frac{p}{n}\right) \quad \Longrightarrow \quad \mathbb{E}\left[\frac{\mathrm{RSS}}{n}\right] = \sigma^2\left(1 - \frac{p}{n}\right). \tag{S124}$$

This expectation mapping justifies the finite-sample correction. Conversely, the classical exact distributional property $\hat{V}/\nu^* = \mathrm{RSS}\ /\sigma^2 \sim \chi^2_{n-p}$ corresponds structurally to the absolute EVD distribution described in Theorem S10, which governs the shape of the test statistic rather than the expected variance topline.

To resolve the expectation bias, we apply the degrees-of-freedom penalty to the total empirical EV ($\hat{V}_j = n\hat{v}_j$) to define the bias-corrected EV:

$$\hat{v}_j^c = \frac{\hat{V}_j}{n - p_j} = 2\pi e\left(\frac{\mathrm{RSS}_j}{n - p_j}\right). \tag{S125}$$

Consequently, the EV-based effect size $R^2_{\mathrm{V}}$ (Eq. (31)) perfectly isolates and recovers the classical Adjusted $R^2$:

$$R^2_{\mathrm{V}} = 1 - \frac{\hat{v}_1^c}{\hat{v}_0^c} = 1 - \frac{2\pi e\left(\frac{\mathrm{RSS}_1}{n-p_1}\right)}{2\pi e\left(\frac{\mathrm{RSS}_0}{n-p_0}\right)} = 1 - \frac{\mathrm{RSS}_1/(n-p_1)}{\mathrm{RSS}_0/(n-p_0)} = R^2_{\mathrm{adj}}. \tag{S126}$$

## C.5 Recovery of Non-Centrality and CIs

When the restricted model is false ($\rho^2_{\mathrm{V}} > 0$), the central test seamlessly transitions to a non-central distribution (Theorem S9). The non-centrality parameter dictating this shift is $\lambda_{\mathrm{V}} = n\phi^2_{\mathrm{V}}$, where the population signal-to-noise ratio is $\phi^2_{\mathrm{V}} = (\nu_0^* - \nu_1^*)/\nu_1^*$.

Substituting the Gaussian variance components yields:

$$\lambda_{\mathrm{V}} = n\left(\frac{2\pi e\sigma_0^2 - 2\pi e\sigma_1^2}{2\pi e\sigma_1^2}\right) = n\left(\frac{\sigma_0^2 - \sigma_1^2}{\sigma_1^2}\right). \tag{S127}$$

This exactly recovers the classical ANOVA non-centrality parameter $\lambda$. Consequently, the numerical inversion procedure (Eq. (35)) for constructing CIs on $\rho_{\mathrm{V}}^2$ perfectly mirrors the classical method of inverting the non-central $F$-distribution for linear models.

## C.6 Recovery of $R^2_{\mathrm{Pred}}$ and MSPE

Finally, we recover the predictive EV and $R^2_{\mathrm{VP}}$. Under the Gaussian log-likelihood (White, 1982), the evaluated cross-entropy at the population optimum $\boldsymbol{\theta}^*$ resolves to:

$$h_j^*(\boldsymbol{\theta}_j^*) = \frac{1}{2}\log(2\pi e\sigma_j^2), \tag{S128}$$

where $\sigma_j^2 = \frac{1}{n}\sum_{i=1}^n \mathbb{E}_{f^*}\left[\,(Y_i - \mathbf{x}_{ij}^\top \boldsymbol{\beta}_j^*)^2\,\right]$ is the true population variance.

To evaluate the predictive performance of a model fitted with parameters $\hat{\boldsymbol{\theta}} = (\hat{\beta}, \hat{\sigma}^2)$, we consider an independent test set $Y^{\text{test}} \sim \mathcal{N}(\mathbf{X}\beta, \sigma^2 I_n)$. We define the cross-entropy $h^*$ as the expected negative log-likelihood of the test data under the model's fitted predictive distribution $\hat{f} = \mathcal{N}(\mathbf{x}^\top\hat{\beta}, \hat{\sigma}^2)$. For a single test point, the model's negative log-density is

$$-\log \hat{f}(y) = \frac{1}{2}\log(2\pi\hat{\sigma}^2) + \frac{(y - \mathbf{x}^\top\hat{\beta})^2}{2\hat{\sigma}^2}. \tag{S129}$$

Averaging over the $n$ test points and taking the expectation under the true distribution $f^*$ leaves the leading term unchanged (it does not depend on $y$) and converts the quadratic term into the mean expected squared prediction error:

$$\frac{1}{n}\sum_{i=1}^{n}\mathbb{E}_{f^*}\left[\,(Y_i^{\text{test}} - \mathbf{x}_i^\top\hat{\beta})^2\,\right] \equiv \mathrm{MSPE}\ (\hat{\beta}). \tag{S130}$$

Because the test noise is independent of $\hat{\beta}$ and mean-zero, the cross-term vanishes and the prediction error separates into irreducible noise plus in-sample estimation error:

$$\mathrm{MSPE}\ (\hat{\beta}) = \sigma^2 + \frac{1}{n}\mid \mathbf{X}(\hat{\beta} - \beta)\mid^2 = \sigma^2 + \frac{1}{n}(\hat{\beta} - \beta)^\top \mathbf{X}^\top\mathbf{X}\,(\hat{\beta} - \beta). \tag{S131}$$

The expected cross-entropy of the fitted model on the test set is therefore:

$$h^*(\hat{\beta}, \hat{\sigma}^2) = \frac{1}{2}\log(2\pi\hat{\sigma}^2) + \frac{\mathrm{MSPE}\ (\hat{\beta})}{2\hat{\sigma}^2}. \tag{S132}$$

To evaluate the true predictive performance across all possible training datasets, we take the expectation $\mathbb{E}[h^*(\hat{\beta}, \hat{\sigma}^2)]$. In Gaussian OLS, the estimator $\hat{\beta}$ and the variance estimator $\hat{\sigma}^2$ are independent. Taking the expectation of the closed-form MSPE over the sampling distribution $\hat{\beta} - \beta \sim \mathcal{N}\left(\,0, \sigma^2(\mathbf{X}^\top\mathbf{X})^{-1}\,\right)$ recovers the exact prediction error directly, since $\mathbb{E}\left[\,(\hat{\beta} - \beta)^\top\mathbf{X}^\top\mathbf{X}(\hat{\beta} - \beta)\,\right] = \sigma^2\,\mathrm{tr}(I_p) = \sigma^2 p$:

$$\text{MSPE} = \mathbb{E}[\text{MSPE }(\hat{\beta})] = \sigma^2\left(1 + \frac{p}{n}\right). \tag{S133}$$

Accounting for the finite-sample variability of the variance estimate $W = n\hat{\sigma}^2/\sigma^2 \sim \chi^2_{n-p}$, and noting that $\mathbb{E}[1/W] = 1/(n-p-2)$, we have:

$$\mathbb{E}\left[\frac{1}{\hat{\sigma}^2}\right] = \frac{n}{\sigma^2}\mathbb{E}\left[\frac{1}{W}\right] = \frac{n}{\sigma^2(n-p-2)} \approx \frac{1}{\sigma^2(1-p/n)}. \tag{S134}$$

We now substitute these moments. Using the independence of $\hat{\beta}$ and $\hat{\sigma}^2$, the expected cross-entropy splits into a log-variance term and a standardized fit term, each evaluated with the moments derived above ($\mathbb{E}[\hat{\sigma}^2] = \sigma^2(1-p/n)$ via the delta method for the log; $\nu^* = 2\pi e\,\sigma^2$ so $\log \nu^* = \log(2\pi\sigma^2) + 1$):

$$\begin{aligned}
2\mathbb{E}[h^*] = \log(2\pi) + &\underbrace{\mathbb{E}[\log \hat{\sigma}^2]}_{\approx \log \sigma^2 + \log(1-p/n)} + \underbrace{\mathbb{E}[\text{MSPE }(\hat{\beta})]\,\mathbb{E}[1/\hat{\sigma}^2]}_{\approx (1+p/n)/(1-p/n)} \\
&\approx \log(2\pi\sigma^2) + \log(1-p/n) + \frac{1+p/n}{1-p/n} && \text{(substitute moments)} \\
&= \log \nu^* - 1 + \log(1-p/n) + \frac{1+p/n}{1-p/n} && \text{(write in terms of } \nu^* \text{)} \\
&= \log \nu^* + \log(1-p/n) + \frac{2p/n}{1-p/n} && \left(\frac{1+p/n}{1-p/n} - 1 = \frac{2p/n}{1-p/n}\right) \\
&\approx \log \nu^* + \log(1-p/n) - 2\log(1-p/n) && \left(\frac{2p/n}{1-p/n} \approx -2\log(1-p/n)\right) \\
&= \log \nu^* - \log(1-p/n) = \log \frac{\nu^*}{1-p/n}.
\end{aligned} \tag{S135}$$

The numerator's predictive inflation and the denominator's variance deflation compound multiplicatively, and to first order in $p/n$ the excess fit cancels the $-1$ while doubling the single log term — so the rational inflation factor structurally emerges from the finite-sample degrees of freedom of the variance estimator. Exponentiating recovers the exact theoretical inflation factor from Theorem S8:

$$\exp\Big(2\mathbb{E}[h^*(\hat{\beta}, \hat{\sigma}^2)]\Big) \approx \nu^*\left(1 - \frac{p}{n}\right)^{-1} = \frac{\nu^*}{1-p/n}(\approx 2\pi e \text{ MSPE}). \tag{S136}$$

To estimate this from training data, we apply the squared penalty to correct for both training deflation and predictive inflation: $\hat{v}_j^{\text{test}} = \hat{v}_j/(1-p_j/n)^2$. In the OLS setting, this recovers the Generalized Cross-Validation (GCV) estimator:

$$\hat{v}_j^{\text{test}} = 2\pi e\left[\frac{\text{RSS}_j/n}{(1-p_j/n)^2}\right] \equiv 2\pi e\widehat{\text{MSPE}}_j. \tag{S137}$$

Because this geometric penalty is asymptotically equivalent to Amemiya's Prediction Cri-terion and Akaike's Final Prediction Error, $R^2_{\text{VP}}$ (Eq. (38)) isolates the classical predictive coefficient of determination using empirical estimators ($\widehat{\text{MSPE}}$):

$$R^2_{\text{VP}} = 1 - \frac{\hat{v}_1^{\text{test}}}{\hat{v}_0^{\text{test}}} = 1 - \frac{2\pi e \widehat{\text{MSPE}}_1}{2\pi e \widehat{\text{MSPE}}_0} = 1 - \frac{\widehat{\text{MSPE}}_1}{\widehat{\text{MSPE}}_0} = R^2_{\text{pred}}. \tag{S138}$$

This establishes $R^2_{\text{VP}}$ not as a heuristic pseudo-$R^2$, but as the strict likelihood-based structural equivalent to the classical predictive coefficient of determination.

## C.7 Reconstruction of RSS of OLS

The structural mechanics of the EVD ($\widehat{\text{EVD}}$) become completely concrete when evaluated under a Gaussian linear model. In this classical Euclidean setting, the EV is directly equivalent to the squared Euclidean distance, meaning our EV Totals ($\widehat{\text{EVT}}$) map directly to the familiar Residual Sums of Squares (RSS).

**Step 1: Defining the EV Totals ($\widehat{\text{EVT}}$)**

Let $\mathbf{y} = \mathbf{X}\boldsymbol{\beta}^* + \boldsymbol{\epsilon}$ be the true DGP. The absolute variance anchor, $\widehat{\text{EVT}}^*$, maps to the squared length of the true, unobservable population errors (RSS$^*$):

$$\widehat{\text{EVT}}^* = |\ \mathbf{y} - \mathbf{X}\boldsymbol{\beta}^*\ |^2 = \text{RSS}^*. \tag{S139}$$

When we fit nested models $\mathcal{M}_0 \subset \mathcal{M}_1$, yielding predictions $\hat{\mathbf{y}}_0 = \mathbf{X}_0\hat{\boldsymbol{\beta}}_0$ and $\hat{\mathbf{y}}_1 = \mathbf{X}_1\hat{\boldsymbol{\beta}}_1$, their empirical variance totals map to their respective observed residual sums of squares:

$$\widehat{\text{EVT}}_0 = |\ \mathbf{y} - \hat{\mathbf{y}}_0\ |^2 = \text{RSS}_0 \tag{S140}$$

$$\widehat{\text{EVT}}_1 = |\ \mathbf{y} - \hat{\mathbf{y}}_1\ |^2 = \text{RSS}_1. \tag{S141}$$

**Step 2: Deriving the Absolute Drops ($\widehat{\text{EVD}}_j^{\text{abs}}$)**

By definition, the absolute drop isolates the variance deflated during model training:

$$\widehat{\text{EVD}}_j^{\text{abs}} = \widehat{\text{EVT}}^* - \widehat{\text{EVT}}_j = |\ \mathbf{y} - \mathbf{X}\boldsymbol{\beta}^*\ |^2 - \left\|\mathbf{y} - \hat{\mathbf{y}}_j\right\|^2. \tag{S142}$$

Under the assumption that the model contains the true DGP, the fitted vector $\hat{\mathbf{y}}_j$ is an orthogonal projection of $\mathbf{y}$. By the Pythagorean theorem, the true error decomposes orthogonally as $\|\mathbf{y} - \mathbf{X}\boldsymbol{\beta}^*\|^2 = \left\|\mathbf{y} - \hat{\mathbf{y}}_j\right\|^2 + \left\|\hat{\mathbf{y}}_j - \mathbf{X}\boldsymbol{\beta}^*\right\|^2$. Substituting this into our definition yields:

$$\widehat{\text{EVD}}_j^{\text{abs}} = \left\|\hat{\mathbf{y}}_j - \mathbf{X}\boldsymbol{\beta}^*\right\|^2. \tag{S143}$$

Geometrically, the absolute drop is the pure estimation error—the variance spuriously absorbed by the $p_j$ parameters "chasing" the noise.

**Step 3: Deriving the Relative Drop ($\widehat{\text{EVD}}$)**

The relative variance drop is defined as the difference between the restricted and full models' variance totals. In OLS, this is the classical Sum of Squares Reduction (SSR):

$$\widehat{\mathrm{EVD}} = \widehat{\mathrm{EVT}}_0 - \widehat{\mathrm{EVT}}_1 = \mathrm{RSS}_0 - \mathrm{RSS}_1. \tag{S144}$$

Because $\mathcal{M}_0 \subset \mathcal{M}_1$, the fitted vectors are orthogonal projections of each other, allowing us to rewrite this Euclidean difference as:

$$\widehat{\mathrm{EVD}} = \|\hat{\mathbf{y}}_1 - \hat{\mathbf{y}}_0\|^2. \tag{S145}$$

### Step 4: Specialization of the Nested Identity and Distributions

We can now map these Euclidean derivations directly into our generalized nested identity ($\widehat{\mathrm{EVD}}_1^{\mathrm{abs}} = \widehat{\mathrm{EVD}}_0^{\mathrm{abs}} + \widehat{\mathrm{EVD}}$):

$$\|\hat{\mathbf{y}}_1 - \mathbf{X}\boldsymbol{\beta}^*\|^2 = \|\hat{\mathbf{y}}_0 - \mathbf{X}\boldsymbol{\beta}^*\|^2 + \|\hat{\mathbf{y}}_1 - \hat{\mathbf{y}}_0\|^2. \tag{S146}$$

This elegantly proves that the total estimation error of the full model ($\widehat{\mathrm{EVD}}_1^{\mathrm{abs}}$) is orthogonally constructed from the baseline model's estimation error ($\widehat{\mathrm{EVD}}_0^{\mathrm{abs}}$) plus the relative variance explicitly explained by the newly added parameters ($\widehat{\mathrm{EVD}}$).

Because these vector components lie in strictly nested orthogonal subspaces, classical normal theory dictates their exact distributions. This perfectly matches our general theoretical framework (setting the population scale factor to $\nu^* = \sigma^2$):

$$\widehat{\mathrm{EVD}}_0^{\mathrm{abs}} \sim \sigma^2 \chi^2_{p_0} \tag{S147}$$

$$\widehat{\mathrm{EVD}}_1^{\mathrm{abs}} \sim \sigma^2 \chi^2_{p_1} \tag{S148}$$

$$\widehat{\mathrm{EVD}} \sim \sigma^2 \chi^2_k. \tag{S149}$$

This geometric alignment confirms that the ubiquitous ANOVA decomposition is simply the Euclidean manifestation of the orthogonal score projections governing the EV manifold.

# D Relationship between $R^2_{\mathrm{V}}$ and Information Criteria

## D.1 Relationship with AIC

Our proposed framework shares a theoretical lineage with Akaike Information Criterion (AIC) through the KL divergence, but they serve distinct inferential goals. In classical OLS and GLM, methodologists have long cautioned against relying purely on information criteria like AIC for model selection, particularly in large datasets (Burnham & Anderson, 2002; Lin et al., 2013). Our framework extends these observations into general likelihood inference by explicitly demonstrating that $\Delta$ AIC functions as a test of statistical significance, whereas its per-observation transformations, $R^2_{\mathrm{V}}$ and $R^2_{\mathrm{VP}}$, measure practical predictive magnitude.

To make this relationship mathematically explicit, let $D = 2n\Delta\hat{h}$ denote the empirical LRT statistic. From the standard definition of AIC, we have the direct identity $D =$ $\Delta$ AIC $+ 2k$. Because $D$ converges asymptotically to a central $\chi^2_k$ distribution under the null hypothesis, the implicit $p$-value corresponding to any observed AIC difference is exactly:

$$p\text{ -value} = P(\chi^2_k > \Delta\text{AIC}^{(\text{obs})} + 2k). \tag{S150}$$

Eq. (S150) shows that $\Delta$ AIC operates as a measure of statistical significance. Any model selection rule requiring $\Delta$ AIC $> c$ for some fixed threshold $c$ is mathematically identical to a classical LRT. For a single parameter addition ($k = 1$), choosing a model simply because $\Delta$ AIC $> 0$ is equivalent to accepting a new predictor based on a rigid $p$-value threshold of $P(\chi^2_1 > 2) \approx 0.157$.

To understand why this equivalence creates a vulnerability in large samples, we must examine the behavior of AIC when a true population effect exists. As derived previously, under the contiguous alternative hypothesis, the asymptotic expectation of the LRT statistic is governed by the log-likelihood non-centrality parameter: $\mathbb{E}[D] \approx k + \lambda_{\text{D}}$. Consequently, the expected value of the AIC difference is:

$$\mathbb{E}[\Delta \text{ AIC}] = \mathbb{E}[D] - 2k \approx (k + \lambda_{\text{D}}) - 2k = \lambda_{\text{D}} - k = -n \log(1 - \rho^2_{\text{V}}) - k. \tag{S151}$$

Eq. (S151) exposes the core limitation of AIC. The metric is driven by a non-centrality parameter $\lambda_{\text{D}}$ that scales linearly with the sample size $n$. If a predictor has a true but practically negligible population effect (e.g., $\rho^2_{\text{V}} = 0.001$), a sufficiently large $n$ will inflate $\lambda_{\text{D}}$ toward infinity. This massive non-centrality ensures that the resulting $D$ yields a near-zero $p$-value, and $\Delta$ AIC will exceed any fixed practical threshold $c$. Because AIC measures the total accumulated statistical evidence rather than the average predictive utility, it inevitably endorses trivial covariates in massive datasets.

To evaluate practical significance, the accumulated evidence must be normalized into a per-observation metric. Substituting the AIC formulation into our SNR-based population and predictive effect sizes yields:

$$R^2_{\text{V}} = 1 - \left(\frac{n - p_0}{n - p_1}\right) \exp\left(-\frac{\Delta \text{ AIC} + 2k}{n}\right). \tag{S152}$$

Furthermore, by applying the first-order approximation $\left(\frac{n-p_0}{n-p_1}\right)^2 \approx \exp(2k/n)$ to the out-of-sample penalty, the predictive effect size simplifies elegantly to:

$$R^2_{\text{VP}} \approx 1 - \exp\left(-\frac{\Delta \text{ AIC}}{n}\right). \tag{S153}$$

**These formulas cleanly decouple statistical evidence from practical magnitude.** Both $R^2_{\text{V}}$ and $R^2_{\text{VP}}$ are governed by the ratio $\Delta$ AIC $/n$. By dividing the total evidence by the sample size inside the exponential, these metrics successfully scale out the $n$ factor from the non-centrality parameter. Because the expectation $\mathbb{E}[\Delta \text{ AIC } /n] \approx -\log(1 - \rho^2_{\text{V}})$, this normalization ensures that the metrics directly estimate the true underlying population effect, remaining completely stable against the sheer sample size $n$.

To empirically demonstrate the mathematical discussions above, we generated a deter-ministic sequence of observed deviances ($2\Delta H_n \in [0, 15]$) for a single parameter addition ($k = 1$) and calculated the exact simultaneous trajectories of $\Delta$ AIC, the F-test $p$-value, $R^2_{\text{V}}$, and $R^2_{\text{VP}}$ across three distinct sample sizes. Fig. S10 plots these results. As illustrated in the rightmost panel (Massive Sample, $n = 10000$), the sheer volume of data inflates the statistical evidence exactly as our asymptotic expectation predicted. Even for minuscule entropy drops, $\Delta$ AIC easily crosses the positive threshold and the $p$-value rapidly collapses to zero, aggressively endorsing the new parameter. However, precisely because $R^2_{\text{V}}$ and $R^2_{\text{VP}}$ normalize this accumulated evidence by $n$, the exact same entropy drops correctly register as practically zero on the practical effect size scale. Ultimately, this demonstrates that $\Delta$ AIC functions essentially as a statistical significance measure. To properly evaluate practical utility, $R^2_{\text{V}}$ and $R^2_{\text{VP}}$—which are approximately equal to $\Delta$ AIC $/n$ for large samples—provide the necessary effect size metrics for regular likelihood families.

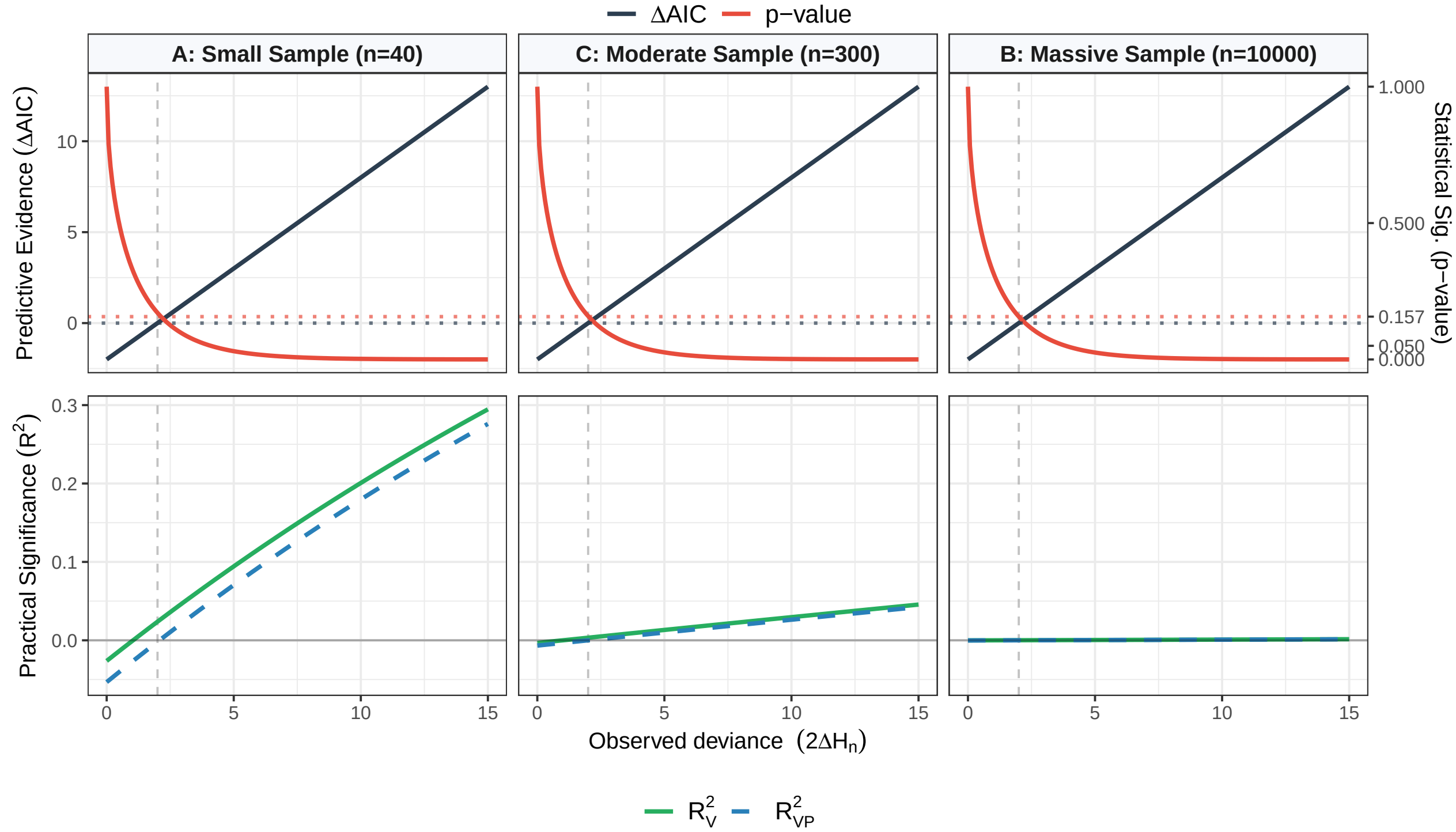

Figure S10: Behavior of inferential metrics across different sample sizes as a function of the observed deviance ($2\Delta H_n$), holding the parameter penalty fixed at $k = 1$.

## D.2 Relationship with BIC

While the Bayesian Information Criterion (BIC) imposes a much stricter complexity penalty than AIC ($k \log n$ rather than $2k$), it remains vulnerable to the same large-sample divergence between statistical evidence and practical utility.

The empirical difference in BIC between the full and restricted models is defined as $\Delta$ BIC $= D - k \log n$. Under the alternative hypothesis, substituting the expected LRT statistic yields:

$$\mathbb{E}[\Delta\ \text{BIC}] \approx \lambda_{\text{D}} + k - k\log n = -n\log(1-\rho_{\text{V}}^2) - k(\log n - 1). \tag{S154}$$

Because BIC evaluates the Laplace approximation of the marginal likelihood, its formulation establishes a race between the empirical signal (which grows at $O(n)$) and the dimensional penalty (which grows at $O(\log n)$). This makes BIC statistically consistent; it is less inclusive than AIC in finite samples, demanding stronger evidence to justify additional complexity.

However, as $n \to \infty$, the linear growth of the sample size mathematically overpowers the logarithmic penalty. Consequently, for any strictly non-zero population effect—no matter how microscopically small $\rho_{\text{V}}^2$ might be—an enormous sample size will eventually drive $\Delta$ BIC to strongly endorse the complex model. BIC reliably answers the existential question of whether an effect is identically zero, but it lacks a natural scale to convey if the effect is practically meaningful.

The predictive effect size $R^2_{\text{VP}}$ bridges this gap by absorbing both the statistical evidence and the Bayesian penalty into a stable metric. By substituting the identity $D = \Delta$ BIC $+ k\log n$ into the asymptotic expansion of the out-of-sample predictive penalty, we map the Bayesian criterion directly into the predictive proportion of variance explained:

$$R^2_{\text{VP}} \approx 1 - \exp\left(-\frac{\Delta\ \text{BIC} + k(\log n - 2)}{n}\right). \tag{S155}$$

Eq. (S155) demonstrates how the framework handles overwhelming statistical evidence. By dividing the total BIC difference by $n$, the predictive metric neutralizes the $O(n)$ inflation of the non-centrality parameter. If a massive dataset produces a large $\Delta$ BIC solely because of sample size leverage on a trivial effect, the $1/n$ scaling will correctly compress $R^2_{\text{VP}}$ back toward zero, warning the practitioner that the variable, while "true", offers negligible out-of-sample predictive utility.

# E Comparisons of EV $R^2_{\text{SV}}$ and Other $R^2$

Historically, various pseudo-$R^2$ statistics have been proposed to extend the classical $R^2$ to GLM. These predictive effect size generally fall into two categories: likelihood-based transformations and variance-partitioning estimators. Our SNR framework situates itself within the likelihood tradition, offering finite-sample corrections and continuous stability that legacy predictive effect size lack.

## E.1 Cox & Snell $R^2$ and Nagelkerke's $R^2$

The most direct ancestor to our framework is the **Cox & Snell $R^2$** ($R^2_{\text{CS}}$), which defines the effect size via the ratio of likelihoods [or, equivalently, the empirical SNR; Cox & Snell (1989)]:

$$R^2_{\text{CS}} = 1 - \exp(-2\Delta\hat{h}). \tag{S156}$$

Because Eq. (S156) relies on raw, unpenalized entropies, it is highly susceptible to training optimism. Our proposed $R^2_{\mathrm{V}}$ (Eq. (31)) serves as the exact bias-corrected analogue to Cox & Snell, incorporating the degrees-of-freedom penalty derived from the sampling distribution.

A well-known limitation of $R^2_{\mathrm{CS}}$ is that its maximum value is strictly less than 1 for discrete models. To resolve this, **Nagelkerke's $R^2$** proposed rescaling the metric by its theoretical maximum (Nagelkerke, 1991):

$$R^2_{\mathrm{Nagelkerke}} = \frac{1 - \exp(-2\Delta\hat{h})}{1 - \exp(-2\hat{h}_{\mathrm{null}})}. \tag{S157}$$

Our standardized statistic, $R^2_{\mathrm{SV}}$ (Eq. (38)), mirrors this logic but introduces vital improvements. Unlike Eq. (S157), whose denominator can become negative or undefined for continuous densities when the log-likelihood is positive, our framework defines the normalization constant via theoretical information capacity, guaranteeing stability and convergence to 1 across both discrete and continuous domains.

## E.2 McFadden's $R^2$

A different likelihood-based approach is **McFadden's $R^2$**, which measures the proportional reduction in total deviance (or empirical entropy):

$$R^2_{\mathrm{McFadden}} = 1 - \frac{\hat{h}_1}{\hat{h}_0}. \tag{S158}$$

While widely used in discrete choice modeling, Eq. (S158) faces catastrophic mathematical limitations as a general-purpose metric for continuous data. Because continuous differential entropy is not bounded below by zero, the raw empirical entropies $\hat{h}_1$ and $\hat{h}_0$ can be negative. This creates severe failure modes for the linear ratio $\hat{h}_1/\hat{h}_0$. For instance, if the restricted model has positive entropy but the full model fits well enough to achieve negative entropy, the ratio becomes negative, yielding an $R^2_{\mathrm{McFadden}} > 1$. Worse, if both models exhibit negative entropies, the metric's logic inverts entirely: a better-fitting model (yielding a more negative $\hat{h}_1$) produces a larger positive ratio, which actively drives $R^2_{\mathrm{McFadden}}$ below zero despite a strict improvement in model fit. Furthermore, the linear ratio of entropies does not map to the quadratic geometry of variance, causing McFadden's metric to systematically produce values much lower than OLS $R^2$ for equivalent effect sizes (Domencich & McFadden, 1975; Hensher et al., 2005). Our SNR framework resolves all of these failures. By exponentiating the entropies before taking the ratio ($v = \exp(2h)$), we map the unbounded entropy space $(-\infty, \infty)$ strictly into the non-negative EV space $[0, \infty)$. This absolute scaling guarantees that $R^2_{\mathrm{V}}$ remains bounded $\leq 1$, preserves the correct directional logic of model improvement, and successfully recovers the classical adjusted $R^2$ in the Gaussian case.

## E.3 $R^2$ based on Decomposition of Variances

Alternatively, some predictive effect size attempt to preserve the "variance explained" interpretation of OLS through variance-partitioning. The most direct empirical approach is **Efron's pseudo-$R^2$**, which applies the classical sum-of-squares formula directly to the raw residuals of generalized models:

$$R^2_{\text{Efron}} = 1 - \frac{\sum_{i=1}^{n}(y_i - \hat{\mu}_i)^2}{\sum_{i=1}^{n}(y_i - \overline{y})^2}. \tag{S159}$$

To formalize this beyond empirical residuals, we can generalize the metric using the model-implied conditional regular variances, $\sigma^2_{j,i} = \text{Var}\,(Y_i \mid \mathbf{x}_i, \boldsymbol{\theta}^*_j)$. By the Law of Total Variance, conditioning on covariates decomposes the marginal outcome variance into the variance of the conditional means (explained variance) and the expectation of the conditional variances (unexplained variance). In a finite sample, this expected unexplained variance is estimated by the arithmetic mean (AM) of the observation-level conditional regular variances. This induces a generalized regular-variance $R^2$ at population level:

$$\rho^2_{\text{variance}} = 1 - \frac{\text{AM}\left(\,[\sigma^2_{1,i}]_{i=1}^{n}\,\right)}{\text{AM}\left(\,[\sigma^2_{0,i}]_{i=1}^{n}\,\right)} = 1 - \frac{\sum_{i=1}^{n}\sigma^2_{1,i}}{\sum_{i=1}^{n}\sigma^2_{0,i}}. \tag{S160}$$

Metrics such as **Gelman's Bayesian $R^2$** (Gelman et al., 2019),recent extensions of variance-based $R^2$ to generalized linear and mixed models (Jaeger et al., 2017; Nakagawa et al., 2017; Zhang, 2017; 2022), as well as domain-specific $R^2$ for binary outcomes (Tjur, 2009), are modern expressions of this decomposition, defining the metric via the variance of the fitted predictive means relative to the modeled residual variance.

### E.3.1 The Failure of Invariance of $\rho^2_{\text{variance}}$ Under Transformation

We take as a guiding principle that a generalized measure of information or effect size should be invariant under strictly monotone transformations. If a researcher models a strictly positive response as $Y$ or applies a log-transformation to model $W = \log(Y)$, the underlying informational strength of the relationship between the predictors and the response has not changed; only the coordinate system has. A rigorous $R^2$ metric should reflect this geometric reality by remaining exactly constant.

Variance-partitioning predictive effect size, however, are tied to the scale of the chosen coordinate system. Suppose we apply a non-linear monotone transformation $W = g(Y)$. By the Delta method, the conditional regular variance of the transformed variable under model $\mathcal{M}_j$ is approximated by scaling the original variance by the squared derivative of the transformation function, evaluated at the model's conditional mean $\mu_{j,i}$:

$$\text{Var}\,(W_i \mid \mathbf{x}_i, \mathcal{M}_j) \approx [g'(\mu_{j,i})]^2\ \text{Var}\,(Y_i \mid \mathbf{x}_i, \mathcal{M}_j). \tag{S161}$$

Let $w_{j,i} = [g'(\mu_{j,i})]^2$ denote this stretch factor. Because the transformation is non-linear, this weight depends on the model's predicted mean. Computing the classical variance-based $R^2$ (Eq. (S160)) in the new coordinate space therefore applies these model-specific localized weights to the additive sum:

$$\rho^2_{\text{variance}}(W) \approx 1 - \frac{\sum_{i=1}^n w_{1,i}\sigma^2_{1,i}}{\sum_{i=1}^n w_{0,i}\sigma^2_{0,i}}. \tag{S162}$$

Because $\mu_{1,i} \neq \mu_{0,i}$ in general, the weights $w_{1,i}$ and $w_{0,i}$ differ, so the non-linear transformation rescales the variances in the numerator and denominator by different amounts. Consequently, $\rho^2_{\text{variance}}(W) \neq \rho^2_{\text{variance}}(Y)$, and the classical metric is not invariant under reparametrization.

### E.3.2 Invariance of $\rho^2_{\text{V}}$ Under Transformation

By contrast, the entropic variance transforms in a controlled, model-independent way: any strictly monotone reparametrization contributes a single global Jacobian factor that cancels in the ratio defining $\rho^2_{\text{V}}$. For a strictly monotone transformation $W = g(Y)$, the change of variables formula gives the predictive density of $W$ as $f_W(w \mid \mathcal{M}_j) = f_Y(g^{-1}(w) \mid \mathcal{M}_j)/ \mid g'(g^{-1}(w)) \mid$; evaluated for the random variable $W = g(Y)$, this becomes $f_W(W \mid \mathcal{M}_j) = f_Y(Y \mid \mathcal{M}_j)/ \mid g'(Y) \mid$.

When we evaluate the cross-entropy of the transformed variable, the logarithm separates the density from the Jacobian:

$$h(W \mid \mathcal{M}_j) = -\mathbb{E}\left[\log\left(\frac{f_Y(Y \mid \mathcal{M}_j)}{\mid g'(Y) \mid}\right)\right] = h(Y \mid \mathcal{M}_j) + \mathbb{E}[\log\mid g'(Y) \mid]. \tag{S163}$$

Let $C = \mathbb{E}[\log\mid g'(Y) \mid]$. Because this term depends only on the transformation $g$ and the true distribution of $Y$, $C$ is independent of the model $\mathcal{M}_j$. Exponentiating to find the EV yields:

$$\nu^*_j(W) = \exp(2[h(Y \mid \mathcal{M}_j) + C]) = \nu^*_j(Y) \cdot e^{2C}. \tag{S164}$$

The non-linear transformation manifests as a single global scalar, $e^{2C}$, applied uniformly to the EV of every model. In the ratio defining the metric, this global scalar appears identically in both the candidate and reference models and cancels:

$$\rho^2_{\text{V}}(W) = 1 - \frac{\nu^*_1(W)}{\nu^*_0(W)} = 1 - \frac{\nu^*_1(Y)\cdot e^{2C}}{\nu^*_0(Y)\cdot e^{2C}} = 1 - \frac{\nu^*_1(Y)}{\nu^*_0(Y)} = \rho^2_{\text{V}}(Y). \tag{S165}$$

By rooting the calculation in the EV of the distribution, the EV $\rho^2_{\text{V}}$ absorbs the local Jacobian, yielding exact invariance across all strictly monotone transformations.

### E.3.3 Numerical Demonstration of Invariance Failure of $\rho^2_{\text{variance}}$

To make this failure concrete, consider a synthetic example with two nested linear models and an exponential transformation of the response variable.

Suppose we have a dataset of two observations ($n = 2$). In the original linear coordinate space, the response is modeled as Gaussian: $Y \sim \mathcal{N}(\mu, \sigma^2)$.

- **Model 1 (Candidate):** Captures a trend with conditional means $\mu_{1,1} = -1$ and

$\mu_{1,2} = 1$, and constant conditional variance $\sigma_1^2 = 1.0$.

- **Model 0 (Null):** By the Law of Total Variance, the collapsed null model absorbs the variance of the omitted conditional means. The marginal mean is $\mu_0 = 0$, the variance of the Model 1 means is Var $(\mu_1) = \frac{(-1)^2+(1)^2}{2} = 1.0$, and so the null variance is $\sigma_0^2 = \text{Var}\ (\mu_1) + \sigma_1^2 = 2.0$.

Now suppose a researcher models the strictly positive exponential of the same data, applying the transformation $W = \exp(Y)$. The variable $W$ follows a Log-Normal distribution, whose exact analytical variance entangles the mean and the scale parameter:

$$\text{Var}\ (W_{j,i}) = \exp(2\mu_{j,i} + \sigma_j^2) \cdot \left( \exp(\sigma_j^2) - 1 \right) . \tag{S166}$$

Because $\mu_{j,i}$ enters the exponent, it acts as an observation-specific multiplier. We can compute the exact conditional variances side-by-side for both coordinate spaces ($Y$ and $W$), along with their arithmetic means (AM).

| Obs ($i$) | $\mu_{0,i}$ | Var $(Y_{0,i})$ | Var $(W_{0,i})$ | $\mu_{1,i}$ | Var $(Y_{1,i})$ | Var $(W_{1,i})$ |
|---|---|---|---|---|---|---|
| **1** | 0 | 2.0 | 47.210 | $-1$ | 1.0 | 0.632 |
| **2** | 0 | 2.0 | 47.210 | 1 | 1.0 | 34.513 |
| **AM** | | **2.0** | **47.210** | | **1.0** | **17.572** |

*Note: Calculations use the exact Log-Normal variance formula. For example,* Var $(W_{0,i}) = \exp(0 + 2.0) \cdot (e^{2.0} - 1) \approx 7.389 \times 6.389 = 47.210$.

Using the arithmetic means (AM) from the table, we compute the classical variance-based $\rho^2$ in both coordinate spaces:

**In the original linear space ($Y$):**

$$\rho^2_{\text{variance}}(Y) = 1 - \frac{\text{AM (Var } (Y_{1,i}))}{\text{AM (Var } (Y_{0,i}))} = 1 - \frac{1.0}{2.0} = 0.500. \tag{S167}$$

**In the transformed exponential space ($W$):**

$$\rho^2_{\text{variance}}(W) = 1 - \frac{\text{AM (Var } (W_{1,i}))}{\text{AM (Var } (W_{0,i}))} = 1 - \frac{17.572}{47.210} \approx 0.628. \tag{S168}$$

The metric fails to be invariant. The underlying probabilistic reality of the models is unchanged—Model 1 still accounts for exactly the same proportion of the underlying uncertainty—yet the variance-based metric shifts from 0.500 to 0.628. The cause is that the arithmetic mean is not log-linear: it cannot separate the genuine reduction in model scale from the observation-specific stretch factor $\exp(2\mu_{j,i})$, which inflates the local variance differently in the numerator and denominator. A metric built on the arithmetic mean of variances therefore conflates model fit with the scale of the chosen coordinate system.

#### E.3.4 Practical Relevance of the Invariance Property

To ground the preceding theoretical analysis, consider a standard empirical scenario. Suppose a log-transformation is applied to a response variable, and a Gaussian linear regression is fitted to $\log(Y)$. By standard probability theory, this is mathematically equivalent to modeling the original response $Y$ with a Log-Normal regression.

Our prior numerical demonstration assumed the evaluation of these two exact, probabilistically equivalent models. The practical relevance of this assumption stems from the standard workflow of applied model selection. Researchers typically do not fit a model in one coordinate space merely to translate its predictions back to the original scale for residual evaluation. Instead, they independently search for the optimal model architecture within whichever coordinate scale they have adopted. If the model selection process is perfectly successful in both domains, a researcher analyzing the transformed scale will identify the Gaussian regression, while a researcher analyzing the original scale will identify the Log-Normal regression.

Evaluating probabilistically equivalent models across two distinct scales therefore represents the "oracle" or ideal scenario of practical model selection. It presupposes that the optimal underlying probability family has been perfectly identified in both the transformed and original domains.

The mathematical failure demonstrated earlier establishes that even under these ideal conditions, the classical variance-based $R^2$ is structurally inconsistent. Because the arithmetic mean of conditional variances is inextricably bound to the localized stretch factors of the coordinate system, the optimal Gaussian model on $\log(Y)$ yields a fundamentally different $R^2_{\text{variance}}$ than its probabilistically identical Log-Normal counterpart on $Y$.

This inconsistency introduces a severe vulnerability into applied statistical workflows. Because practitioners routinely rely on $R^2$ metrics to adjudicate between candidate models, the lack of metric invariance actively distorts this identification process. A researcher comparing the two domains might erroneously conclude that modeling $\log(Y)$ provides a strictly superior statistical fit compared to modeling $Y$. In reality, the underlying predictive accuracy is identical; the metric has simply been confounded by coordinate warping.

By guaranteeing absolute invariance across strictly monotone transformations, the EV-based $\rho_V^2$ outputs identical values for probabilistically equivalent models. It effectively isolates the substantive statistical question of model fit from the arbitrary nuisance of coordinate scale, permitting researchers to accurately adjudicate between competing data-generating architectures without scale-induced bias.

# F A Numerical Comparison of $\rho_{\text{V}}^2$ with $\rho^2$ based on Variance and Average Precision (AUPRC) in Bernoulli Models

For binary outcomes, classical variance predictive effect size often falter. Here, we evaluate the population information gain for a general Bernoulli process, demonstrating that the EV-based framework provides a natural, link-function-agnostic measure of explained heterogeneity.

### The Standardized EV Predictive Effect Size ($\rho^2_{\mathrm{SV}}$)

Let the true DGP be independent Bernoulli trials, $Y_i \sim \text{Bern}\,(\pi_i^*)$. We compare a correctly specified full model $\mathcal{M}_1$ (with true success probabilities $\pi_i^*$) against an intercept-only reduced model $\mathcal{M}_0$ (projecting the marginal mean $\overline{\pi}^*$).

The population information gain, $2\Delta h^*$, can be elegantly expressed as twice the average Kullback-Leibler (KL) divergence between the true individual probabilities and the marginal mean:

$$2\Delta h^* = \frac{2}{n}\sum_{i=1}^{n} D_{\mathrm{KL}}(\pi_i^* \parallel \overline{\pi}^*). \tag{S169}$$

This raw information gain is strictly bounded from above by $2h_0^*$, the cross-entropy of the null model. Because the null entropy $h_0^* = -[\overline{\pi}^* \log(\overline{\pi}^*) + (1-\overline{\pi}^*)\log(1-\overline{\pi}^*)]$ approaches 0 as the marginal prevalence $\overline{\pi}^*$ approaches 0 or 1, original $\Delta h^*$ is forced to be vanishingly close to 0 in extreme prevalence settings—even for a perfectly separating model.

To ensure the metric properly scales to the true limits of binary predictability—allowing a perfect model to yield an effect size of exactly 1 regardless of prevalence—we divide the raw SNR by its theoretical ceiling, yielding the standardized global effect size:

$$\rho^2_{\mathrm{SV}} = \frac{1-\exp(-2\Delta h^*)}{1-\exp(-2h_0^*)}. \tag{S170}$$

### Variance and Precision-Based Alternatives

To contextualize $\rho^2_{\mathrm{SV}}$, we compare it against two primary alternatives used in practice:

- **Variance-Based ($\rho^2_G$):** Based on the Bayesian $R^2$ proposed by Gelman et al. (Gelman et al., 2019), this formulation simplifies exactly to the explained signal variance scaled by the total marginal variance of the binary outcome:

$$\rho^2_G = \frac{\mathrm{Var}\,(\pi_i^*)}{\overline{\pi}^*(1-\overline{\pi}^*)}. \tag{S171}$$

- **Population-Level Average Precision ($\rho^2_{\mathrm{AP}}$):** To capture tail discrimination rather than global variance, we formalize the population-level counterpart to the empirical Average Precision (AUPRC). Normalizing this by the topline prevalence yields the rank-driven effect size:

$$\rho^2_{\mathrm{AP}} = \frac{\mathrm{AUPRC}_{\mathrm{pop}} - \overline{\pi}^*}{1-\overline{\pi}^*}. \tag{S172}$$

### Numerical Demonstration of Effect Sizes

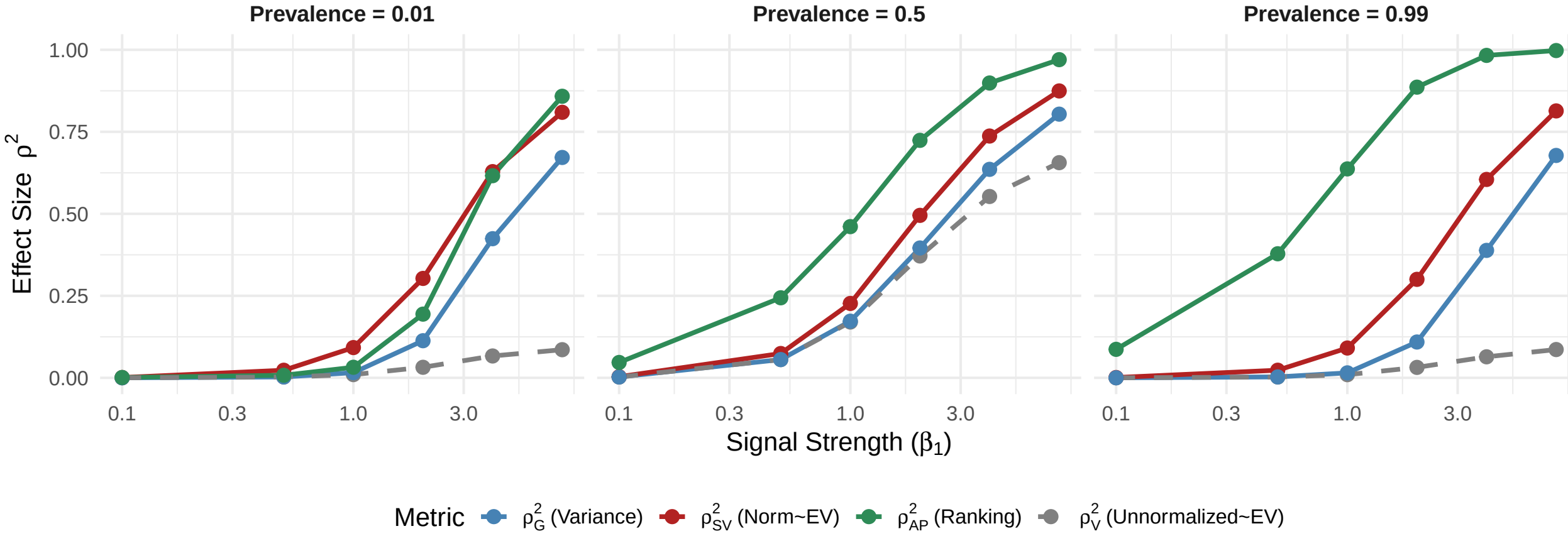

Figure S11: Comparison of EV-based, Variance-based, and Precision-based measures across prevalences.

Because analytical approximations between these metrics break down as signal strength increases or prevalence becomes highly imbalanced, we demonstrate their exact theoretical relationships via numerical simulation. We simulate a population ($n = 50,000$) with a single normally distributed predictor, varying the true signal strength ($\beta_1$) across three fixed marginal prevalences: 0.01 (rare), 0.50 (balanced), and 0.99 (ubiquitous). Fig. S11 illustrates several fundamental properties regarding how these metrics evaluate predictive performance:

1. **The Necessity of Normalizing $\rho^2_{\mathrm{V}}$:** The raw information gain ($\rho^2_{\mathrm{V}}$, gray dashed line) suffers from a severe artificial ceiling dictated entirely by the dataset's prevalence. At maximum uncertainty ($p = 0.50$), a perfectly separating model caps out at 0.75. In extreme cases ($p = 0.01$ and $p = 0.99$), it flatlines at an abysmal $\approx 0.106$. Applying the discrete normalization is absolutely necessary to restore the $[0, 1]$ scale.
2. **Global Alignment:** Once appropriately standardized, the three primary methods ($\rho^2_{\mathrm{SV}}$, $\rho^2_{\mathrm{G}}$, and $\rho^2_{\mathrm{AP}}$) perform remarkably close to one another in measuring the predictive effect size. They track the same global trajectory, providing strong empirical support for the use of the EV-based $\rho^2_{\mathrm{SV}}$ as a robust counterpart to classical variance.
3. **Entropy as an Amplifier:** While $\rho^2_{\mathrm{SV}}$ and $\rho^2_{\mathrm{G}}$ share similar paths, $\rho^2_{\mathrm{SV}}$ consistently yields a larger value than $\rho^2_{\mathrm{G}}$, particularly when the signal $\beta_1$ is small. This reflects the logarithmic nature of cross-entropy, which heavily penalizes uncertainty compared to the quadratic penalty of classical variance. Consequently, entropy acts as an amplifier, rewarding early signal detection more aggressively.
4. **The Asymmetry of $\rho^2_{\mathrm{AP}}$:** A fundamental property of a generalized $R^2$ is label invariance: the effect size should remain identical regardless of whether a biological state is coded as $Y = 1$ or $Y = 0$. Both variance ($\rho^2_{\mathrm{G}}$) and entropy ($\rho^2_{\mathrm{SV}}$) satisfy this symmetry, treating $p = 0.01$ and $p = 0.99$ as mathematically identical challenges. In contrast, the precision metric ($\rho^2_{\mathrm{AP}}$) exhibits severe asymmetry. It is highly conservative for rare events ($p = 0.01$) but becomes artificially inflated when the target event is ubiquitous ($p = 0.99$), rising rapidly even with weak signals. This violation of label symmetry fundamentally disqualifies AUPRC as a universal measure of explained heterogeneity.

# G Supplementary Tables and Figures

## G.1 Supplementary Materials for Sec. 4.1 for Illustrating the Performance of ER-based $R^2$ When Class Separation Occurs

This section presents supplementary simulation results for extreme class imbalance (target prevalence of 0.1). Under such severe imbalance, datasets are prone to perfect or near-perfect separation, particularly when noise variables (like the $z_5$ group) are included. When separation occurs, the log-likelihood's quadratic approximation degrades, rendering estimates of the true population $\rho^2_{\mathrm{SV}}$ unreliable. In practice, this instability serves as a critical diagnostic warning, easily detected when the raw $R^2_{\mathrm{SV}}$ artificially inflates to 1. Fig. S12 evaluates this effect by detailing variable selection and confidence interval (CI) coverage rates for strong ($x_1$), weak ($w$), and noise ($z_5$) groups. While EV-based estimators and CIs remain stable across most prevalences, the class separation induced at a 0.1 prevalence significantly reduces CI coverage and inflates Type I errors for $z_5$. This section provides a barplot, the $R^2$ trajectory, and a complete summary of variable selection performance and confidence interval (CI) coverage rates to illustrate the broader impact of extreme class imbalance.

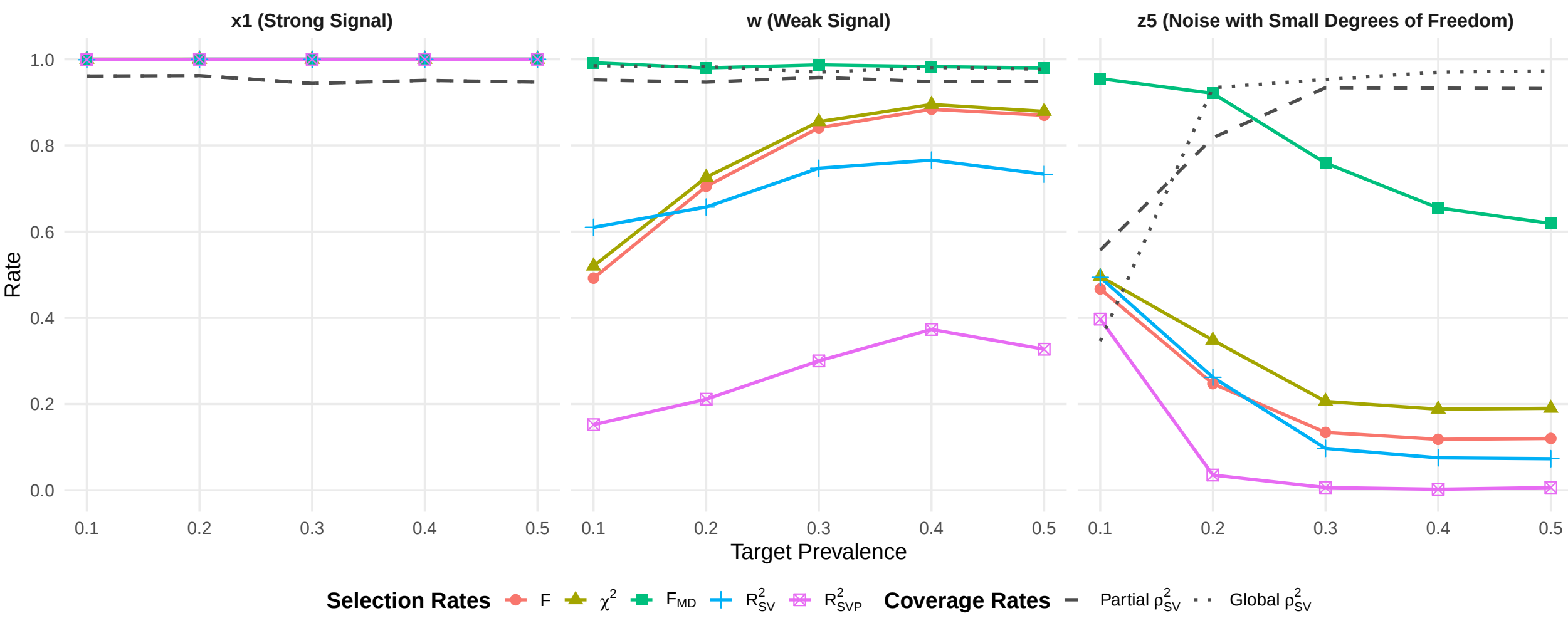

Figure S12: Variable Selection Rates and CI Covarage Rates Across Five Different Prevalence Rates.

Table S5: Summary of inferences for $N_{\text{sim}} = 1000$ datasets simulated from a logistic regression model with prevalence = 0.1. **Variable Selection** reports selection rates using five different methods. To summarize the performance of partial and global $R^2_{\text{SV}}$, the table reports their true value (True), median (Med), inter-quartile range (IQR), and coverage rate (CR) of their CIs.

| | | Variable Selection | | | | | Estimate of Partial $\rho^2_{\text{SV}}$ | | | | Estimate of Global $\rho^2_{\text{SV}}$ | | | |
|---|---|---|---|---|---|---|---|---|---|---|---|---|---|---|
| Term | $p$ | $F$ | $\chi^2$ | $F_{\text{MD}}$ | $R^2_{\text{SV}}$ | $R^2_{\text{SVP}}$ | True | Med | IQR | CR | True | Med | IQR | CR |
| x1 | 1 | 1.000 | 1.000 | 1.000 | 0.999 | 0.999 | 0.178 | 0.180 | 0.063 | 0.961 | 0.178 | 0.180 | 0.063 | 0.961 |
| x2 | 1 | 1.000 | 1.000 | 1.000 | 1.000 | 1.000 | 0.239 | 0.233 | 0.075 | 0.973 | 0.374 | 0.374 | 0.079 | 0.974 |
| x3 | 1 | 1.000 | 1.000 | 1.000 | 1.000 | 1.000 | 0.361 | 0.353 | 0.085 | 0.966 | 0.600 | 0.596 | 0.074 | 0.987 |
| w | 10 | 0.492 | 0.520 | 0.992 | 0.610 | 0.152 | 0.062 | 0.062 | 0.071 | 0.952 | 0.625 | 0.623 | 0.077 | 0.985 |
| z1 | 10 | 0.091 | 0.116 | 0.942 | 0.162 | 0.006 | 0.000 | 0.006 | 0.052 | 0.958 | 0.625 | 0.627 | 0.082 | 0.983 |
| z2 | 10 | 0.173 | 0.214 | 0.950 | 0.261 | 0.032 | 0.000 | 0.017 | 0.066 | 0.887 | 0.625 | 0.634 | 0.079 | 0.968 |
| z3 | 10 | 0.324 | 0.381 | 0.978 | 0.423 | 0.145 | 0.000 | 0.036 | 0.097 | 0.743 | 0.625 | 0.651 | 0.093 | 0.889 |
| z4 | 10 | 0.522 | 0.573 | 0.990 | 0.592 | 0.355 | 0.000 | 0.074 | 0.347 | 0.539 | 0.625 | 0.689 | 0.220 | 0.642 |
| z5 | 10 | 0.467 | 0.496 | 0.955 | 0.494 | 0.397 | 0.000 | 0.046 | 0.688 | 0.557 | 0.625 | 0.824 | 0.168 | 0.346 |

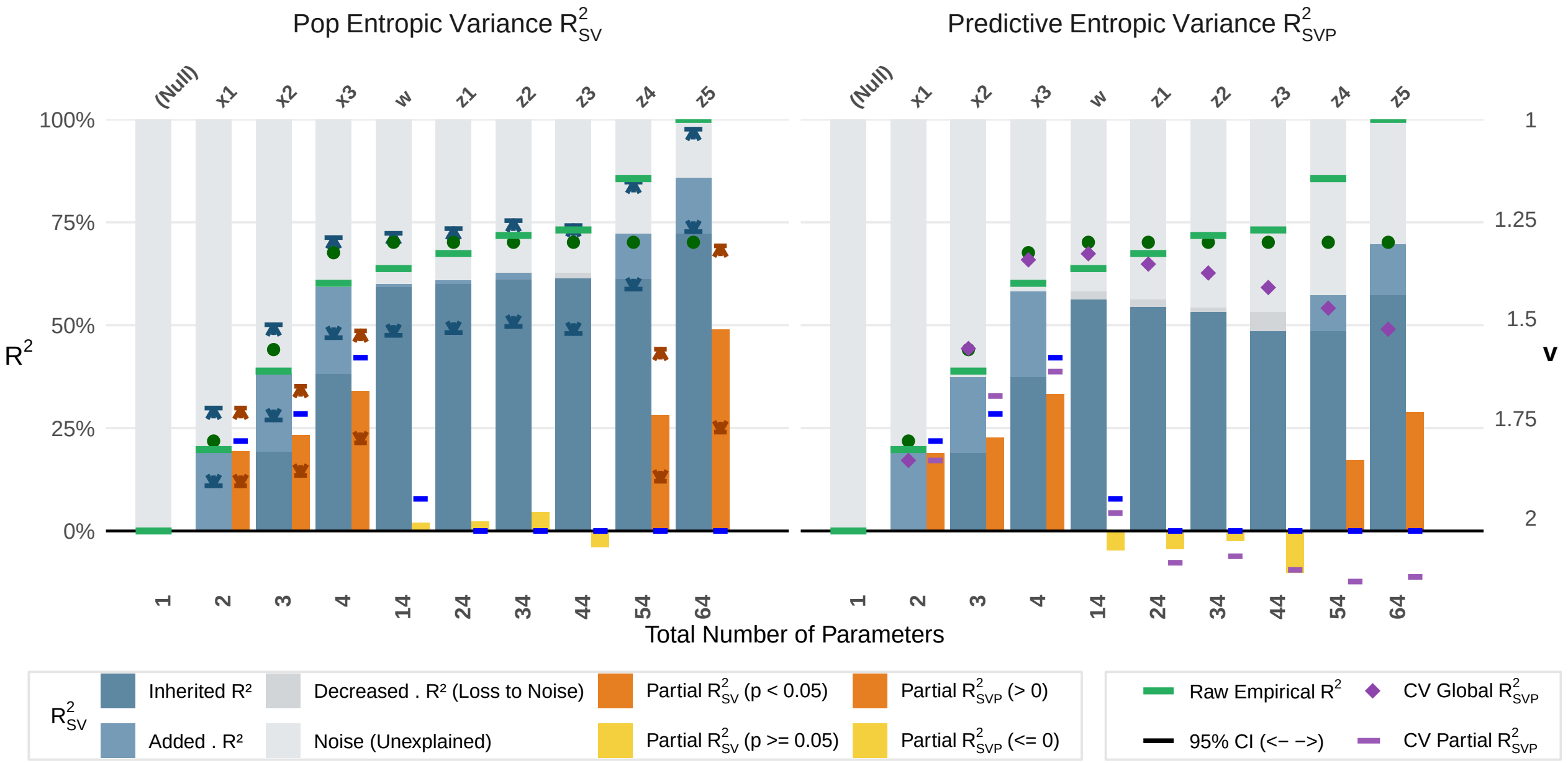

Figure S13: Comparison of estimated vs. true partial and global $\rho^2_{\text{SV}}$ for a single simulated dataset with prevalence 0.1.

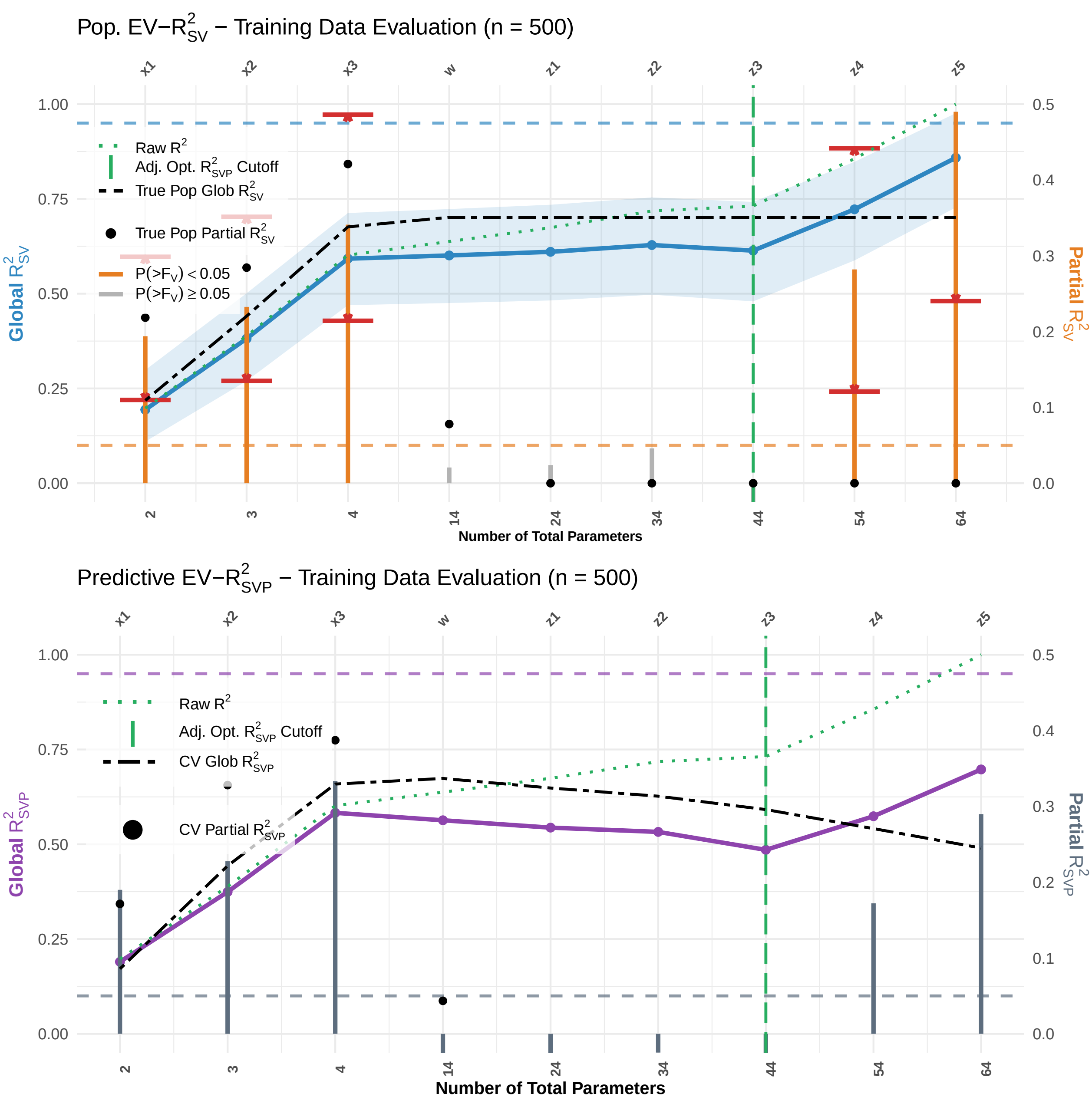

Figure S14: Comparison of estimated vs. true partial and global $\rho^2_{SV}$ for a single simulated dataset with prevalence 0.1.

## G.2 Supplementary Materials for Sec. 4.2 for Illustrating Internal Featuer Selection Bias

This trajectory illustrates the persistent, gradual inflation of explained variance characteristic of internal LASSO ordering. As the model sequentially incorporates predictors, the global EV-$R^2$ climbs steadily, reaching an upward bias of 0.95, while the raw $R^2$ converges to 1. The inability of standard degree-of-freedom corrections to penalize these metrics back to the true underlying $R^2_{SV}$ (approx. 0.75) highlights the systematic failure of internal validation. Each step in the selection process harvests incremental noise correlations that mirror the response, and because these corrections assume valid model-building rather than data-driven selection, they prove insufficient to counteract the cumulative optimistic bias.

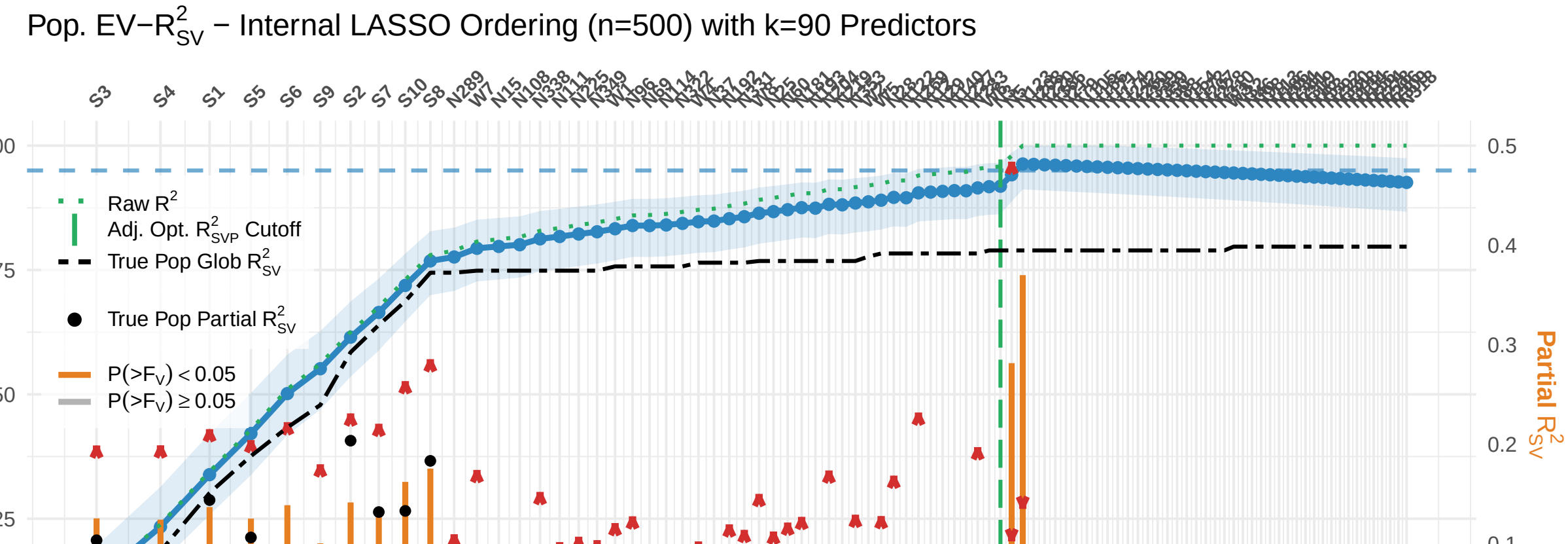

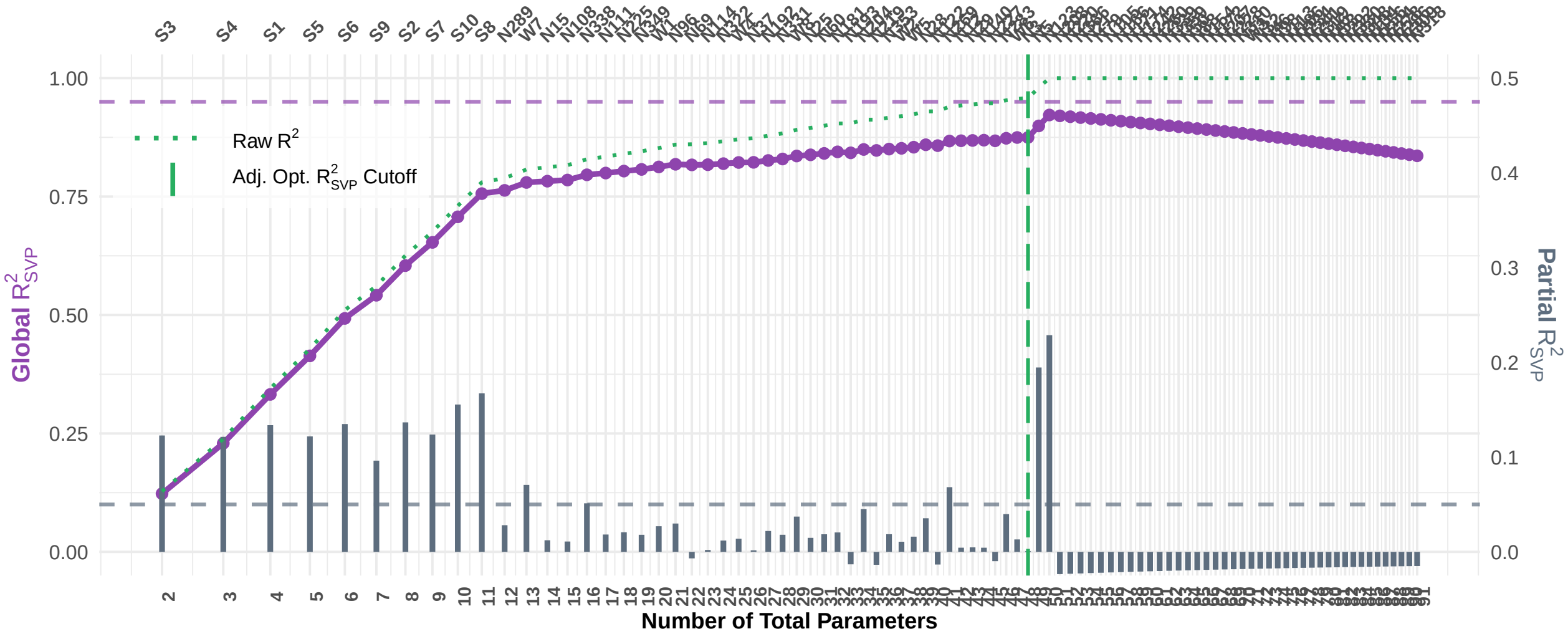

Figure S15: Trajectory of $R^2_{\text{SV}}$ metrics under internal LASSO ordering for a simulated dataset. The global EV-$R^2$ climbs gradually toward 0.95, while the raw $R^2$ approaches 1.0. Despite degrees-of-freedom (df) corrections, the $R^2_{\text{SV}}$ is still significantly higher than the truth ($\approx 0.75$), with failed CI coverage.

## G.3 Supplementary Materials for Sec. 5 to Show the Results of Applying Internal Evaluation to Parkinson Microbiome Data

The following supplementary figures detail the variable selection results when applying the internal LASSO ordering procedure to the Parkinson gut microbiome data. By utilizing the full dataset to both determine the predictor entry sequence and evaluate the corresponding selection criteria, this internal approach yields the specific microbial feature compositions illustrated below. As established in the primary text, empirical metrics derived under this single-sample regime are presented to illustrate the uncorrected, optimistic selection bias inherent to the internal high-dimensional search.

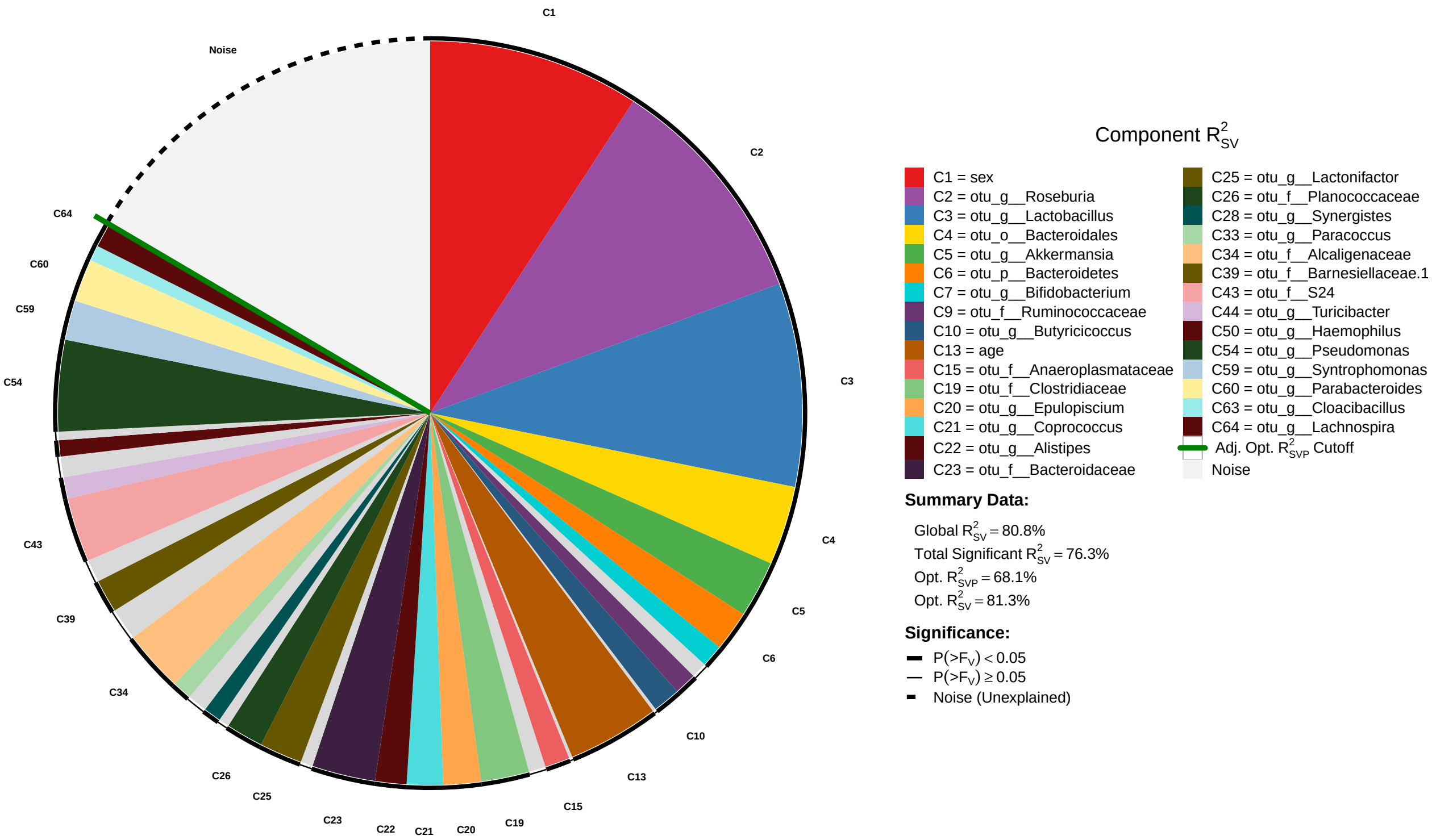

Figure S16: Component $R^2_{\text{SV}}$ contributions based on internal LASSO ordering. Note the inflated $R^2_{\text{SV}}$ (76%) and the high number of selected predictors, reflecting optimistic feature selection bias compared to the strict data-splitting framework.

The trajectory of internal LASSO-ordered $R^2_{SV}$ metrics (Fig. S17) reveals a persistent, gradual upward trend, rather than a plateau. This continuous, incremental gain in explained variance—even as the model complexity increases—is a hallmark of optimistic feature selection bias. In this internal framework, the selection process repeatedly exploits residual noise correlations that appear significant, leading to an artificially inflated performance ceiling of 76% that fails to replicate in out-of-sample data.

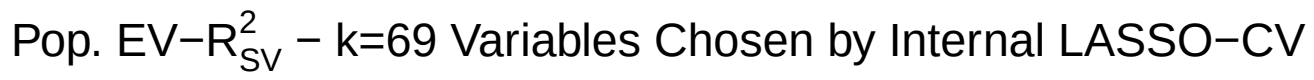

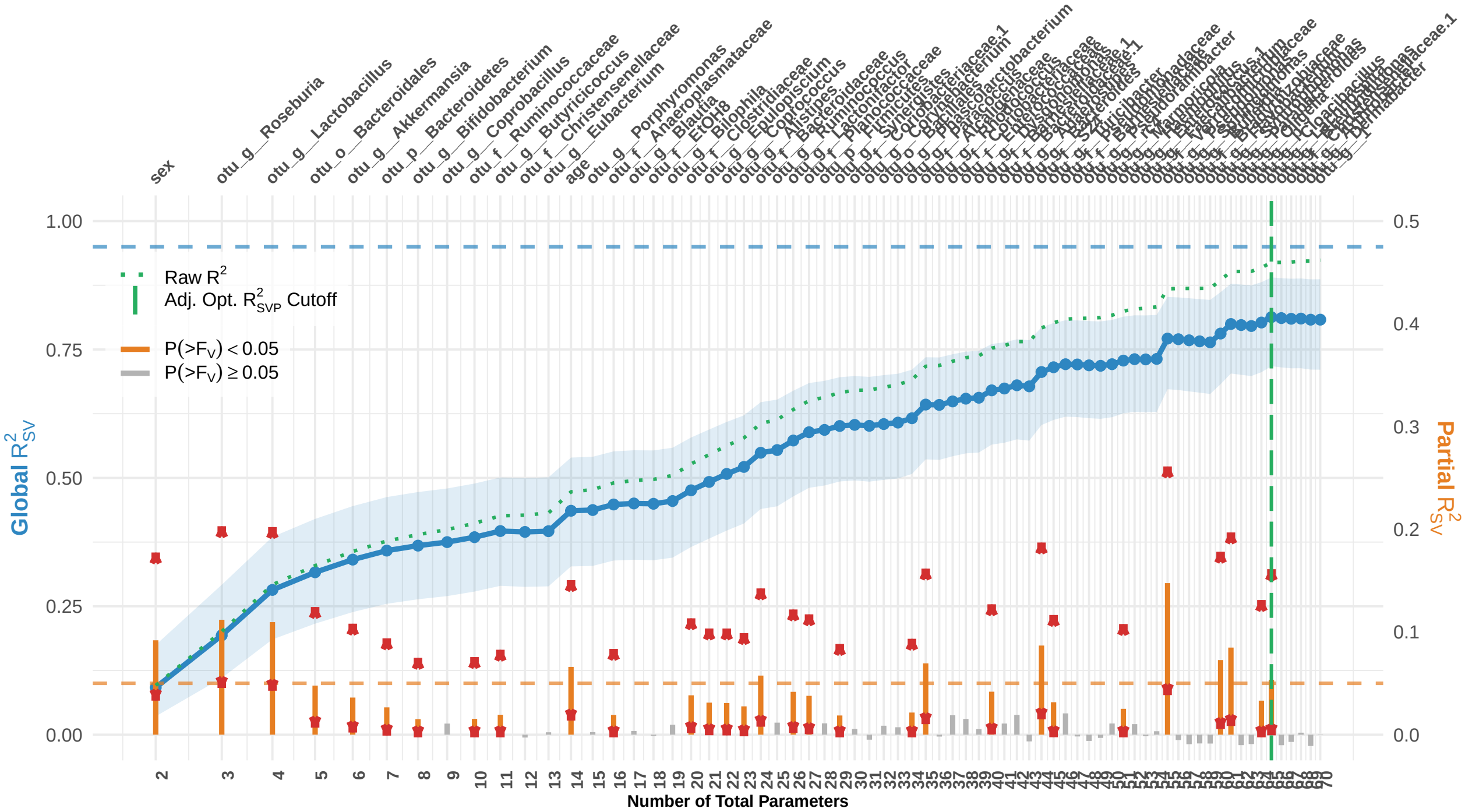

Predictive EV−$R^2_{SVP}$ − k=69 Variables Chosen by Internal LASSO−CV

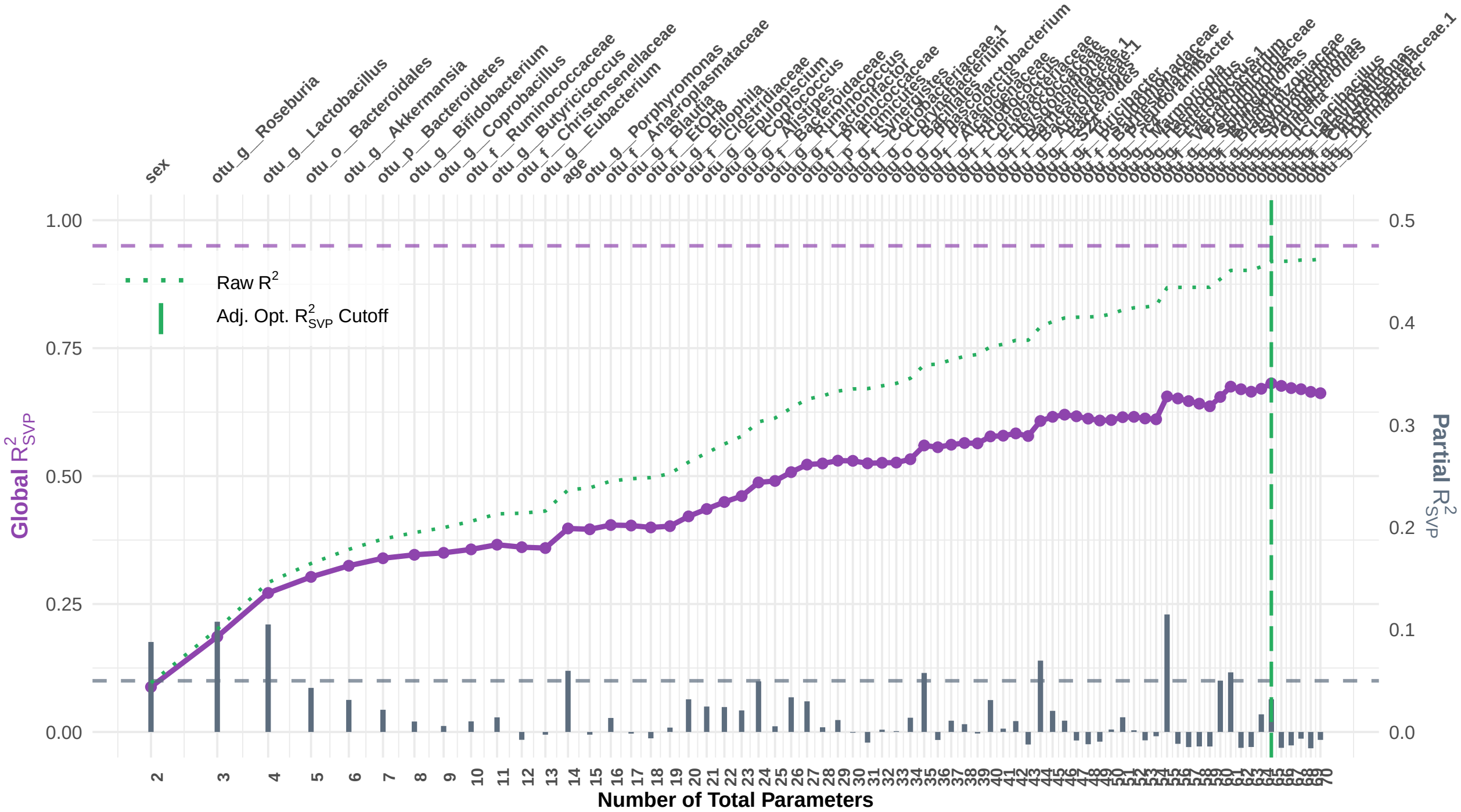

Figure S17: Trajectory of internal LASSO-ordered $R^2_{SV}$ metrics. The gradual increase in $R^2_{SV}$ toward 76% demonstrates the characteristic optimistic feature selection bias inherent in non-split internal ordering compared to the strict data-splitting framework.